\documentclass[a4paper,11pt]{article}
\pdfoutput=1 
\usepackage{jheppub} 
\usepackage[T1]{fontenc} 
\usepackage{multirow}
\usepackage{color}
\usepackage{ulem}
\usepackage{rotating}
\usepackage{booktabs}
\usepackage{dcolumn}
\usepackage{array} 
\usepackage{multirow}
\usepackage{lineno} 
\usepackage{siunitx}
\usepackage{amsmath}
\usepackage{bm}
\usepackage{mathrsfs}
\usepackage{graphicx}
\usepackage{comment}
\usepackage{float}
\DeclareSIUnit{\bmm}{\bm{m}}

\DeclareSIUnit{\clight}{\textnormal{\textit{c}}}

\newcolumntype{d}{D{.}{.}{-1}}
\newcolumntype{e}{D{.}{.}{8}}
\newcolumntype{f}{D{.}{.}{18}}
\newcolumntype{h}{D{.}{.}{13}}
\newcolumntype{g}{D{.}{.}{12}}

\title{\protect\boldmath Measurement of the cross sections of $\rm \EE\ar K^{-}\bar\Xi^{+}\Lambda/\Sigma^{0}$ at center-of-mass energies between $\mathbf{3.510}$ and \si{\mathbf{4.914}\,{\textbf{GeV}}}}

\collaborationImg{\includegraphics[width=0.25\textwidth]{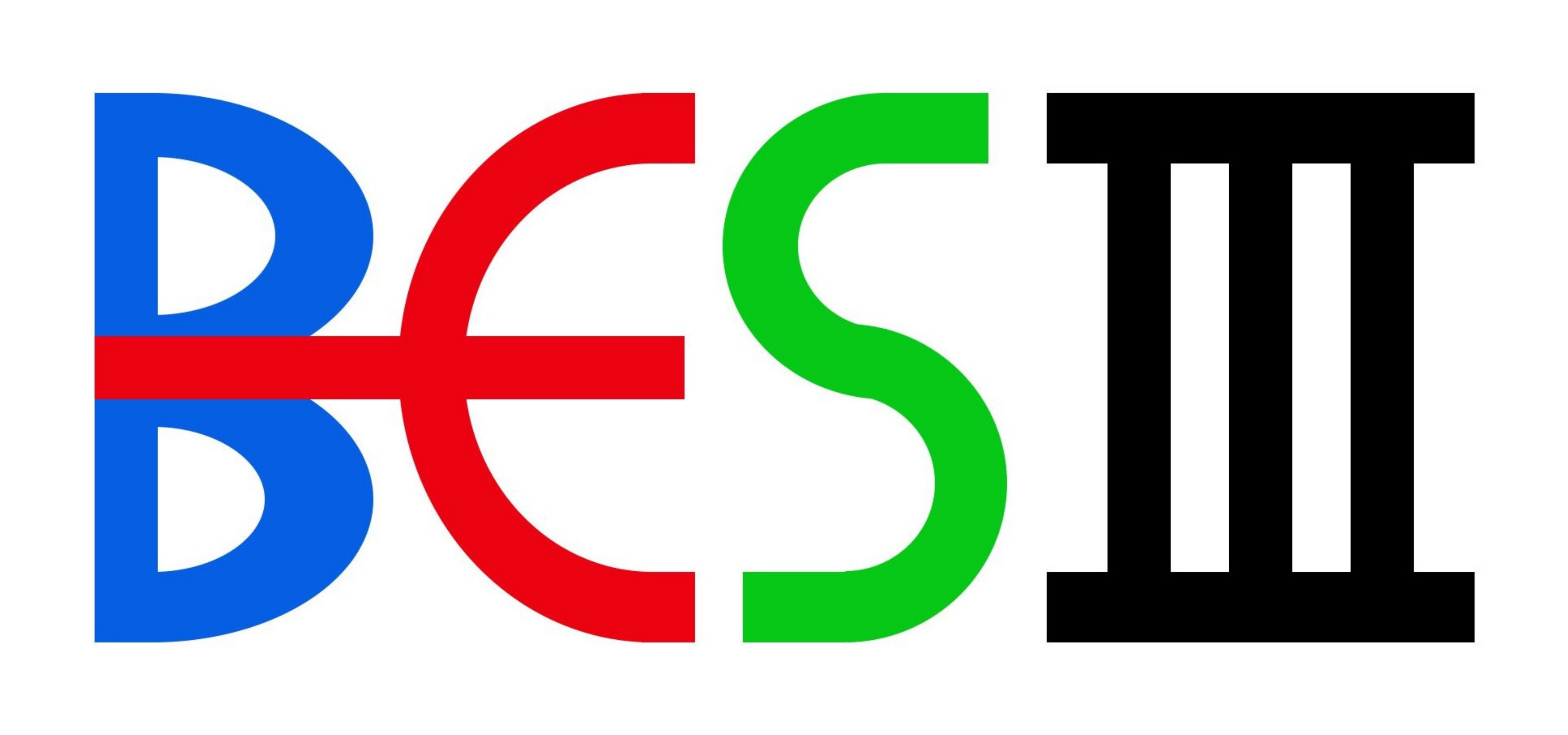}}
\collaboration{The BESIII Collaboration}

\emailAdd{besiii-publications@ihep.ac.cn}

\begin{document} 
\abstract{
Using $e^+e^-$ collision data collected with the BESIII detector at the BEPCII collider at center-of-mass energies between 3.510 and \SI{4.914}{GeV}, corresponding to an integrated luminosity of 25 fb$^{-1}$, we measure the Born cross sections for the process $\EE\ar K^-\bar{\Xi}^+\Lambda/\Sigma^{0}$ at thirty-five energy points with a partial-reconstruction strategy.
By fitting the dressed cross sections of $\EE \ar K^-\bar{\Xi}^+\Lambda/\Sigma^{0}$, evidence for $\psi(4160) \ar K^{-}\bar\Xi^{+}\Lambda$ is found for the first time with a significance of 4.4$\sigma$, including  systematic uncertainties. No evidence for other possible resonances is found.
In addition, the products of electronic partial width and branching fraction for all assumed resonances decaying into $K^{-}\bar\Xi^{+}\Lambda/\Sigma^{0}$ are determined.}

\clearpage{}
\newcommand{\Xib}{\bar\Xi^{+}}
\newcommand{\KXL}{\EE\ar K^{-}\bar\Xi^{+}\Lambda}
\newcommand{\KXS}{\EE\ar K^{-}\bar\Xi^{+}\Sigma^{0}}
\newcommand{\KXX}{\EE\ar K^{-}\bar\Xi^{+}\Lambda/\Sigma^{0}}
\newcommand{\EE}{e^+e^-}
\newcommand{\ar}{\rightarrow}
\newcommand{\bbt}{\bibitem}

\clearpage{}
\maketitle
\flushbottom
%\linenumbers     
\section{Introduction}
\label{sec:intro}
Studies of baryon-pair decays from vector ($J^{PC}=1^{--}$) charmonium(-like) resonances provide a testing ground for quantum chromodynamics~\cite{Brambilla:2010cs, Briceno:2015rlt}.
Below the open-charm threshold, the mass spectrum of the observed charmonium states is well-matched to the predictions of the potential quark model~\cite{Barnes:2005pb}. Above the open-charm threshold, the quark model predicts six vector charmonium states between the threshold to \SI{4.9}{GeV/{\clight}^{2}}, namely, the $1D$, $3S$, $2D$, $4S$, $3D$, and $5S$ states. 
However, the experimentally observed vector states in this 
energy region are overpopulated.
The decays of the three states, $\psi(4040)$, $\psi(4160)$, and $\psi(4415)$, observed from the inclusive hadronic cross sections are dominated by open-charm processes~\cite{BES:2001ckj}. The other states, such as $Y(4230)$, $Y(4360)$, and $Y(4660)$, have been observed through  hidden-charm final states, via initial-state radiation (ISR) processes at \textsc{BaBar} and Belle~\cite{BaBar:2005hhc, Belle:2007dxy, BaBar:2006ait, Belle:2007umv, Belle:2008xmh, BaBar:2012hpr, Belle:2013yex, Belle:2014wyt, BaBar:2012vyb} or direct-production processes in  $e^+e^-$ annihilation at CLEO~\cite{CLEO} and BESIII~\cite{BESIIIAB, BESIII:2023cmv}.
These $Y$ states do not appear to be resonances with simple $c\bar{c}$ quark content, and many theoretical models, such as hybrid, multiple-quark state, and molecule, etc, have been proposed to interpret them~\cite{Farrar,Briceno,Chen:2016qju, Wang:2019mhs, Close:2005iz,
Qian:2021neg}. However, no solid conclusion has yet emerged and the true nature of these states remains a puzzle. 
This status reflects our poor understanding of  the behaviour of the strong interaction in the non-perturbative regime. To make progress, more high-precision  measurements  are required.  Among these measurements, studies of the baryonic decays of charmonium (-like) states hold particular promise due to the simple topologies of the final states and relatively well understood mechanisms.
Although many experimental studies of baryonic processes have been performed by the Belle and BESIII experiments~\cite{Belle:2008xmh,BESIII:2021ccp, Ablikim:2019kkp, Ablikim:2013pgf,  BESIII:2017kqg,Wang:2021lfq,Wang:2022bzl,BESIII:2023rse, BESIII:2024umc, BESIII:2022kzc, Wang:2022zyc}, only one observation of $\psi(4660)\to\Lambda^{+}_{c}\bar\Lambda^{-}_{c}$~\cite{Belle:2008xmh} and two evidences for the decays $\psi(3770)\to\Lambda\bar\Lambda$ and $\Xi^-\bar\Xi^+$~\cite{BESIII:2021ccp, BESIII:2023rse} were reported by Belle and BESIII experiments.
More precise measurements on the cross sections of the $e^+e^-\to $ baryonic exclusive processes above the open-charm threshold are desirable as they may provide additional information to understand the nature of these vector charmonium (-like) states.

In this article, a measurement of the Born cross sections for the processes $\EE\ar K^-\bar{\Xi}^+\Lambda/\Sigma^{0}$ is presented using $e^+e^-$ collision data corresponding to a total integrated luminosity of \SI{25}{fb^{-1}} collected at center-of-mass (CM) energies $\sqrt{s}$ between 3.510 and \SI{4.914}{GeV}~\cite{ene1, ene3} with the BESIII detector~\cite{besiii} at the BEPCII collider~\cite{BEPCII}. In addition, vector resonances are searched for by fitting the dressed cross sections of $\EE\ar K^-\bar{\Xi}^+\Lambda/\Sigma^{0}$.  

\section{BESIII detector and Monte Carlo simulation}
The BESIII detector~\cite{besiii} records symmetric $e^+e^-$ collisions 
provided by the BEPCII storage ring~\cite{BEPCII}
in the CM energy range of 2.00 to \SI{4.95}{GeV},
with a peak luminosity of \SI{1e33}{\per\centi\meter\squared\per\second}
achieved at $\sqrt{s} =$ \SI{3.77}{GeV}. 
BESIII has collected large data samples in this energy region~\cite{Ablikim:2019hff, EcmsMea, EventFilter}. The cylindrical core of the BESIII detector covers 93\% of the full solid angle and consists of a helium-based
 multilayer drift chamber~(MDC), a plastic scintillator time-of-flight
system~(TOF), and a CsI(Tl) electromagnetic calorimeter~(EMC),
which are all enclosed in a superconducting solenoidal magnet
providing a \SI{1.0}{T} magnetic field. The solenoid is supported by an
octagonal flux-return yoke with resistive plate counter muon
identification modules interleaved with steel. 
The charged-particle momentum resolution at \SI{1}{GeV/\clight} is
$0.5\%$, and the 
${\rm d}E/{\rm d}x$
resolution is $6\%$ for electrons
from Bhabha scattering. The EMC measures photon energies with a
resolution of $2.5\%$ ($5\%$) at \SI{1}{GeV} in the barrel (end-cap)
region. The time resolution in the TOF barrel region is \SI{68}{ps}, while
that in the end-cap region is \SI{110}{ps}. The end-cap TOF
system was upgraded in 2015 using multigap resistive plate chamber
technology, providing a time resolution of \SI{60}{ps}~\cite{etof1, etof2, etof3}.

Simulation samples produced with a {\sc geant4}-based~\cite{GEANT4} Monte Carlo (MC) package, which includes the geometric description of the BESIII detector~\cite{Huang:2022wuo} and the detector response, are used to determine detection efficiencies and estimate backgrounds. The simulation models the beam-energy spread and ISR in the $e^+e^-$ annihilation using the generator {\sc kkmc}~\cite{KKMC}.
The inclusive MC sample includes the production of hadron processes, ISR production of the $J/\psi$, and the continuum processes incorporated in {\sc kkmc}~\cite{KKMC}.
The detection efficiency of $\EE\ar K^-\bar{\Xi}^+\Lambda/\Sigma^{0}$ is determined by MC simulations. A sample of 200,000 signal events is simulated with a phase-space (PHSP) distribution for each energy point, where the $\bar\Xi^{+}$ baryon and its subsequent decays to $\bar\Lambda \pi^+$ are described by the {\sc evtgen} program~\cite{evtgen2,EVTGEN} with a PHSP model.

 \section{Event selection}
A partial-reconstruction technique is employed to select the $\KXX$ candidate events, where the $\Xib$ baryon is reconstructed by the $\bar\Lambda\pi^+$ mode with the subsequent decay $\bar\Lambda\to \bar p\pi^+$, and
the $\Lambda/\Sigma^{0}$ is inferred from the recoiling system against the reconstructed $K^-\Xib$ system. Throughout this article, unless explicitly stated, the charge-conjugate state is always implied.

The selection criteria for charged particle tracks in the MDC are as follows: the charged tracks detected in the MDC are required to be within a polar angle range of $|\!\cos\theta| < 0.93$, where $\theta$ is defined with respect to the $z~$axis which is the symmetry axis of the MDC. Due to the implementation of the partial-reconstruction strategy, at least two positively charged tracks and two negatively charged tracks are required. 
These tracks are required to be well reconstructed in the MDC with good helix fits.
In order to identify charged particles, a likelihood-based particle identification (PID) method is employed. This method combines measurements of the energy loss in the MDC (d$E$/d$x$) and the time of flight in the TOF to form likelihoods ${\cal L}(h)$ $(h = p, K, \pi)$ for each hadron $h$ hypothesis. Tracks are identified as protons when ${\cal L}(p) > {\cal L}(K)$ and ${\cal L}(p) > {\cal L}(\pi)$, while charged kaons and pions are identified if ${\cal L}(K) > {\cal L}(\pi)$ and ${\cal L}(\pi) > {\cal L}(K)$ are satisfied, respectively. Only the events that contain at least two $\pi^+$, one $\bar{p}$ and one $K^-$ are retained for further analysis. 

The reconstruction of $\bar \Lambda$ and $\Xib$ decays follows the procedures reported in refs.~\cite{BESIII:2012ghz, BESIII:2021cvv, BESIII:2023euh}. Briefly, to reconstruct $\bar\Lambda$ candidates and suppress non-$\bar\Lambda$ background, a secondary-vertex fit~\cite{vtxfit} is implemented for the $\bar p\pi^+$ combinations, and
the decay length of the $\bar \Lambda$ candidate from the fit, i.e. the distance between its production and decay positions, is required to be greater than zero to suppress the background from non-$\bar \Lambda$ events.
The $\bar{p}\pi^+$ invariant mass is required to be within a window of \SI{\pm8}{MeV/\clight^{2}} of the known $\bar \Lambda$ mass. This criterion is determined by optimizing the figure of merit (FOM) after choosing the best candidates, defined as $S/\sqrt{S+B}$, where $S$ is the number of signal MC events and $B$ is the number of the background events estimated with the inclusive MC simulation.  The $\Xib$ candidates are reconstructed using a similar secondary-vertex fit from each combination of the remaining $\pi^+$ and reconstructed $\bar \Lambda$. The best $\Xib$ and $\bar \Lambda$ candidates are kept by minimizing the combined-mass difference $|M_{\pi^+\bar \Lambda}-m_{\Xib}| + |M_{\pi^+\bar p}-m_{\bar \Lambda}|$, where $M_{\pi^+\bar\Lambda}$ and $M_{\pi^+\bar p}$ are the invariant masses of the $\pi^+\bar\Lambda$ and $\pi^+\bar p$ combinations, respectively, and $m_{\Xib}$ and $m_{\bar \Lambda}$ are the known masses of the $\Xib$ and $\bar \Lambda$ baryons from the Particle Data Group (PDG)~\cite{PDG2020}. Moreover, the $\Xib$ signal region in the $M_{\pi^+\bar \Lambda}$ distribution is determined by optimizing the FOM and defined as lying within a window of \SI{\pm6}{MeV/\clight^{2}} of the known $\Xib$ mass. The decay length of the $\Xib$ candidate also needs to be greater than zero.

To be sensitive to the presence  of signal candidates, a kinematic variable of mass recoiling against the selected $K^-\Xib$ is defined as
\begin{equation}
	M^{\rm recoil}_{K^-\Xib} = \sqrt{(\sqrt{s}-E_{K^-\Xib})^{2} - |\vec{p}_{K^-\Xib}|^{2}},
\end{equation}
where $E_{K^-\Xib}$ and $\vec{p}_{K^-\Xib}$ are the energy and momentum of the selected $K^-\Xib$ candidates in the $e^+e^-$ CM frame, and $\sqrt{s}$ is the CM energy. 

\section{Born cross-section measurement}
\subsection{Extraction of signal yields}
\begin{figure}[!htbp]
    \begin{center}
    \includegraphics[width=0.22\textwidth]{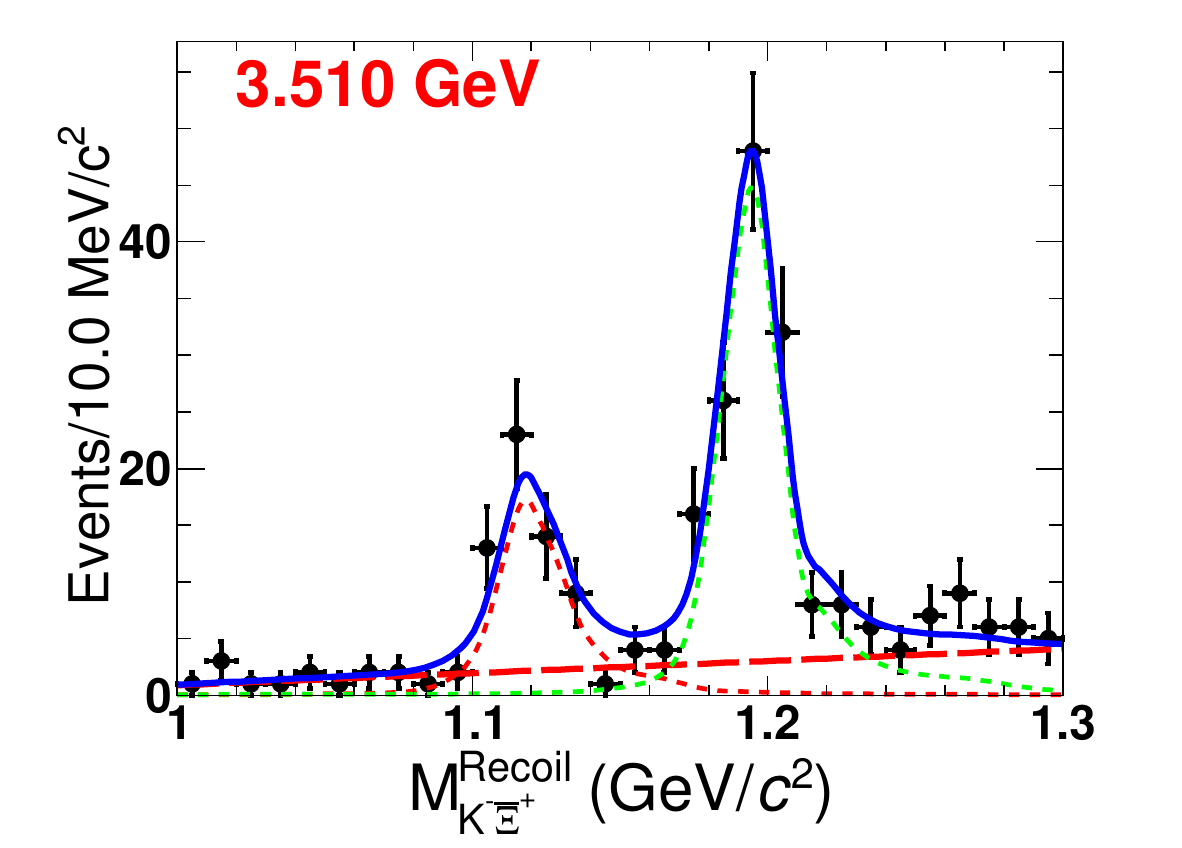}
    \includegraphics[width=0.22\textwidth]{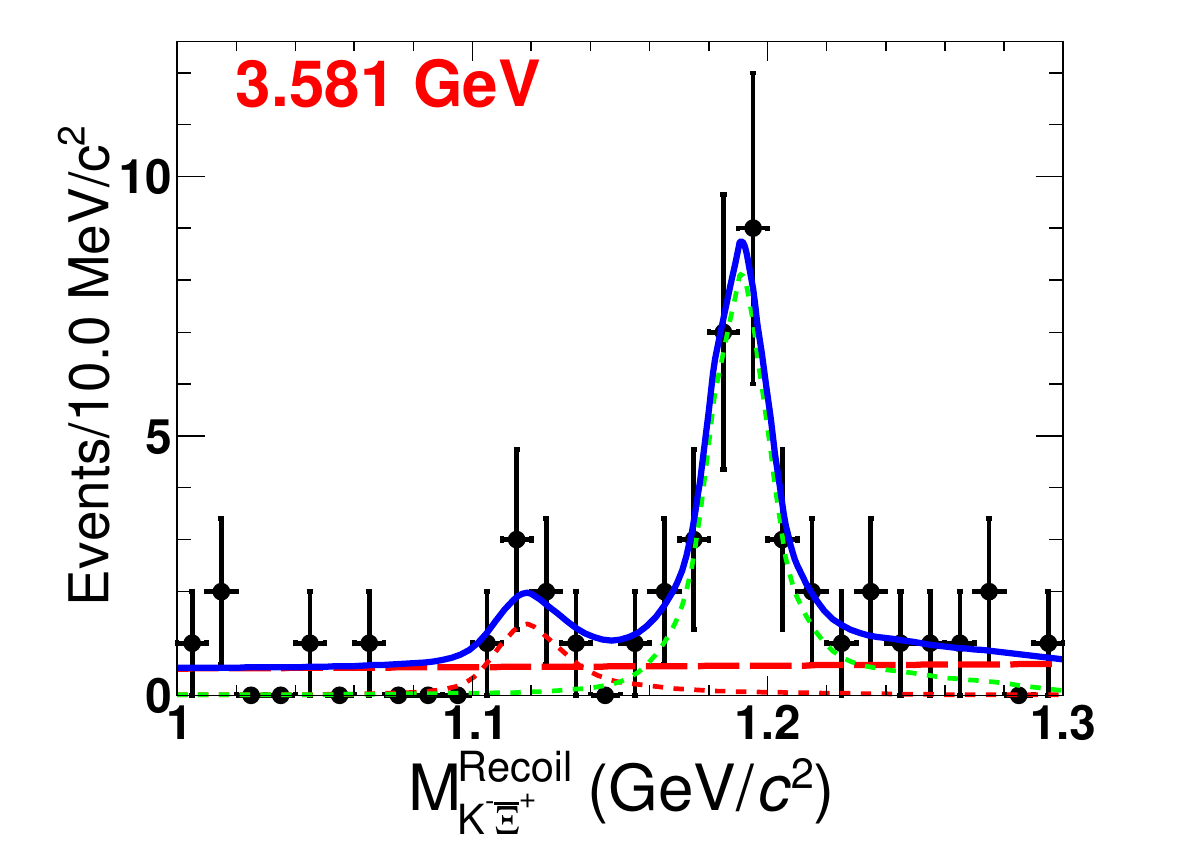}
    \includegraphics[width=0.22\textwidth]{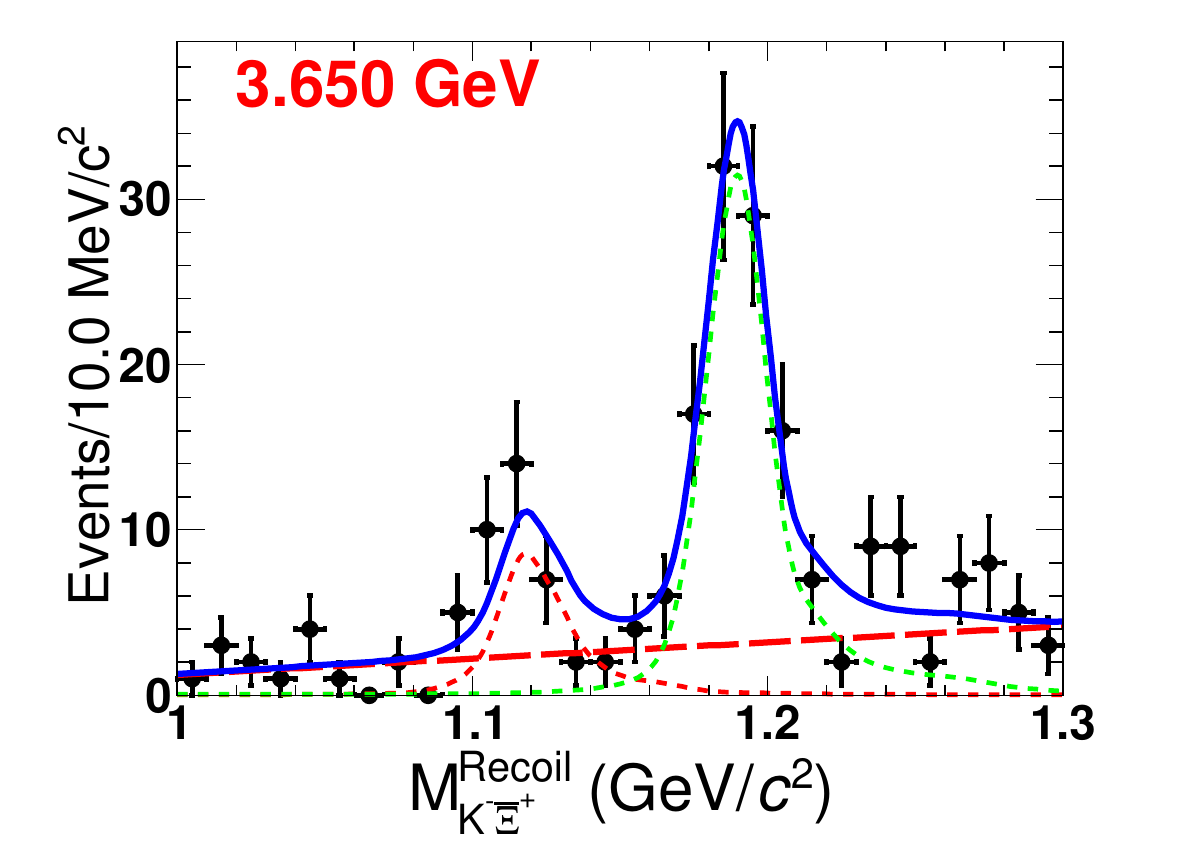}
    \includegraphics[width=0.22\textwidth]{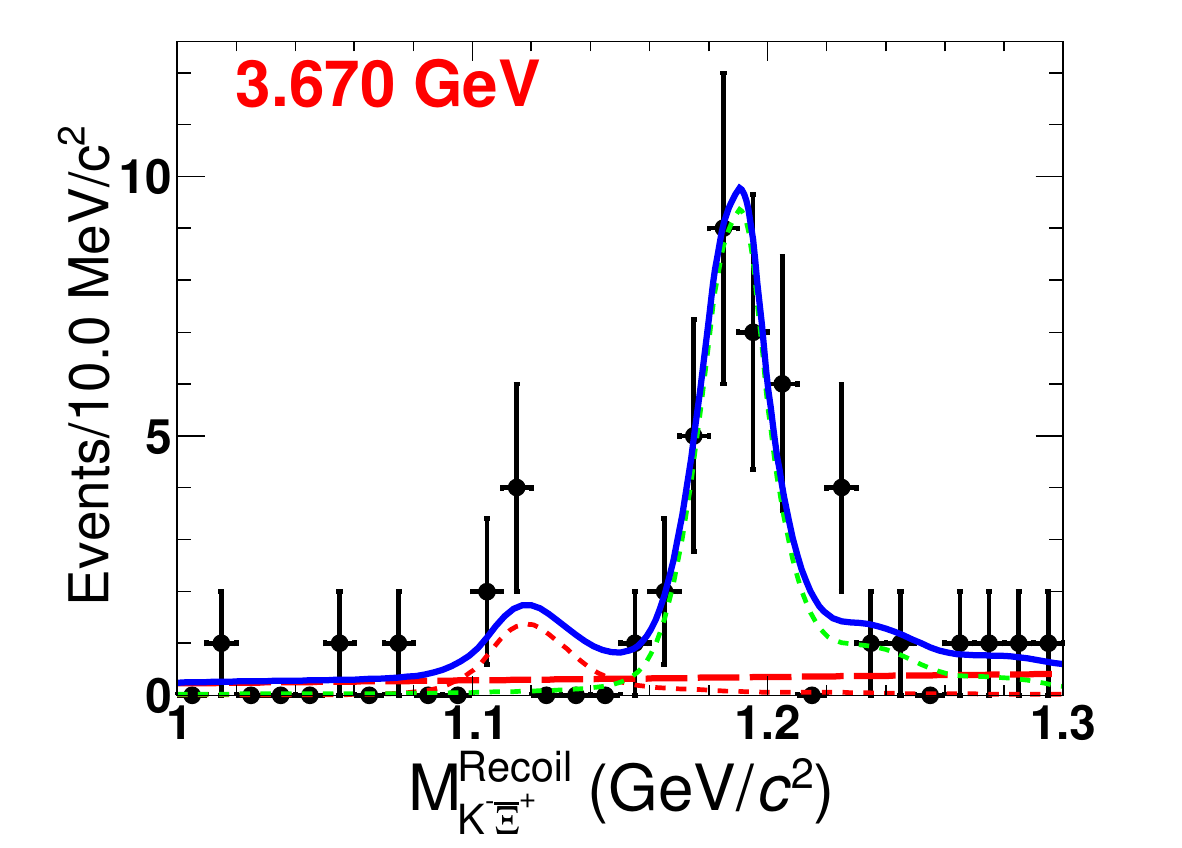}\\
    \includegraphics[width=0.22\textwidth]{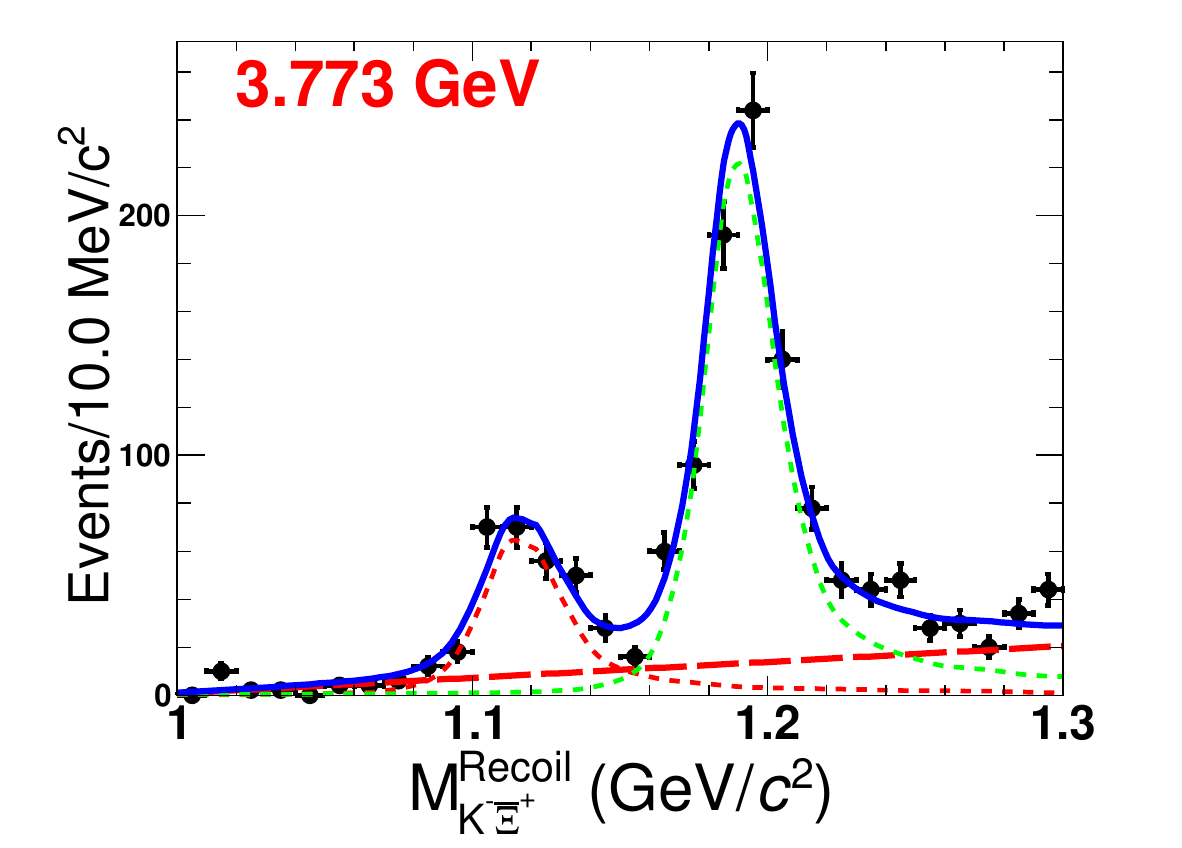}
   \includegraphics[width=0.22\textwidth]{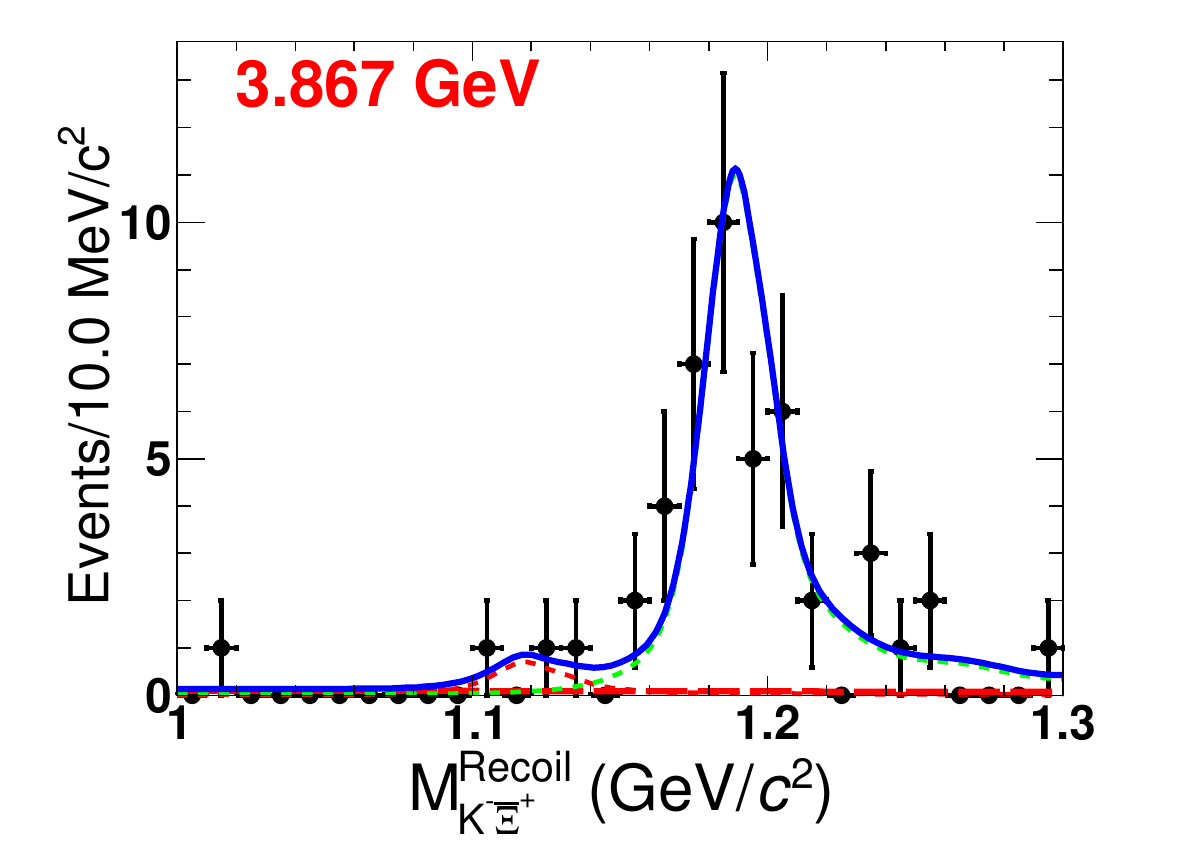}
    \includegraphics[width=0.22\textwidth]{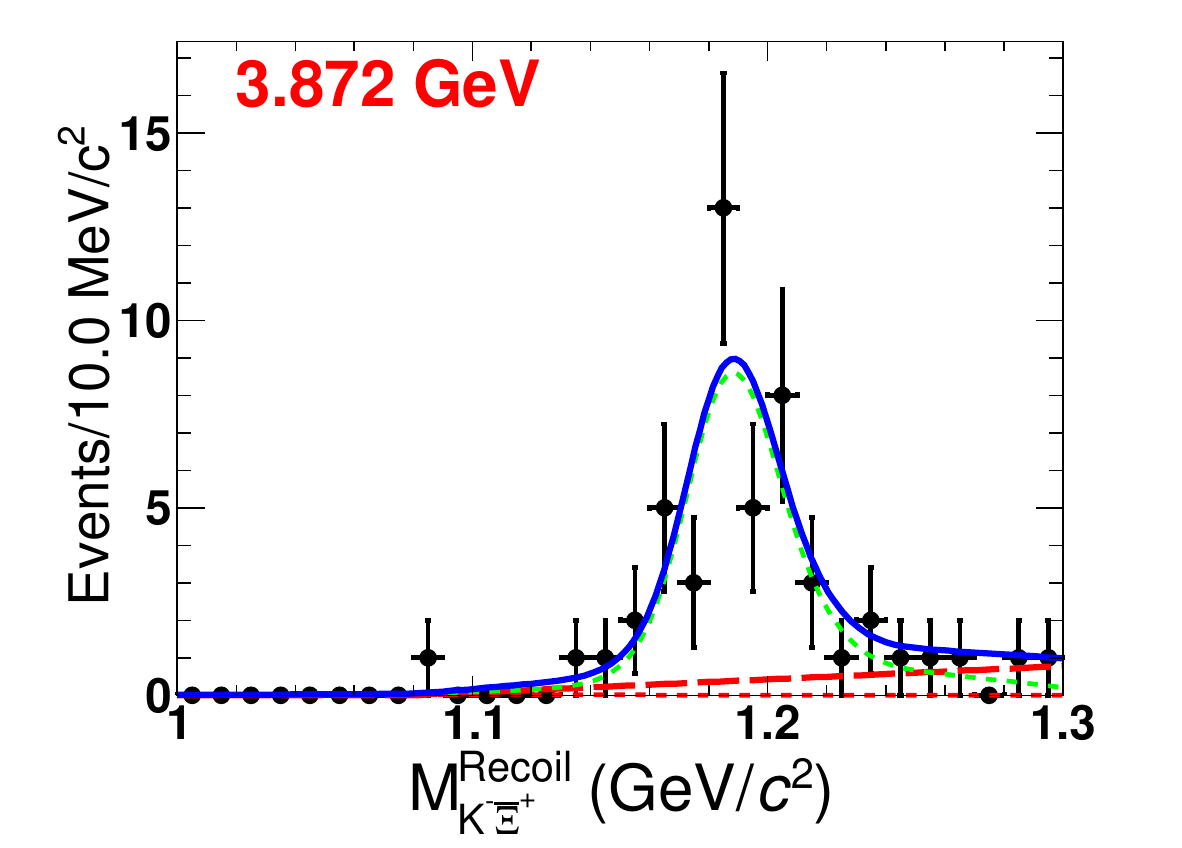}
    \includegraphics[width=0.22\textwidth]{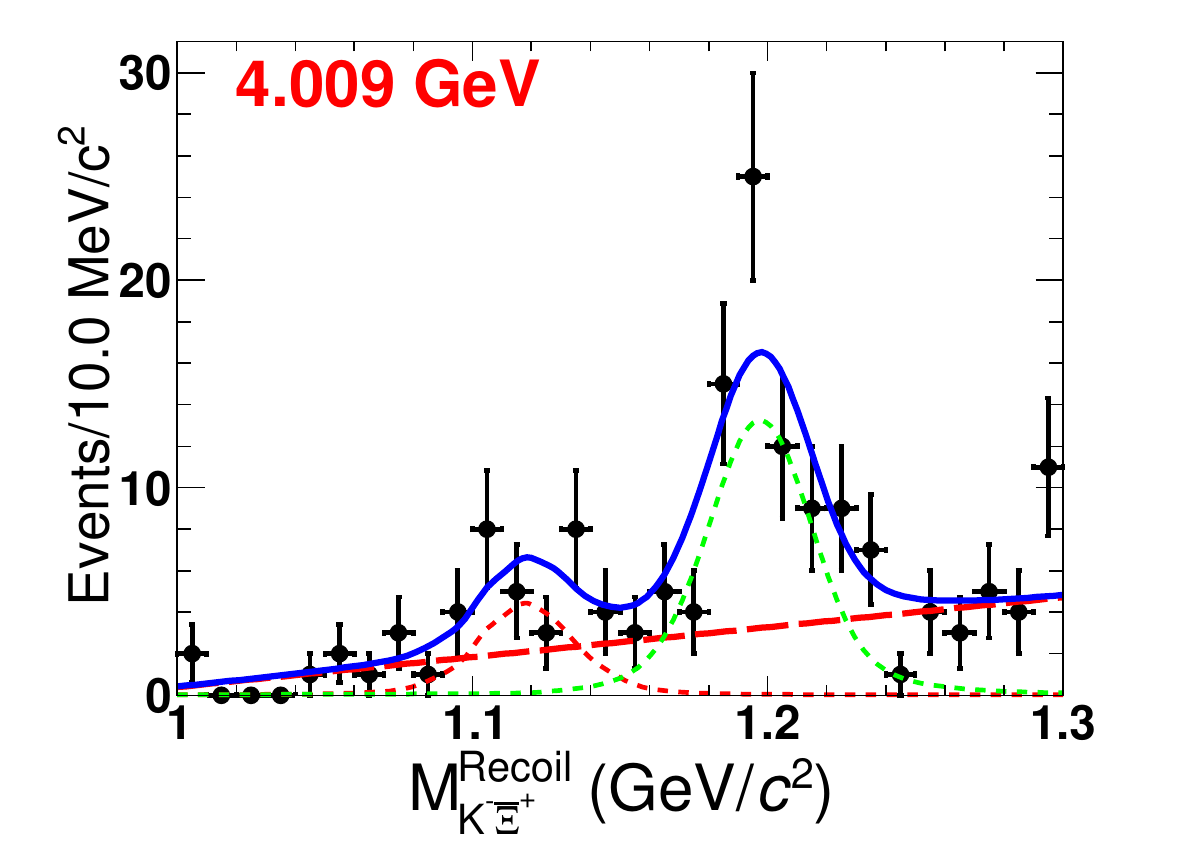}\\
    \includegraphics[width=0.22\textwidth]{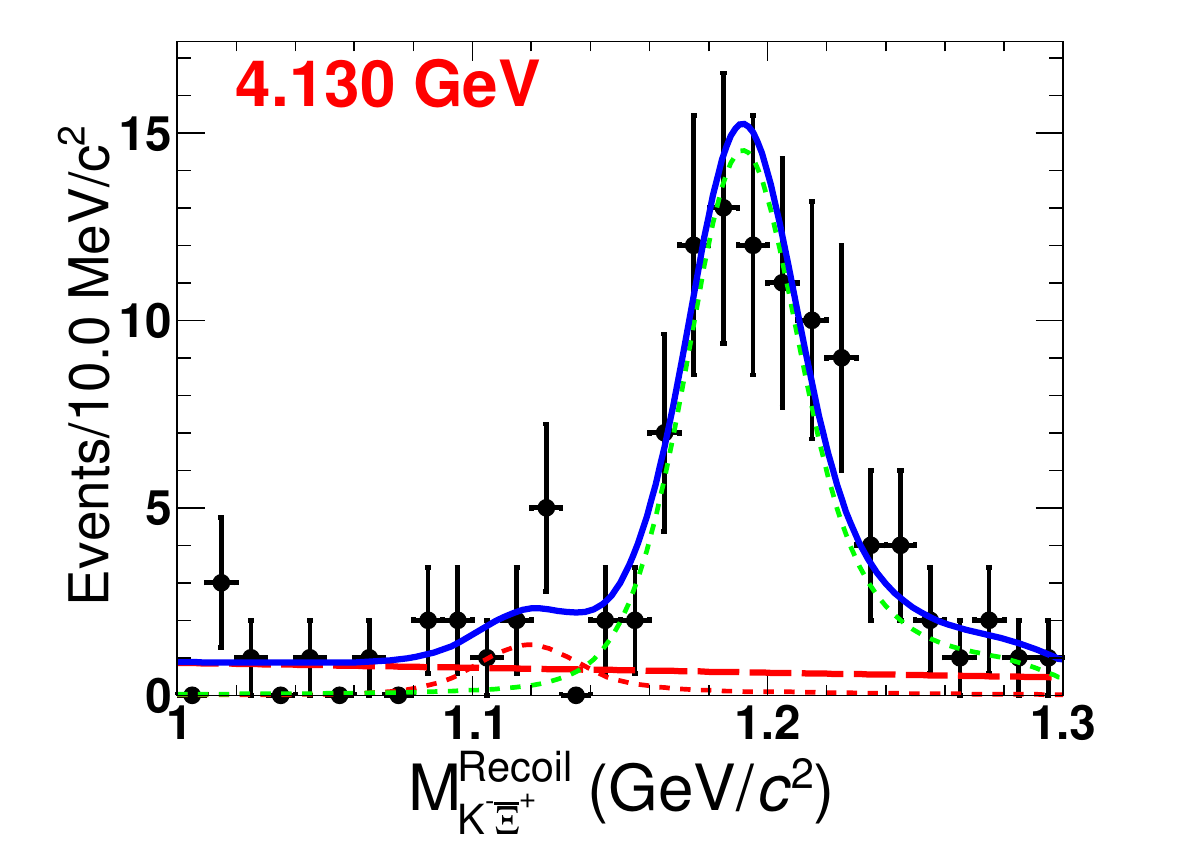}
    \includegraphics[width=0.22\textwidth]{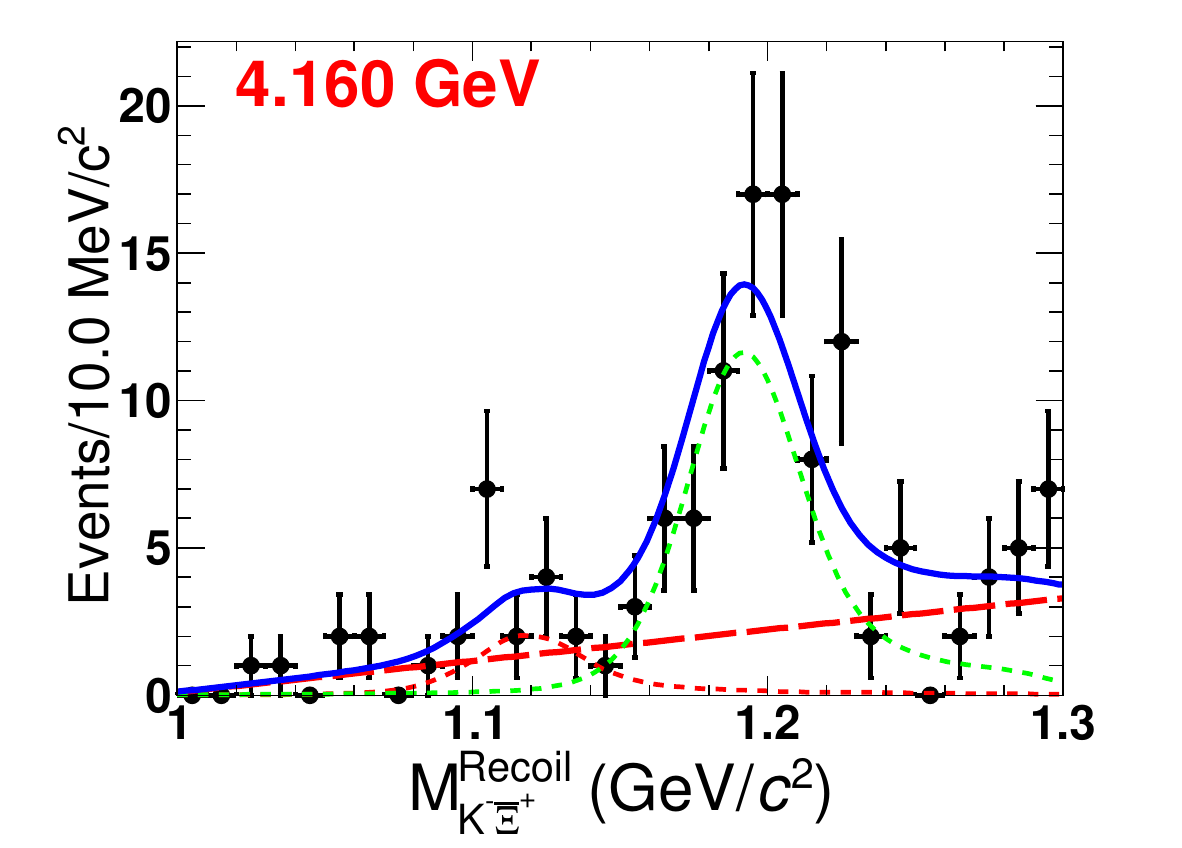}
    \includegraphics[width=0.22\textwidth]{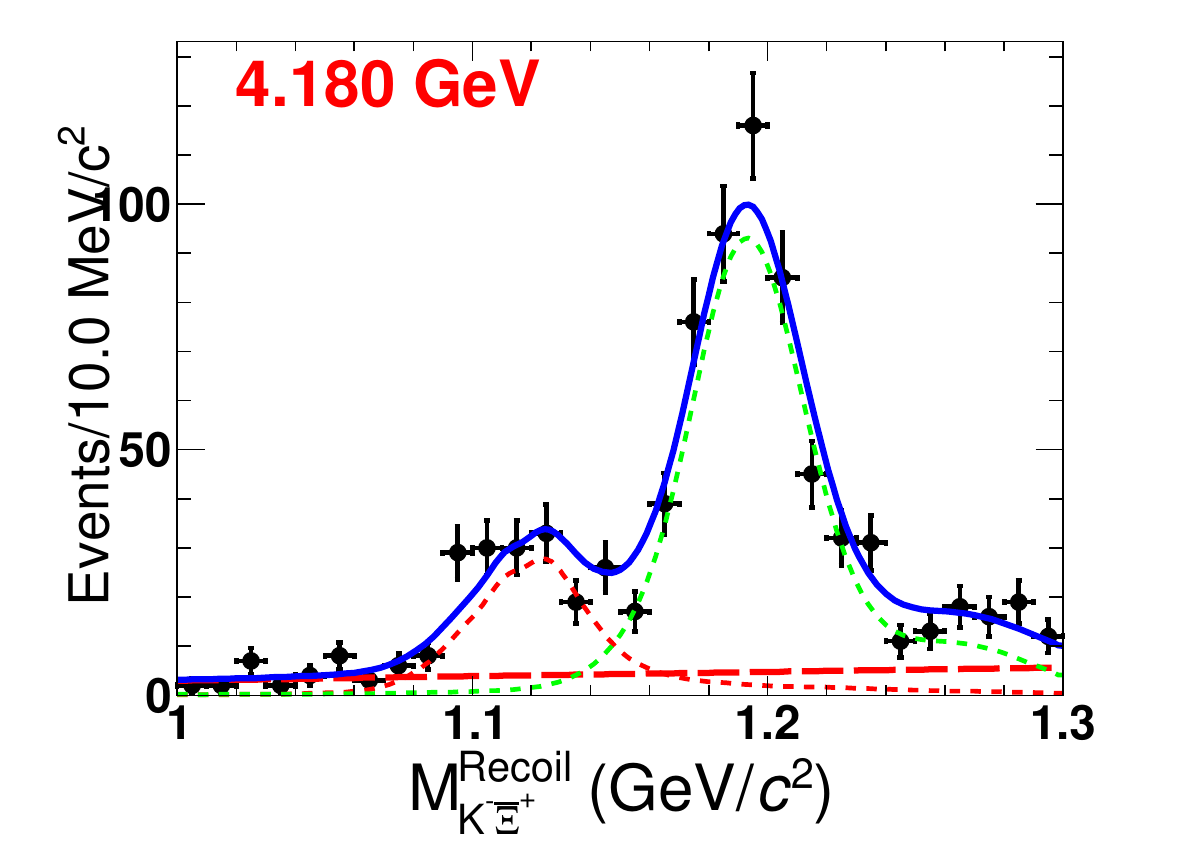}
    \includegraphics[width=0.22\textwidth]{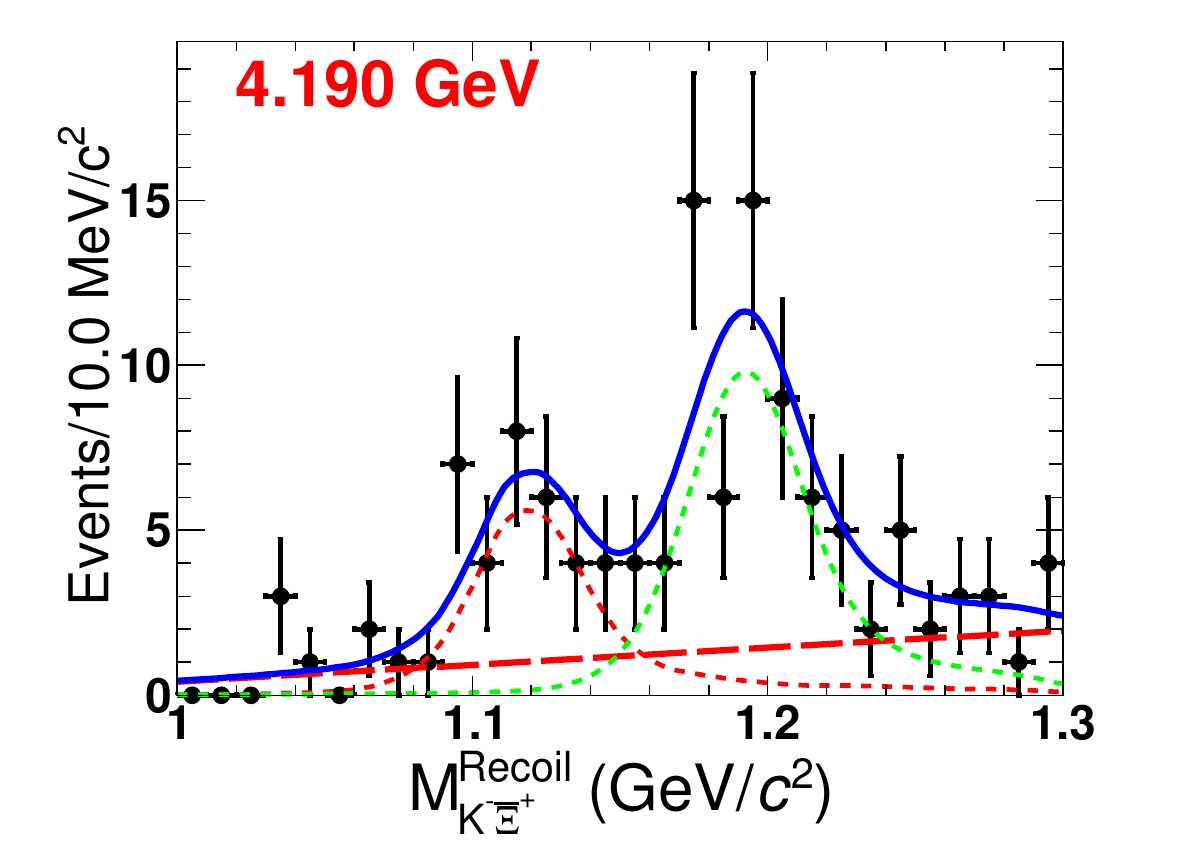}\\
    \includegraphics[width=0.22\textwidth]{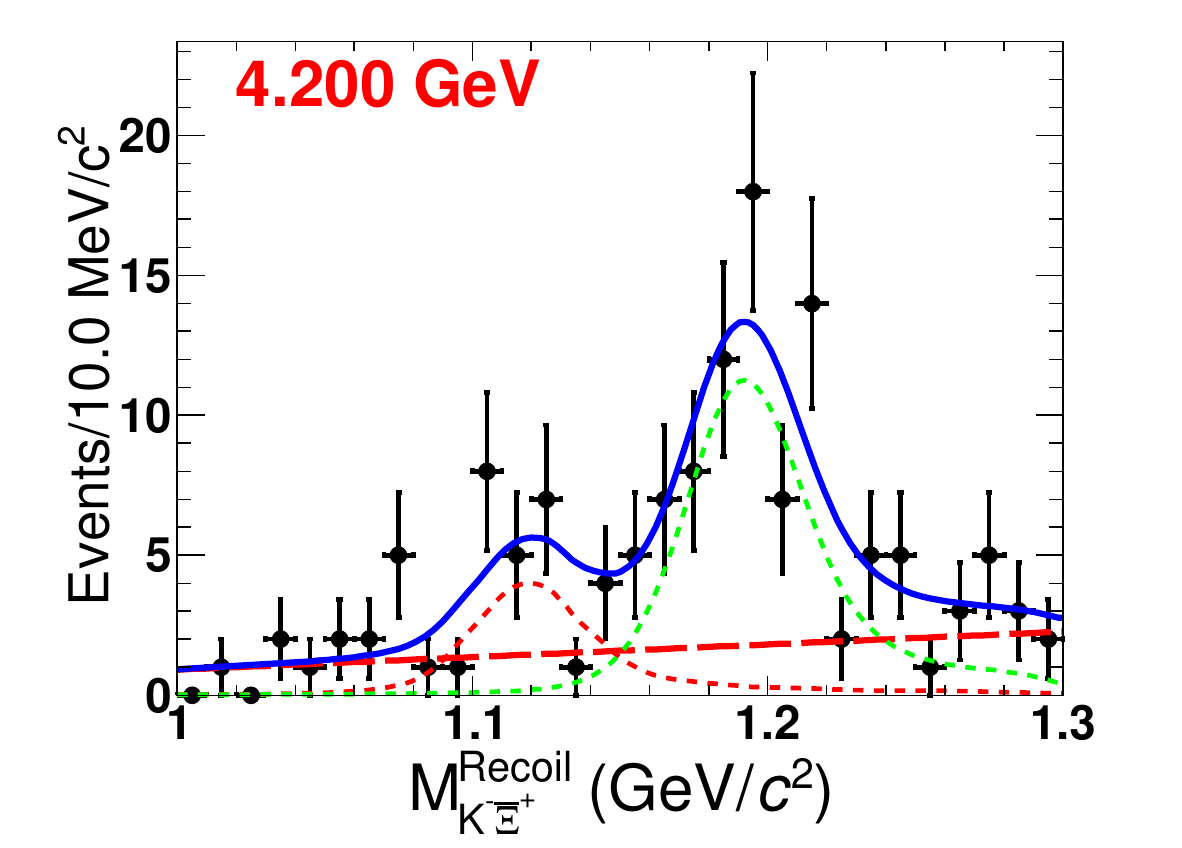}
    \includegraphics[width=0.22\textwidth]{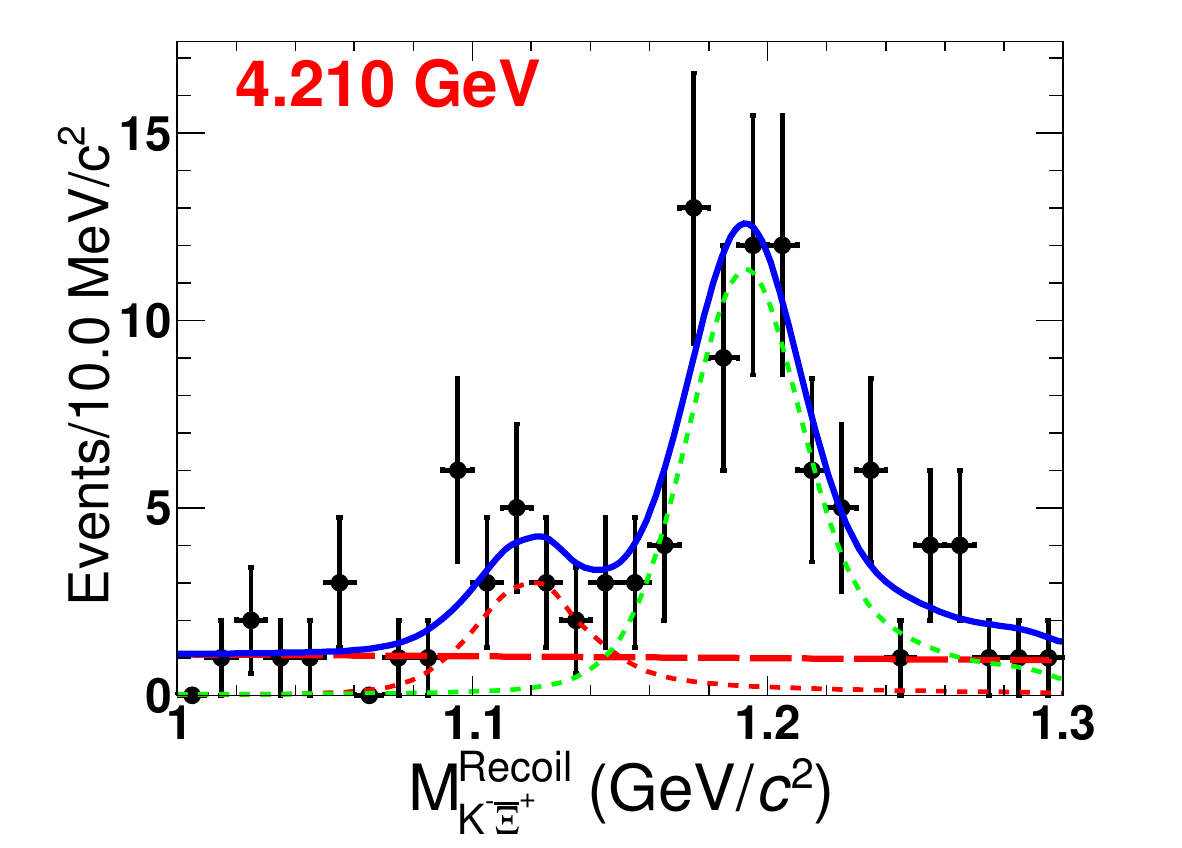}
    \includegraphics[width=0.22\textwidth]{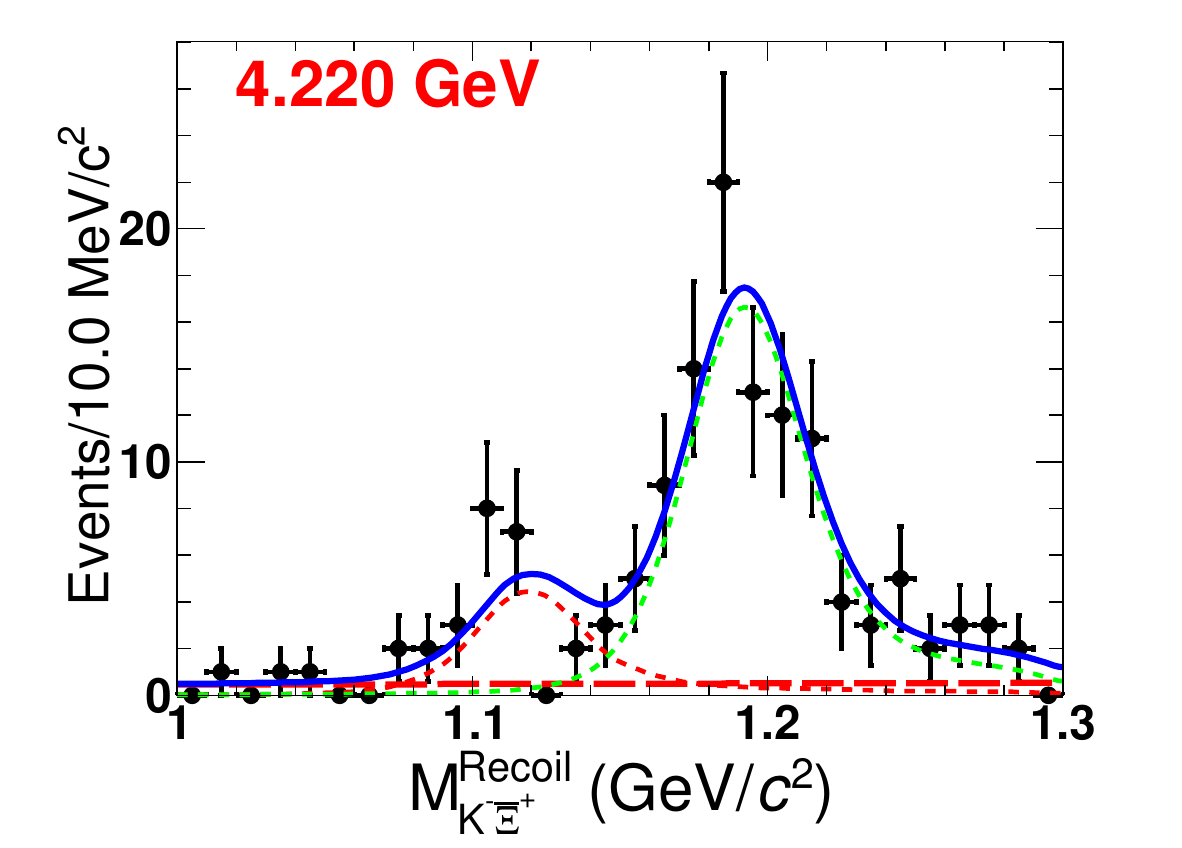}
    \includegraphics[width=0.22\textwidth]{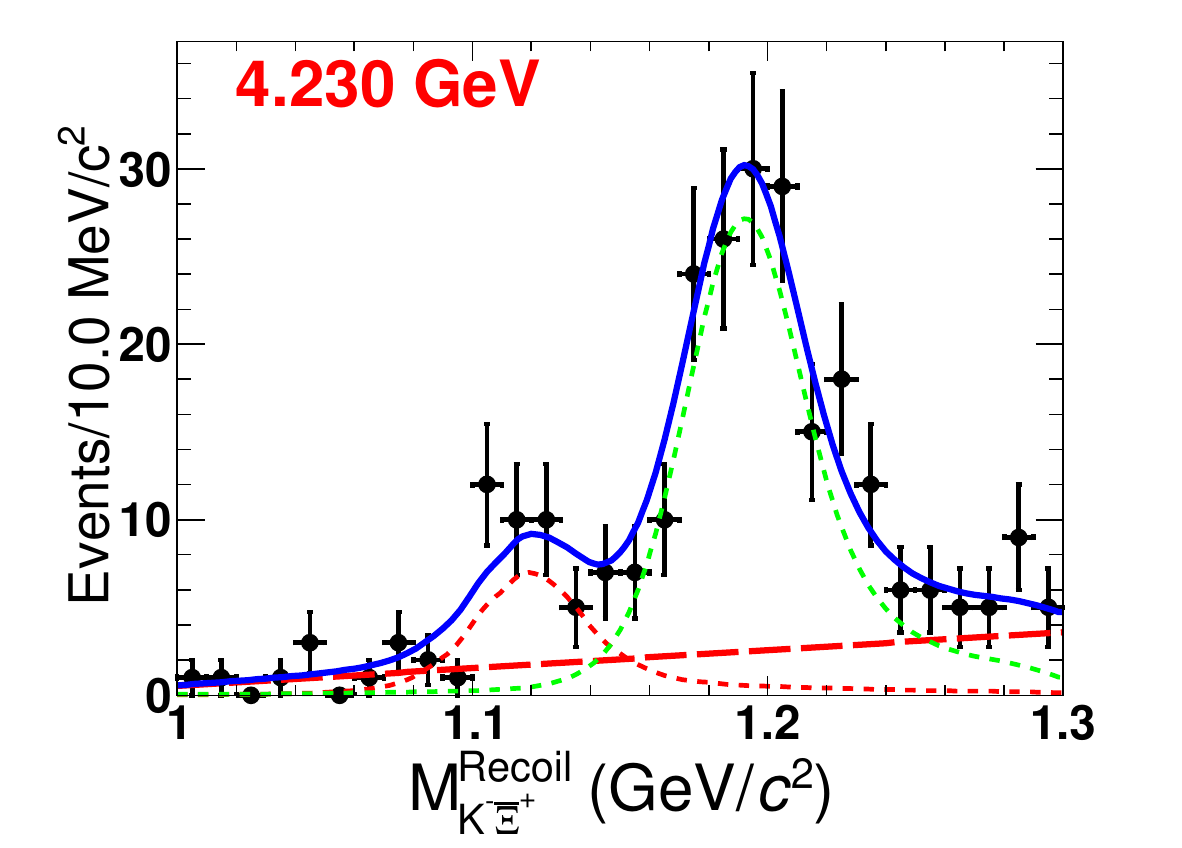}\\
    \includegraphics[width=0.22\textwidth]{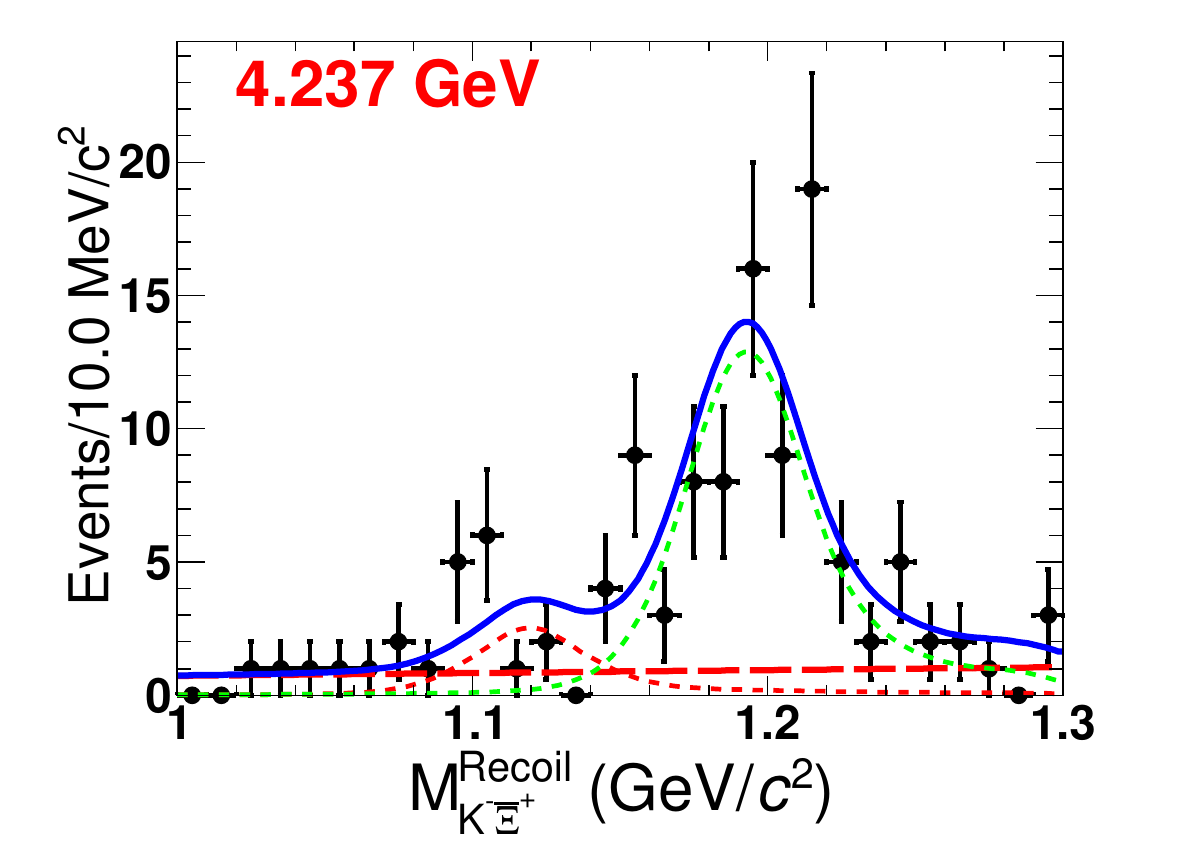}
    \includegraphics[width=0.22\textwidth]{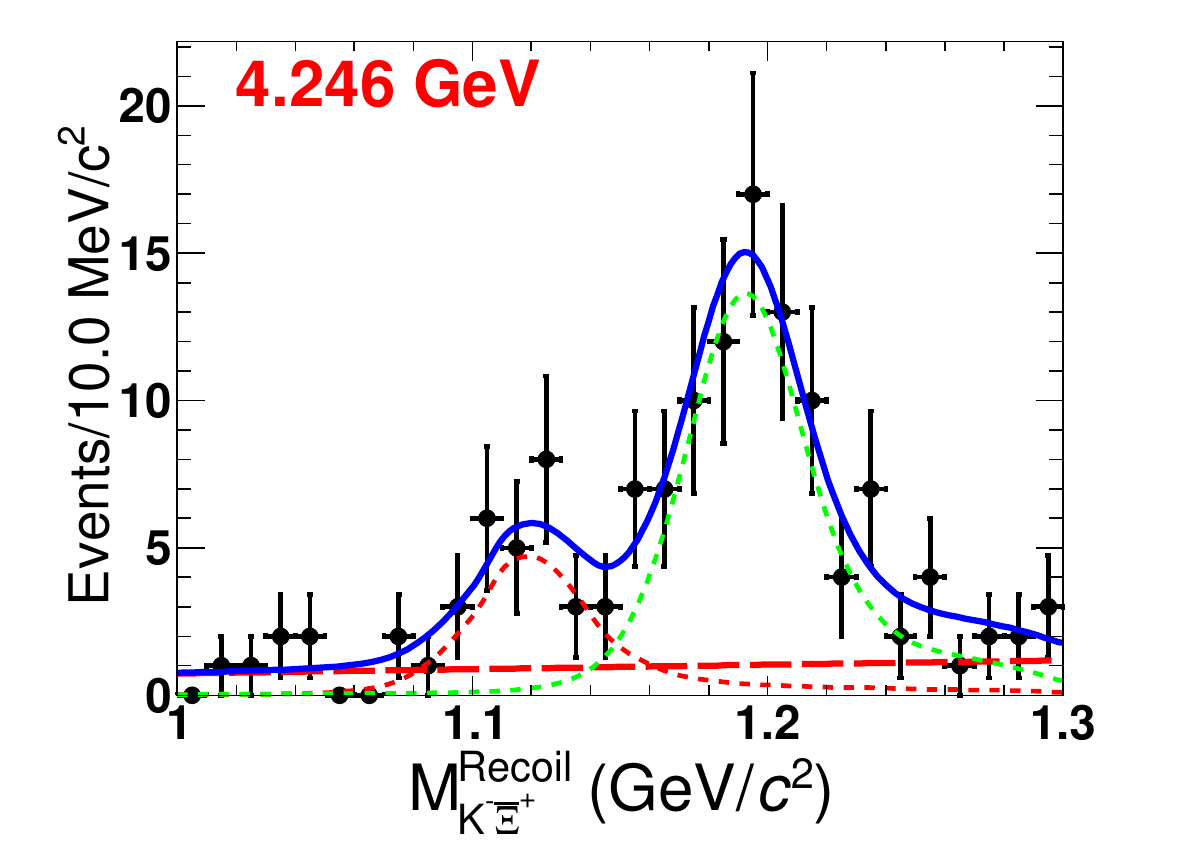}
    \includegraphics[width=0.22\textwidth]{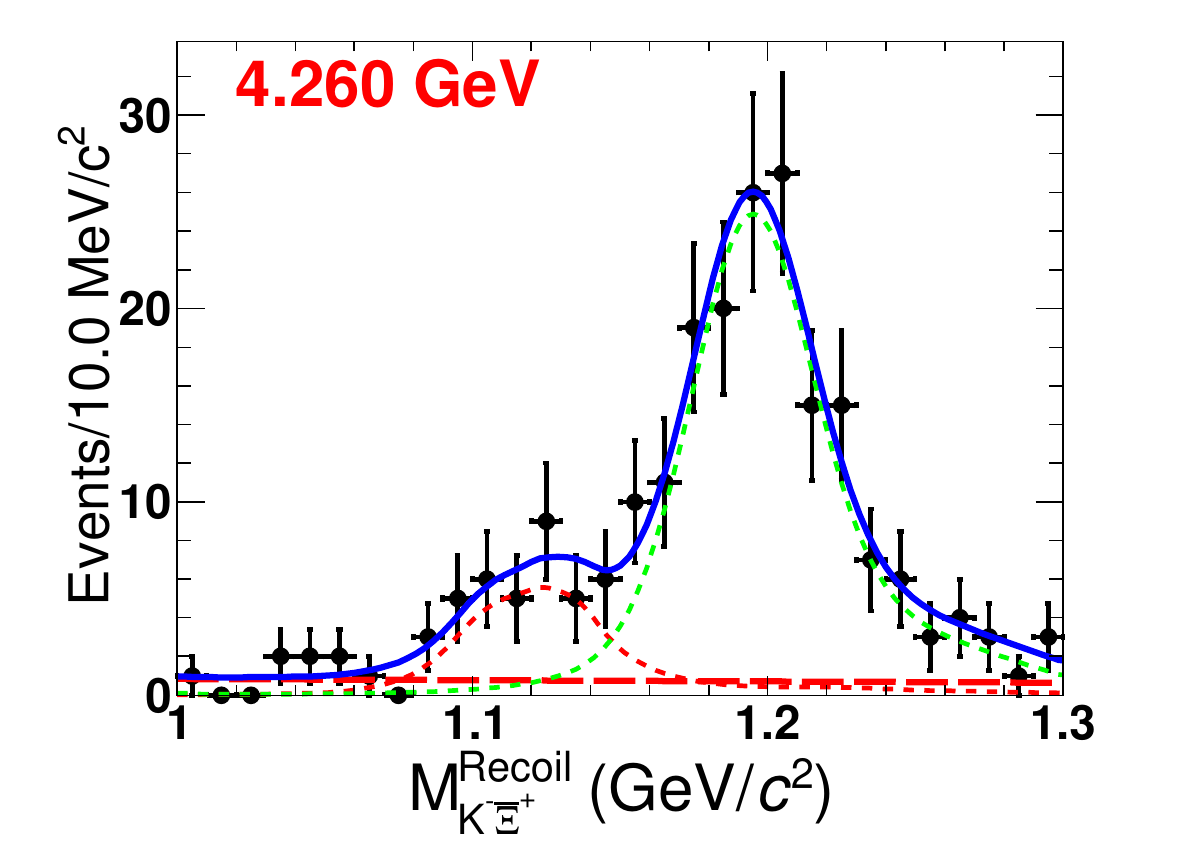}
    \includegraphics[width=0.22\textwidth]{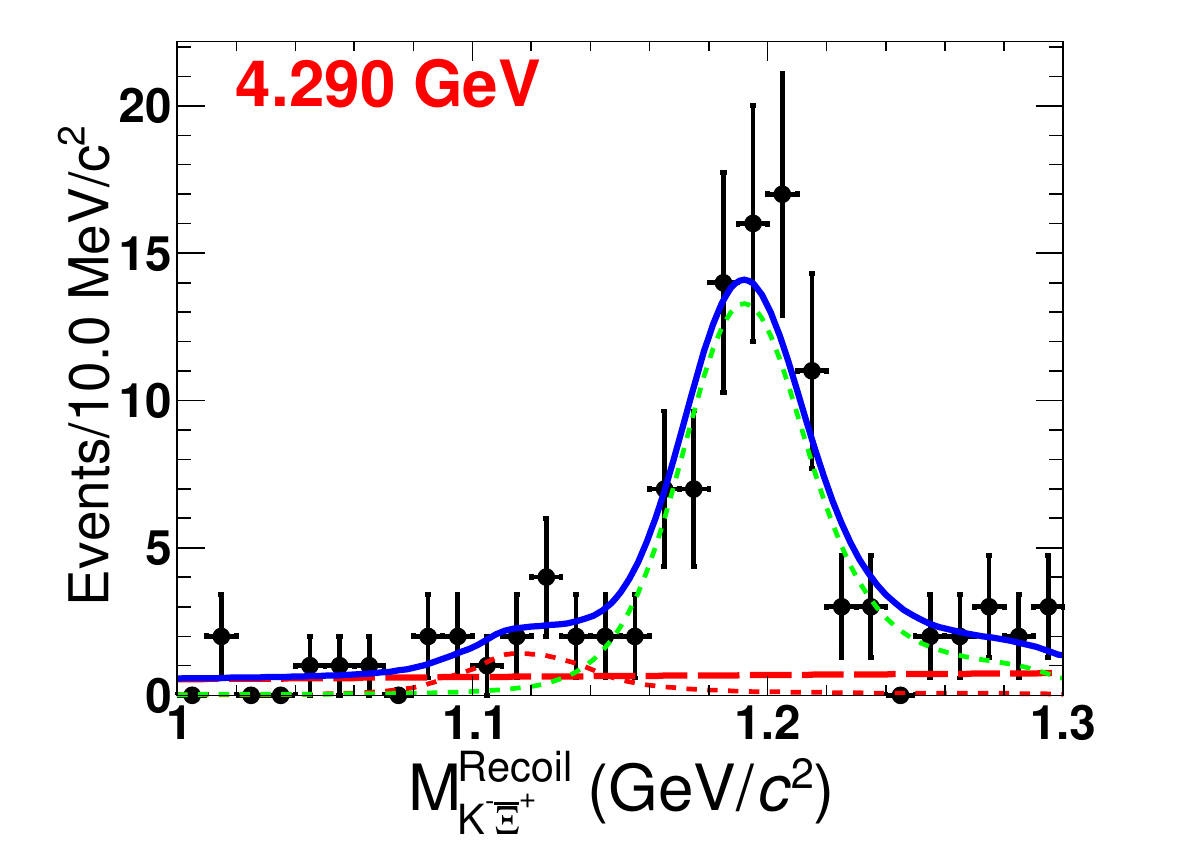}\\
    \includegraphics[width=0.22\textwidth]{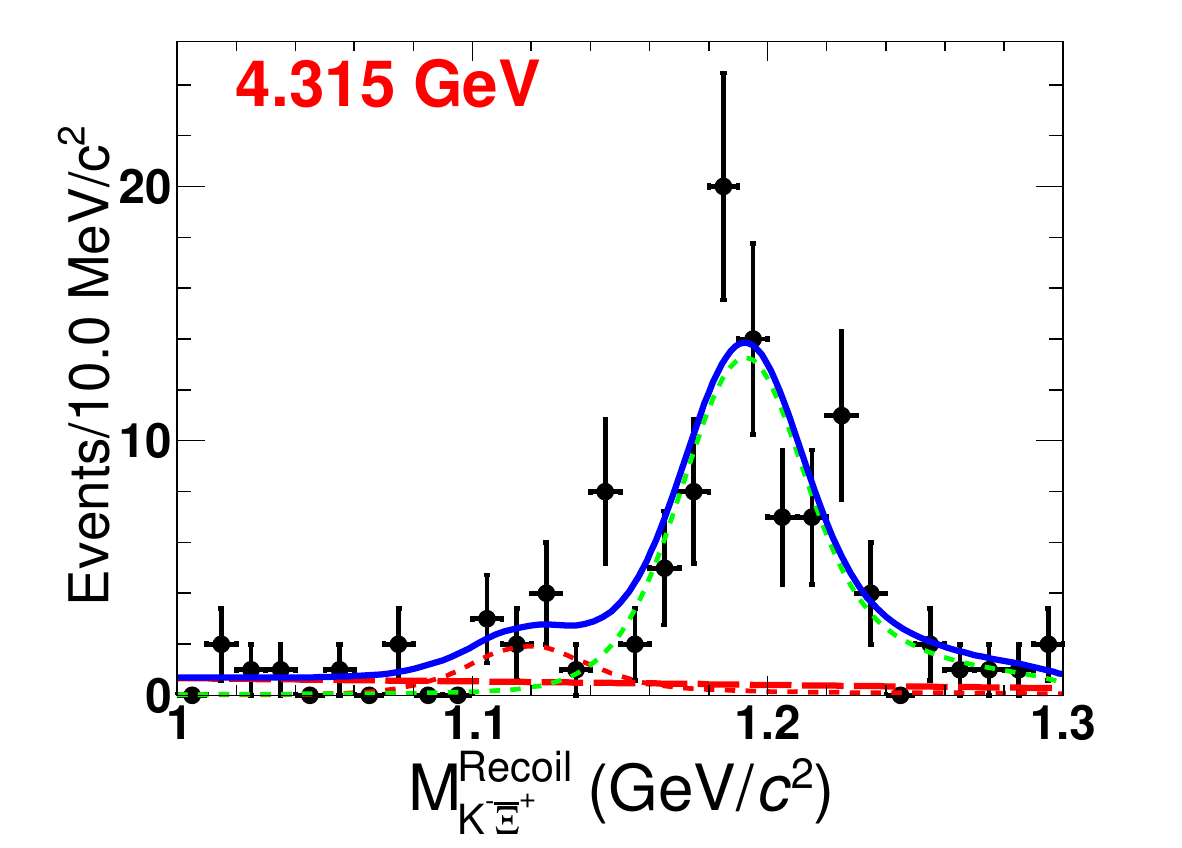}
    \includegraphics[width=0.22\textwidth]{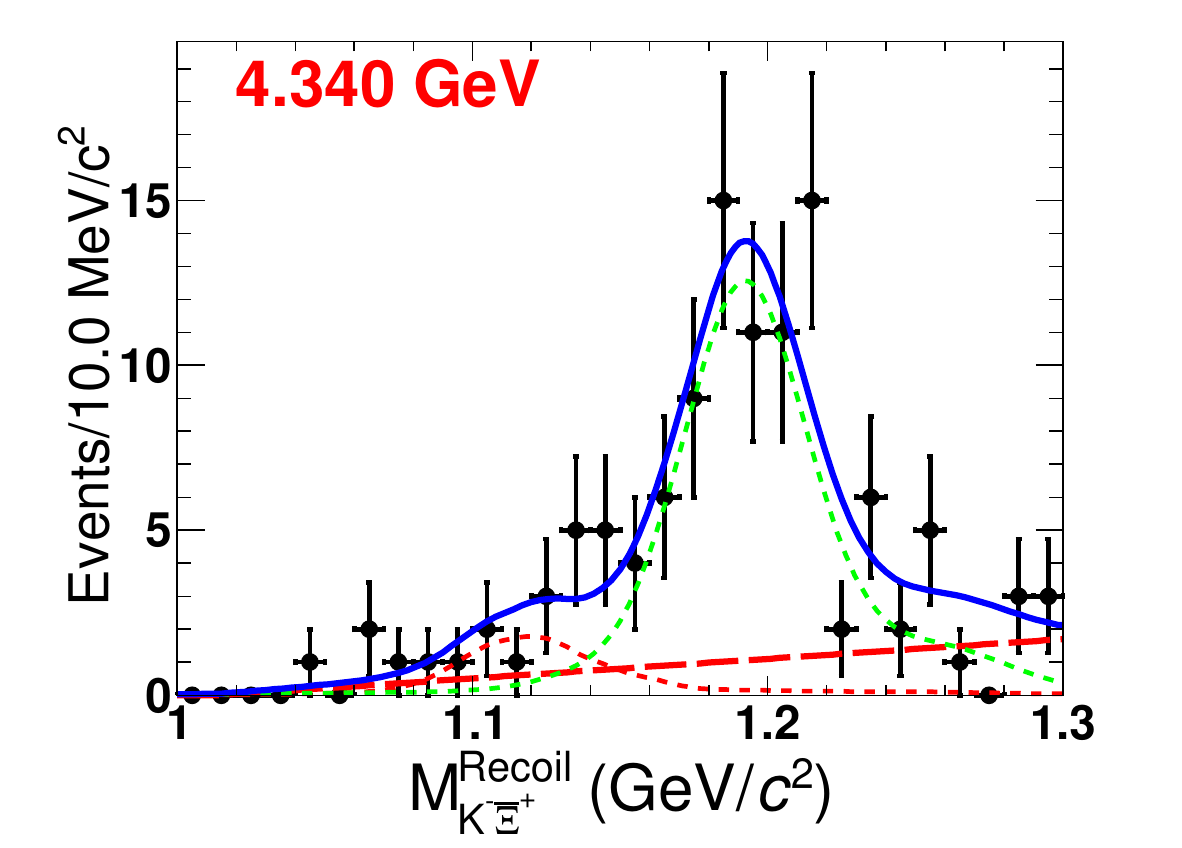}
    \includegraphics[width=0.22\textwidth]{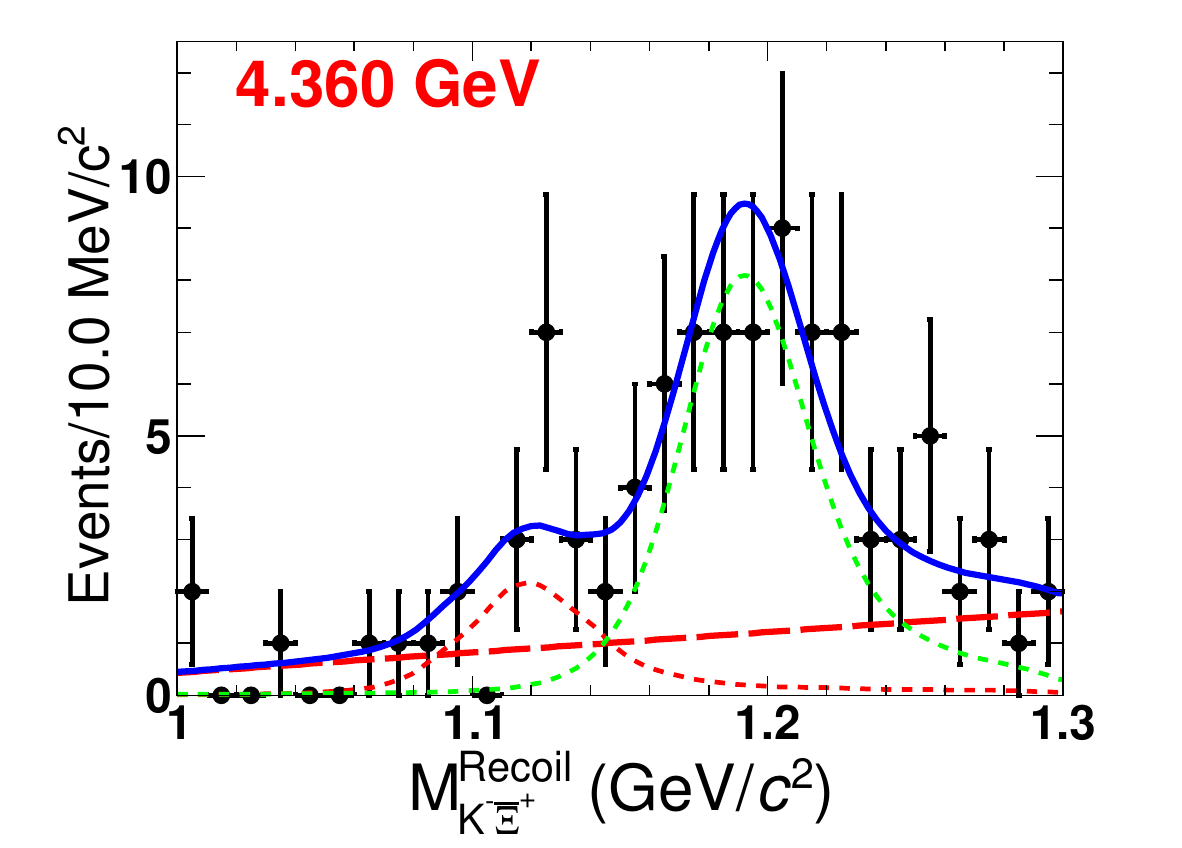}
    \includegraphics[width=0.22\textwidth]{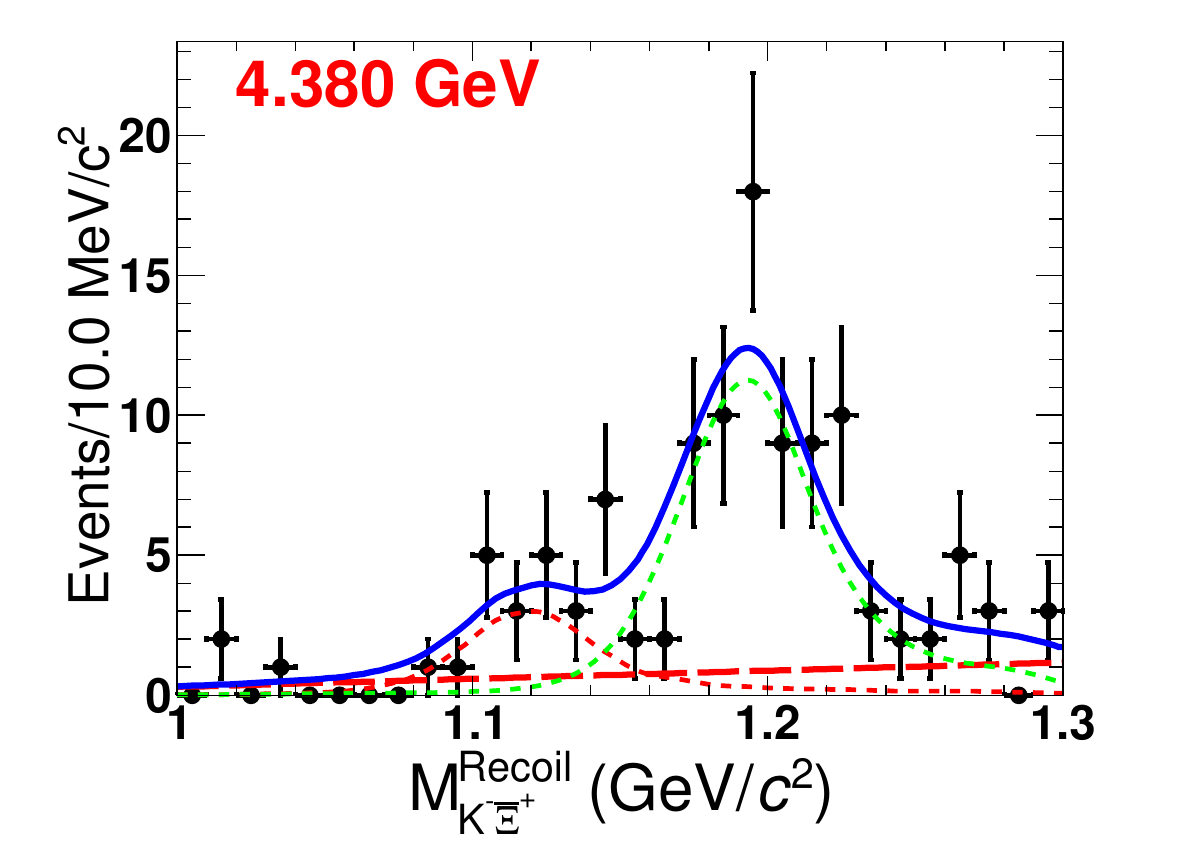}\\
    \includegraphics[width=0.22\textwidth]{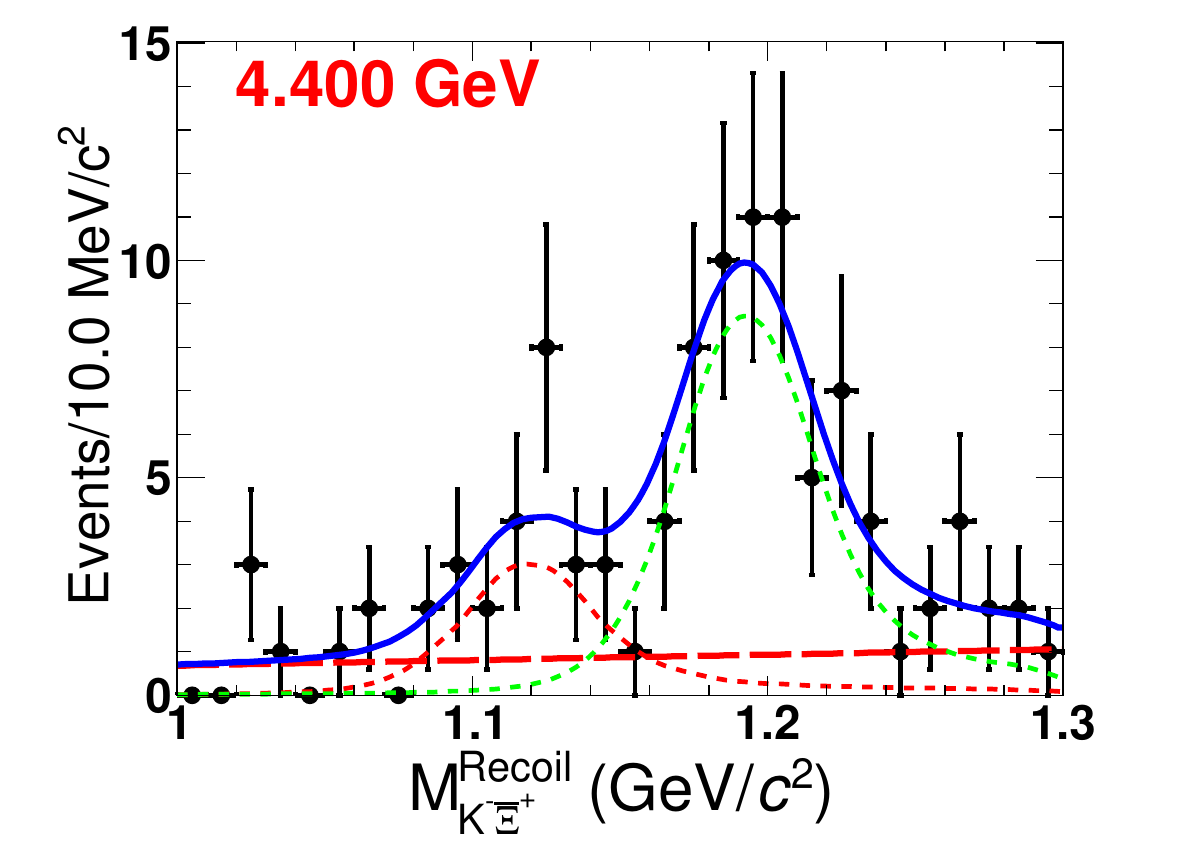}
    \includegraphics[width=0.22\textwidth]{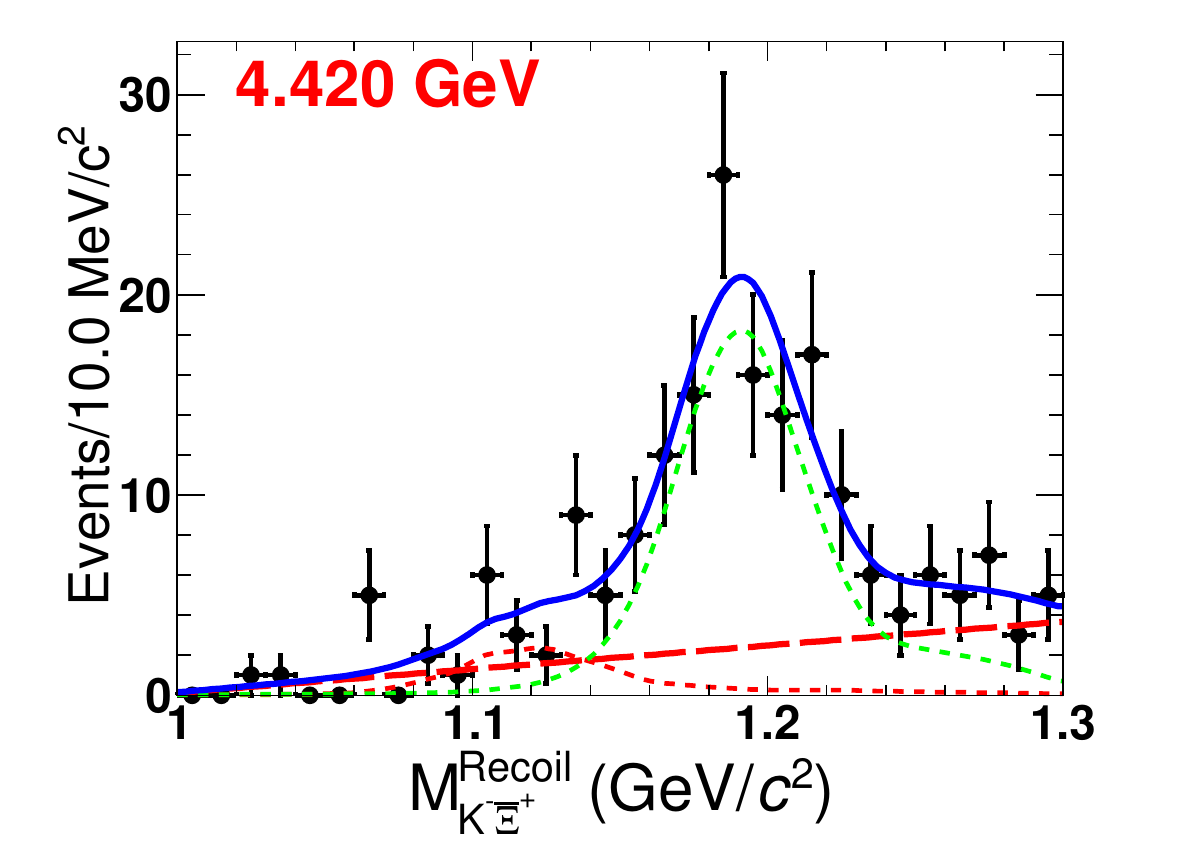}
    \includegraphics[width=0.22\textwidth]{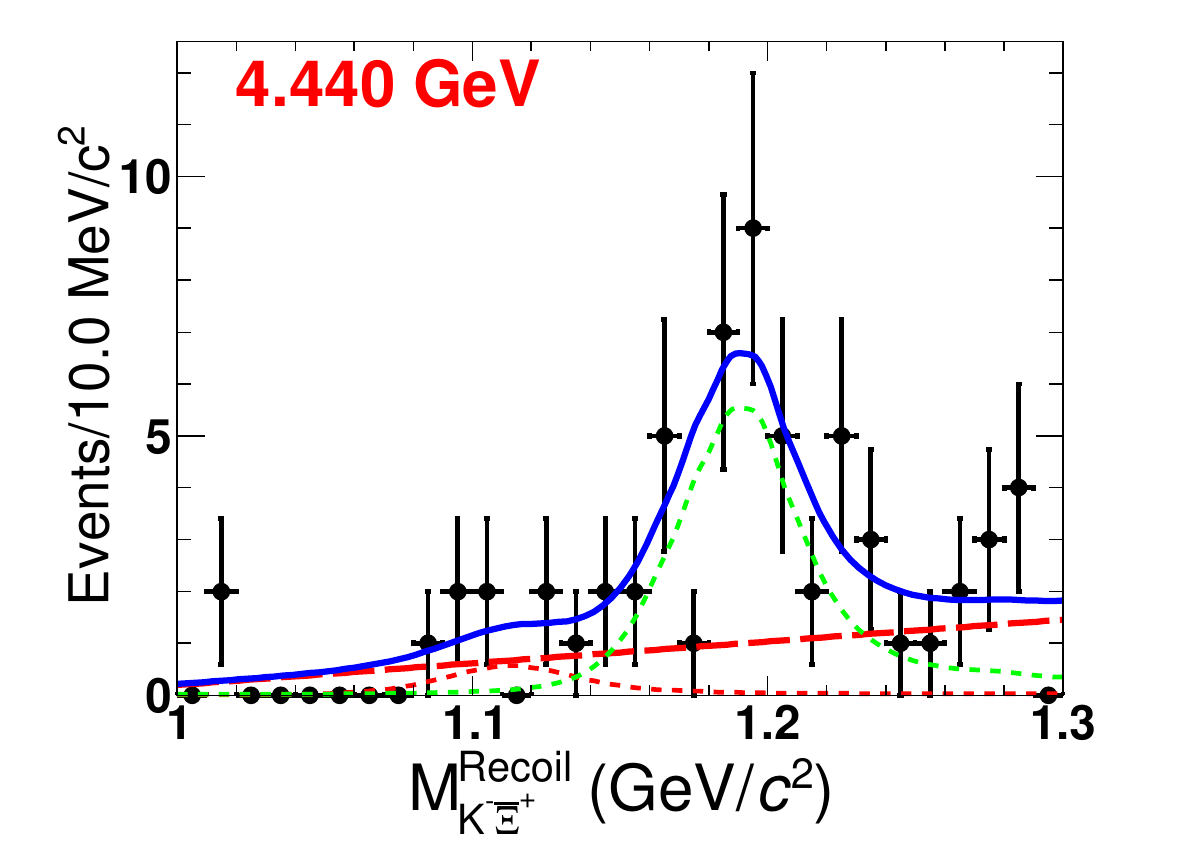}
    \includegraphics[width=0.22\textwidth]{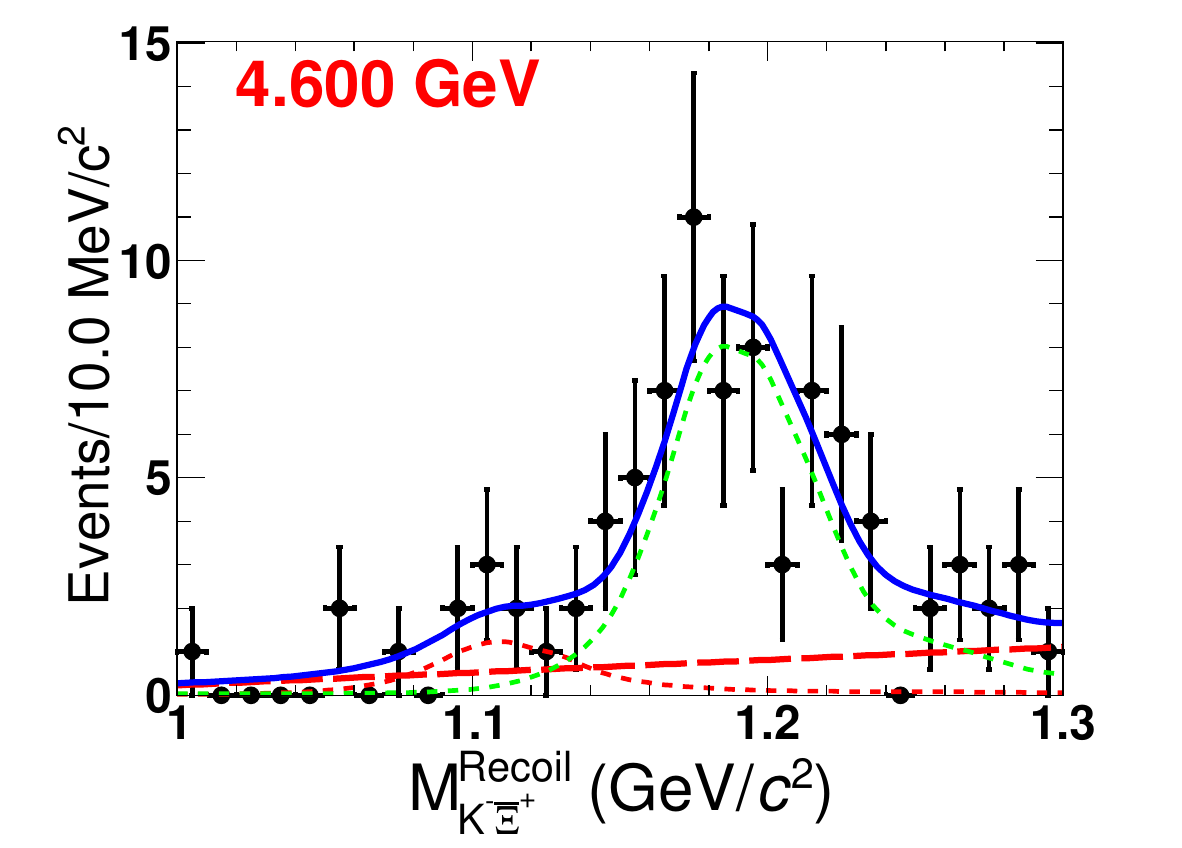}\\
    \includegraphics[width=0.22\textwidth]{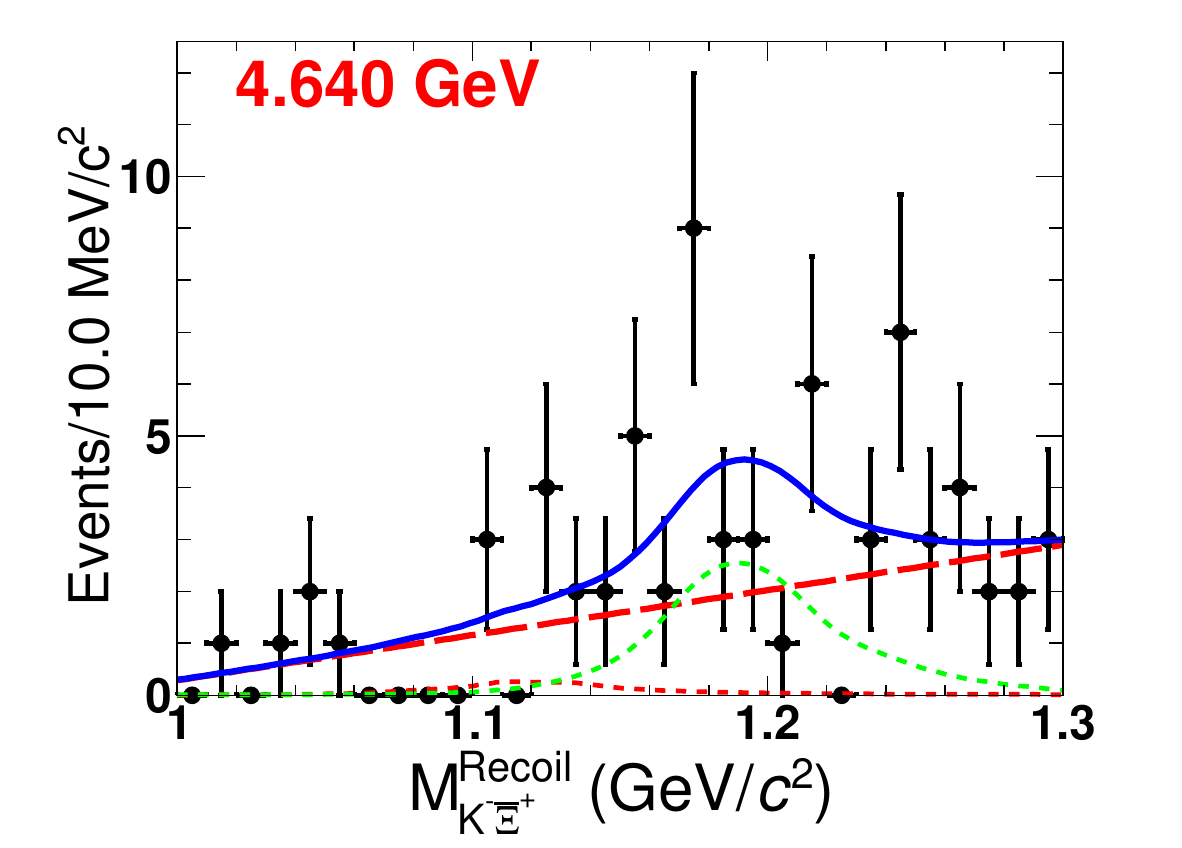}
    \includegraphics[width=0.22\textwidth]{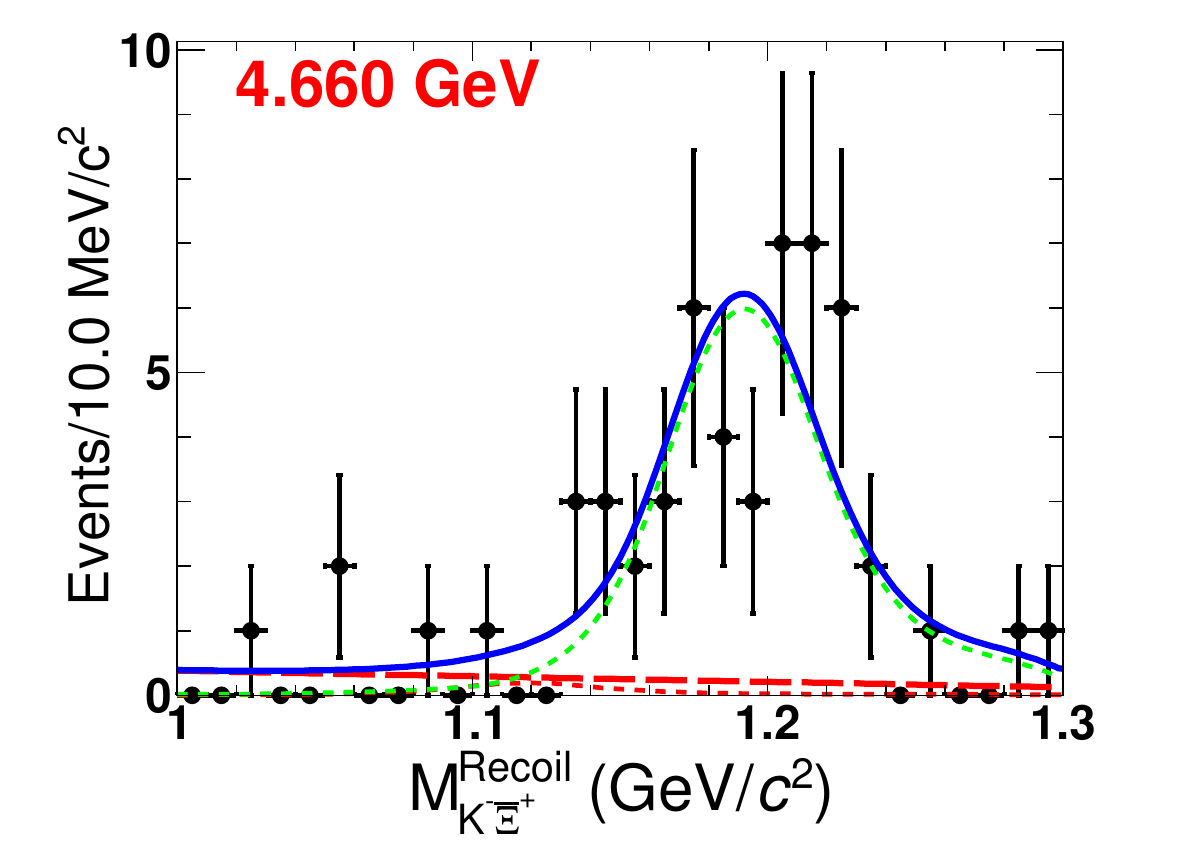}
    \includegraphics[width=0.22\textwidth]{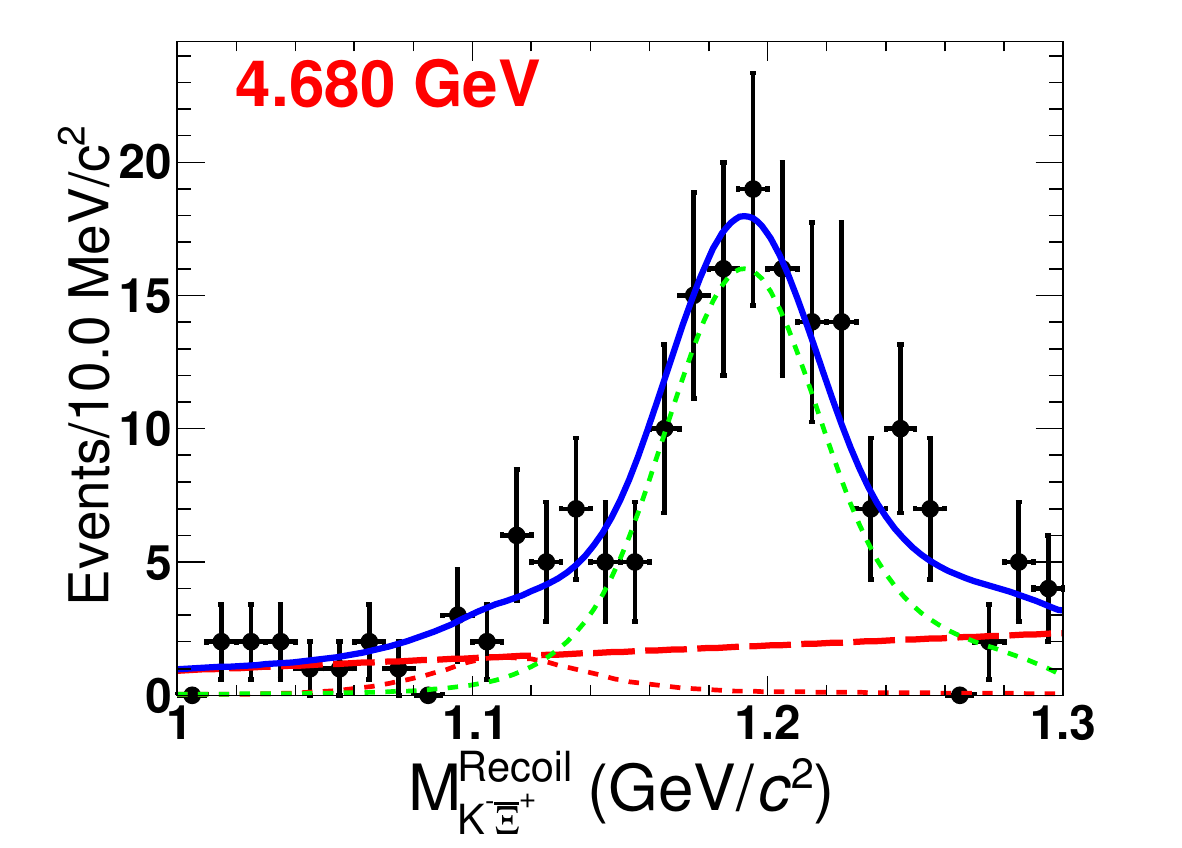}
    \includegraphics[width=0.22\textwidth]{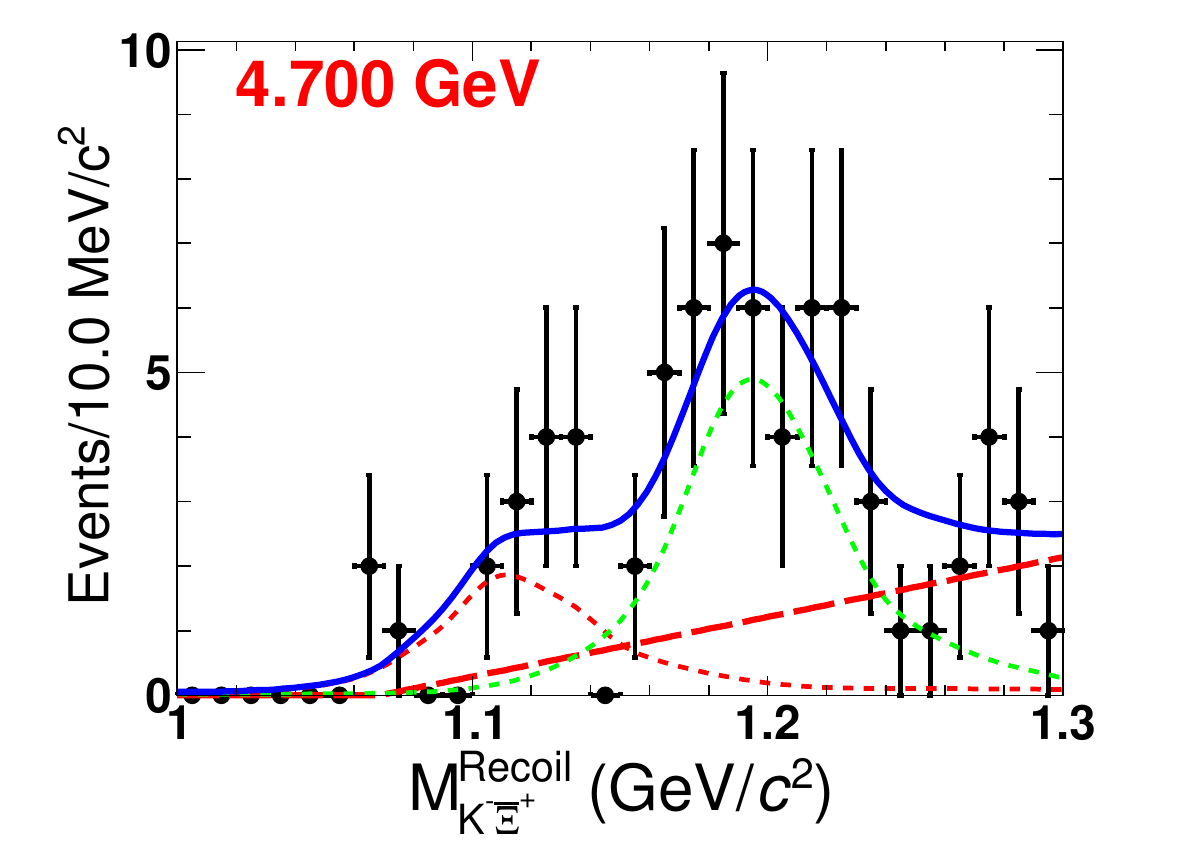}\\
    \includegraphics[width=0.22\textwidth]{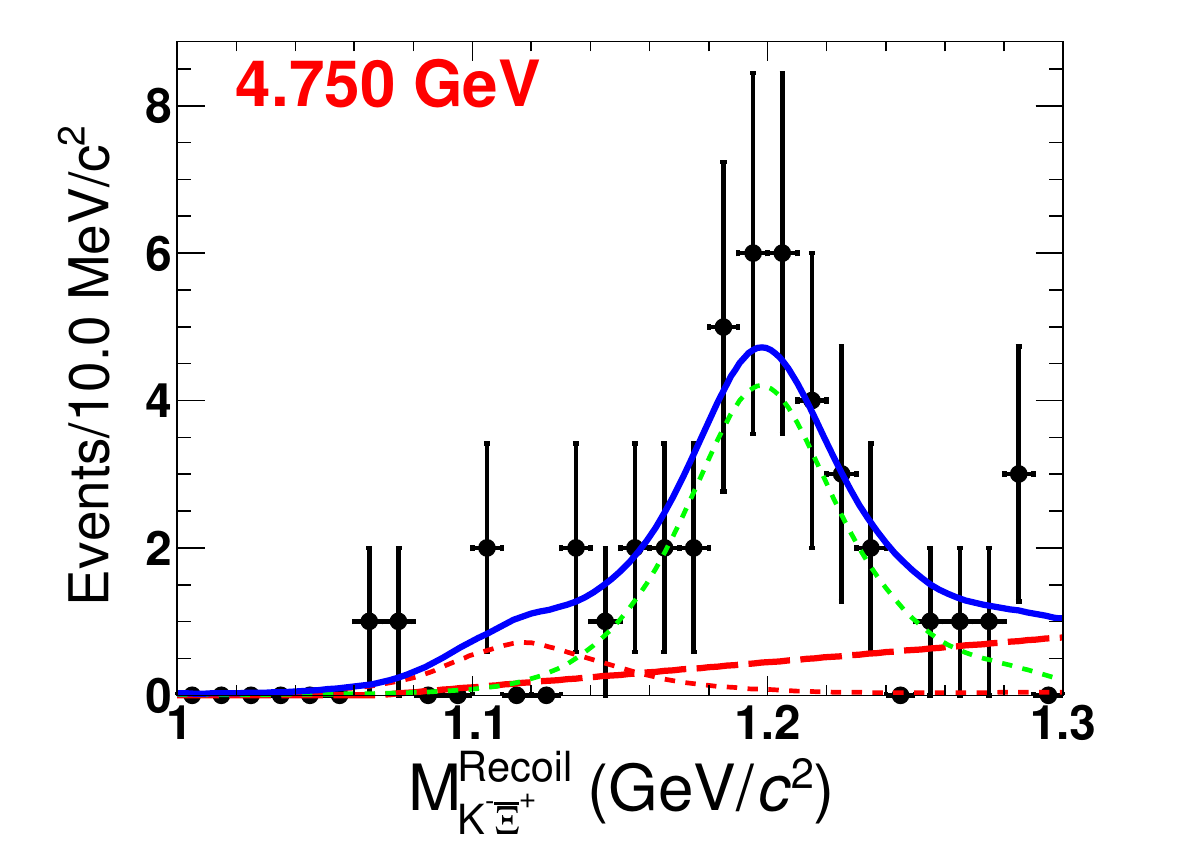}
    \includegraphics[width=0.22\textwidth]{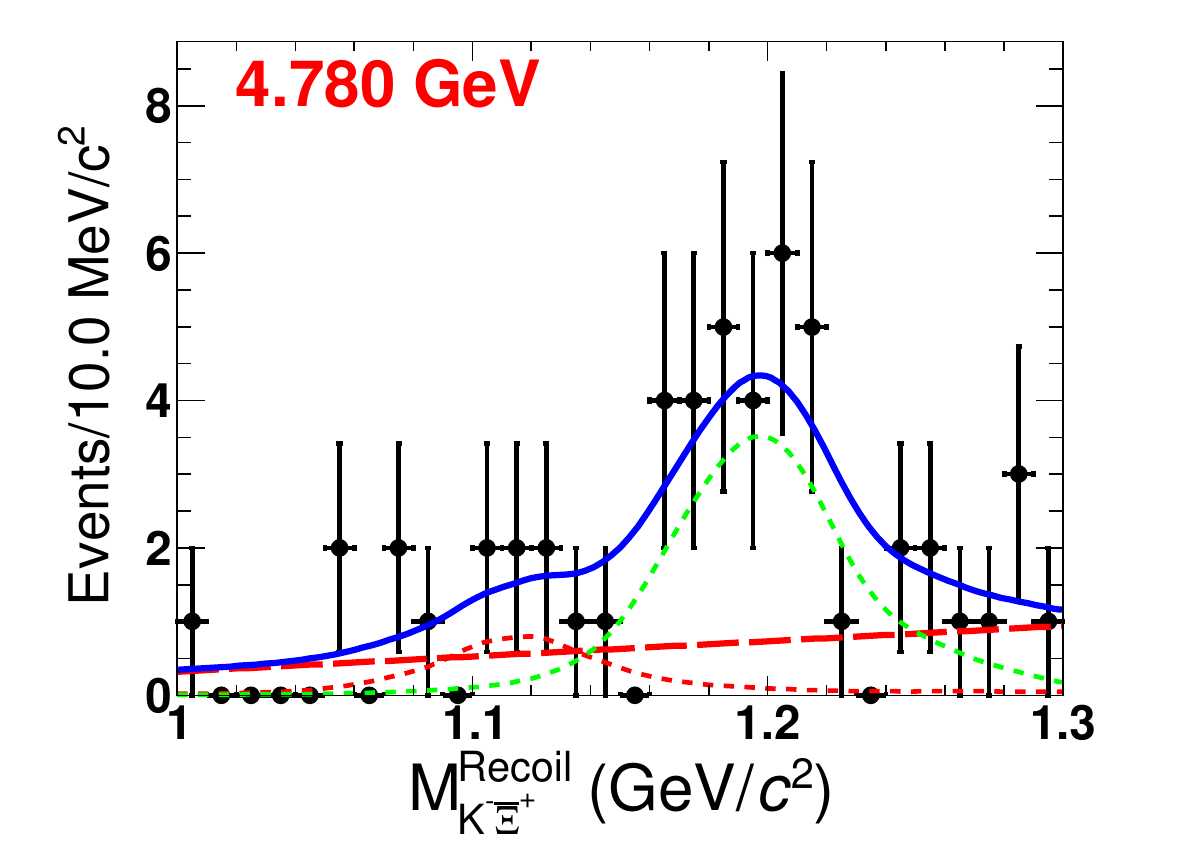}
    \includegraphics[width=0.22\textwidth]{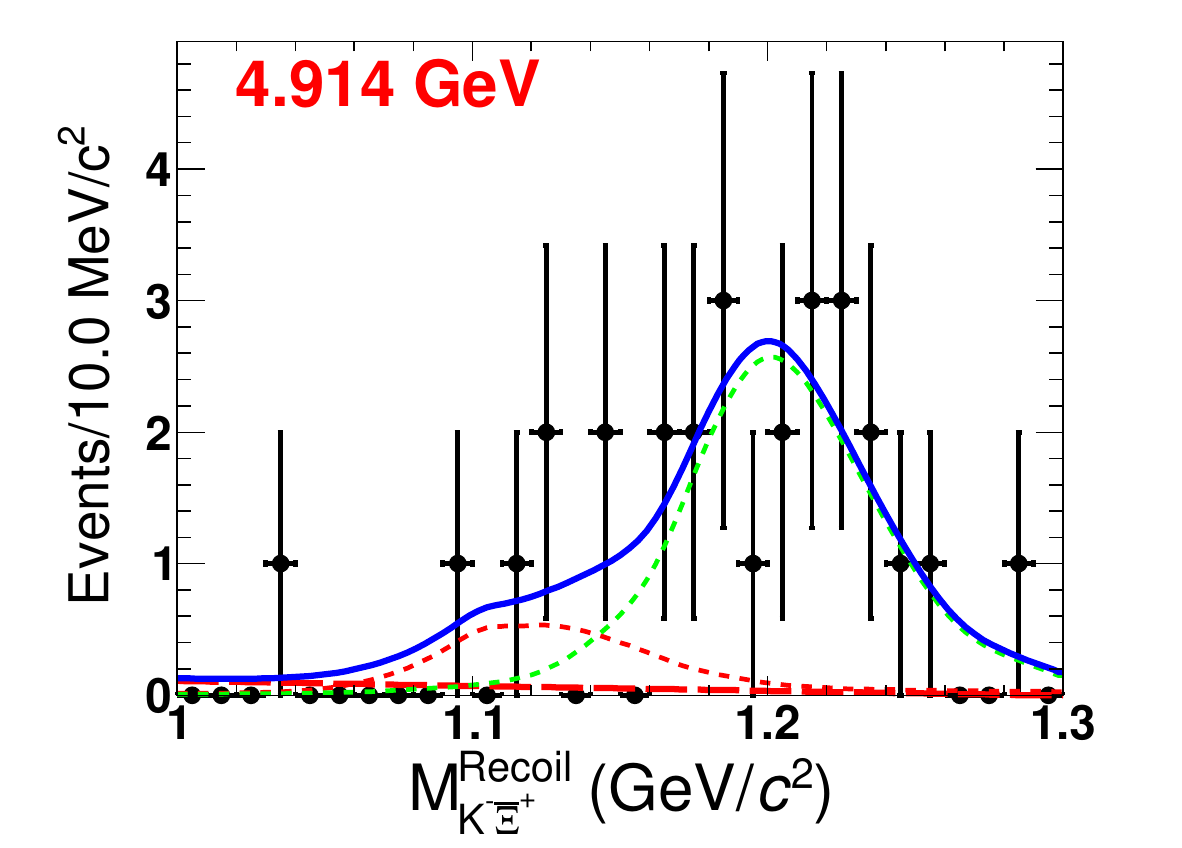}    
    \vspace*{-0.5cm}
    \end{center}
    \caption{
Fits to the $M^{\rm recoil}_{K^-\bar\Xi^+}$ distributions from data at CM energies from 3.510 and 4.914\,{GeV}. The data are the dots with error bars. The blue lines represent the total fit, the red dashed lines represent the background, and the red and green dotted lines represent the $\Lambda$ and $\Sigma^0$ signals, respectively.}
    \label{Fig:SP:DATA:fitting}
    \end{figure}

The signal yields for the processes $\EE\ar K^-\bar{\Xi}^+\Lambda/\Sigma^{0}$ at each energy point are determined by performing an extended maximum-likelihood fit to the $M^{\rm recoil}_{K^-\Xib}$ spectra in the range of 1.0 to \SI{1.3}{GeV/\clight^{2}} as shown in figure~\ref{Fig:SP:DATA:fitting}. In the fit, the signal shapes for $\EE\ar K^-\bar{\Xi}^+\Lambda/\Sigma^{0}$ at each energy point are represented by the MC simulated shapes, and the background shapes are represented by a linear function. 
The inclusive MC indicates that background arises from  $\pi^+\pi^- J/\psi$,  $J/\psi\to K^-p\bar\Lambda$ events, the distribution of which is smooth in the region of interest.  
Tables~\ref{tab:NUM_BCSL} and~\ref{tab:NUM_BCSS} summarize the signal yields at each energy point.
The significance at some of the energy points is less than 3.0$\sigma$, as assessed by comparing the change of likelihood with and without the signal contribution in the fits. The upper limits of signal yields including additive part of systematic uncertainty at the 90\% confidence level (C.L.) at these energy points are determined with the Bayesian method~\cite{Zhu:2008ca}. 
The additive uncertainties are accounted for by extracting the likelihood distributions $\cal{L}$,  and the signal shapes corresponding to the maximum upper limits among all additive items are chosen. Then the upper limit on the signal yield $(N^{\rm UL})$ at the $90\%$ C.L. is determined from the condition $\int_{0}^{N^{\rm UL}} \cal{L}\,{\rm d}$$N_{\rm obs}/\int_{0}^{\infty} \cal{L}\,{\rm d}$$N_{\rm obs}=0.9$. The upper limits for cross sections based on these likelihood distributions and incorporating the multiplicative systematic uncertainties in the calculation are obtained by smearing the likelihood distribution by a Gaussian function with a mean of zero and a width equal to $\sigma_{\rm multi}$, where $\sigma_{\rm multi}$ is the multiplicative part of systematic uncertainty mentioned in section~\ref{SYS_UN}.

\subsection{Determination of Born cross section}
The Born cross section for $\EE \ar K^-\bar{\Xi}^+\Lambda/\Sigma^{0}$ is calculated by 
\begin{equation}
\sigma^{B} =\frac{N_{\rm obs}}{2 \cdot {\cal{L}}\cdot(1 + \delta)\cdot\frac{1}{|1 - \prod|^{2}}\cdot\epsilon\cdot {\cal B}(\bar\Xi^+\ar\pi^+\bar\Lambda)\cdot {\cal B}(\bar\Lambda\ar \bar p\pi^+)},
\label{bcs}
\end{equation}
where $N_{\rm obs}$ is the number of the observed signal events, the factor of 2 represents the average for both modes by considering the charge-conjugate channel, ${\cal{L}}$ is the integrated luminosity, $(1 + \delta)$ is the ISR correction factor, $\frac{1}{|1-\Pi|^2}$ is the vacuum polarization (VP) correction factor, $\epsilon$ is the detection efficiency, 
and ${\cal B}(\bar\Xi^+\ar\pi^+\bar\Lambda)$ and ${\cal B}(\bar\Lambda\ar \bar p\pi^+)$ are the branching fractions taken from the PDG~\cite{PDG2020}. Note that the cross section corresponds to only one charge mode. The ISR correction factor is obtained using the QED calculation as described in ref.~\cite{Kuraev:1985hb}. The VP correction factor is calculated according to ref.~\cite{Jegerlehner:2011ti}.

\begin{figure}
    \centering
    \includegraphics[width=1\textwidth]{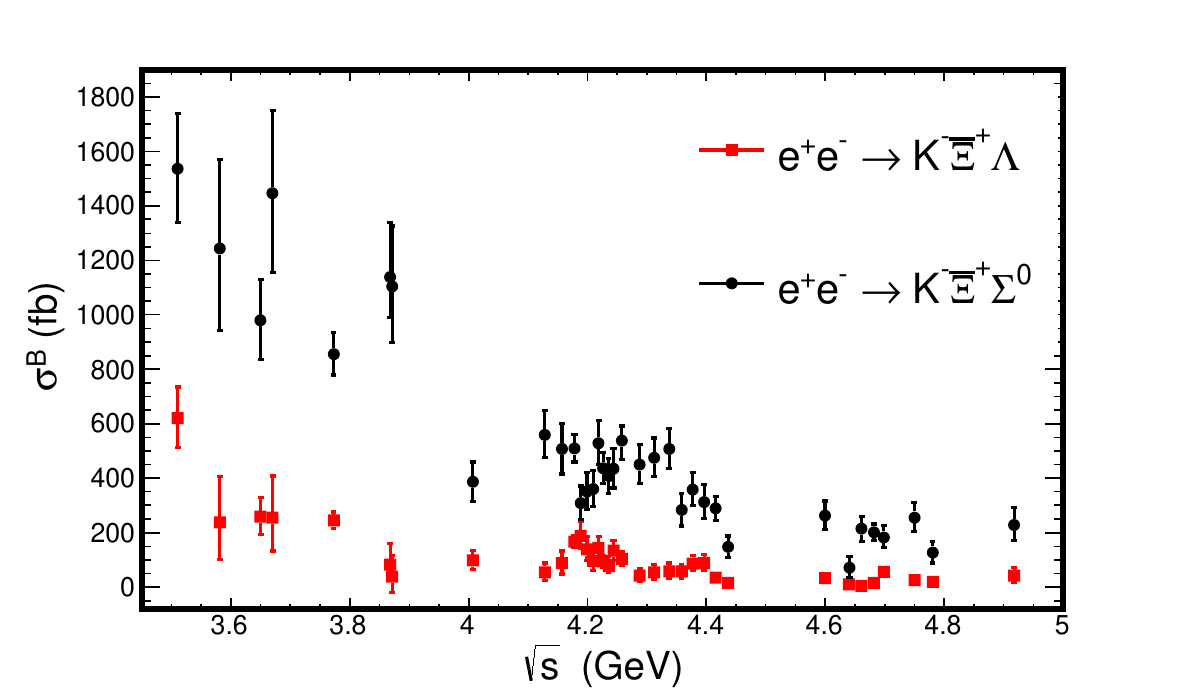}
    \caption{
    Measured Born cross section for $\KXS$ (black) and $K^-\bar\Xi^+\Lambda$ (red) for each energy point, where the uncertainties include both the statistical and systematic contributions.}
    
    \label{fig:cs}
\end{figure}

Initially, the cross section is measured without any ISR correction. Using this initial measured line shape of the cross sections, signal MC samples are regenerated to obtain revised values of the efficiencies and ISR correction factors, and the Born cross sections are updated subsequently. The Born cross sections are calculated iteratively until the values converge, defined by when the $(1 + \delta)\epsilon$ difference between last two iterations is less than 0.1\%. The values of the efficiency, ISR correction factor, and Born cross section are obtained through this iterative process~\cite{Sun:2020ehv}. The Born cross section at each energy point is shown in figure~\ref{fig:cs}.
\begin{table}[!htp]
    \centering
    \caption{\small
    Numerical results for $\EE\ar K^-\bar\Xi^+\Lambda$, where $\frac{1}{|1 - \prod|^{2}}$ is the VP correction factor, $1 + \delta$ is the ISR correction factor, $\epsilon$ is the detection efficiency, $N_{\rm obs}$ denotes the number of the observed signal event, $N^{\rm UL}$ is the upper limit of the signal event, $\sigma^{B}$ represents the Born cross section, and $\sigma^{\rm UL}$ is the upper limit of Born cross section, which take multiplicative and additive systematic uncertainties into account. The first and second uncertainties for $\sigma^{B}$ are statistical and systematic, respectively.
    }
    \resizebox{\textwidth}{!}{
     \renewcommand\arraystretch{1.1}
    \begin{tabular}{cccccllc}   \hline \hline
    \multicolumn{1}{c}{$\sqrt{s}$ (GeV)}  & \multicolumn{1}{c}{$\int{\cal L}dt$ (pb$^{-1})$} & \multicolumn{1}{c}{$\frac{1}{|1 - \prod|^{2}}$}&\multicolumn{1}{c}{$1+\delta$ }&\multicolumn{1}{c}{$\epsilon$ (\%) } & \multicolumn{1}{l}{$N_{\rm obs}$ ($N^{\rm UL}$)}&\multicolumn{1}{c}{$\sigma^{B}$ ($\sigma^{\rm UL}$) (fb)} & $S~(\sigma)$\\ \hline
        3.510   &405.4      & 1.04  & 0.74& 23.87    &$58.7 ^{+9.4 }_{-8.7 }$               & $ 621.2^{+99.5}_{-92.1} \pm43.5      $           & 6.2\\   
        3.581   &85.7       & 1.05  & 0.79& 24.88     &$5.4^{+3.8}_{-3.1 }~(< 9.4)$        & $ 237.9  ^{+167.4}_{-136.6} \pm18.1      ~(<424.7)$  & 1.3\\
        3.650   &410.0        & 1.02  & 0.86 & 24.20  &$28.7^{+7.6}_{-6.9 }$               & $ 258.3  ^{+68.4}_{-62.1}  \pm16.7      $            & 3.6\\   
        3.670   &84.7       & 1.02  & 0.86& 25.07    &$6.2^{+3.7}_{-2.9 }~(<9.0)$         & $ 254.2  ^{+151.7}_{-118.9}   \pm 17.8      ~(<398.9)$  & 1.7\\
        3.773   &2931.8     & 1.06  & 0.92& 27.10    &$239.2 ^{+20.0 }_{-19.1}$           & $  245.2  ^{+20.5}_{-19.6}   \pm 17.2      $           & 12.1\\   
        3.867   &108.9      & 1.06  & 0.97& 27.30    &$3.2^{+2.9}_{-2.0}~(<7.6)$          & $  83 .0  ^{+75.3}_{-52.7}  \pm 6.1   ~(<209.6)$      & 1.0\\
        3.871   &110.3      & 1.05  & 0.97& 23.33     &$1.4^{+2.8}_{-2.1}~(<5.0)$         & $ 38 .4  ^{+76.8}_{-56.2}  \pm 3.1~  ~(<137.1)$      & 0.3\\
        4.009   &482.0        & 1.05  & 1.04& 28.57  &$18.9 ^{+6.9}_{-6.3 }~(<28.2)$      & $ 97 .5  ^{+35.6}_{-32.5}  \pm 7.0   ~(<158.4)$        & 2.4\\   
        4.130   &401.5      & 1.05  & 1.04& 26.97     &$8.3^{+5.3}_{-4.6 }~(<15.6)$        & $ 54 .5  ^{+34.8}_{-30.2}  \pm 3.9   ~(<106.8)$      & 1.2\\   
        4.160   &408.7      & 1.05  & 0.97& 27.18     &$12.8^{+6.4}_{-5.7 }~(<21.6)$       & $ 87 .9  ^{+44.0}_{-39.2} \pm6.2   ~(<142.5)$      & 1.5\\   
        4.180   &3189.0       & 1.05  & 0.91& 25.93  &$170.2^{+18.6}_{-17.9}$               & $  167.1  ^{+18.3}_{-17.6}  \pm 11.6    $               & 7.7\\   
        4.190   &526.7      & 1.06  & 0.90& 28.24    &$34.2^{+8.6}_{-7.9 }$                 & $  188.8  ^{+47.5}_{-43.6} \pm 13.4      $           & 3.5\\   
        4.200   &526.0        & 1.06  & 0.92& 28.55  &$25.7^{+8.6}_{-6.8 }~(<36.1)$      & $ 137.6  ^{+46.0}_{-36.4}   \pm10.0   ~(<181.4)$       & 2.5\\   
        4.210   &517.1      & 1.06  & 0.96& 27.95     &$17.9^{+7.1}_{-6.5 }~(<28.7)$       & $ 95 .5  ^{+37.9}_{-34.7}  \pm 6.8   ~(<153.3)$      & 2.1\\   
        4.220   &514.6      & 1.06  & 1.00& 27.81     &$27.6^{+7.7}_{-7.0}$                  & $  143.6  ^{+40.1}_{-36.4}  \pm 10.3   $              & 3.1\\
        4.230   &1100.9     & 1.06  & 1.04& 28.96    &$41.8^{+9.7}_{-9.1 }$                  & $  94 .5  ^{+21.9}_{-20.6} \pm 6.6   $               & 3.7\\
        4.237   &530.3      & 1.06  & 1.07& 28.53    &$17.0^{+7.2}_{-5.6 }~(<26.4)$       & $ 78 .0  ^{+33.1}_{-25.7}  \pm 5.5   ~(<131.7)$      & 1.7\\
        4.246   &538.1      & 1.06  & 1.10& 28.40    &$30.2^{+8.2}_{-7.6 }$                 & $  133.9  ^{+36.4}_{-33.7} \pm 9.7    $              & 3.1\\
        4.260   &828.4      & 1.05  & 1.14& 29.01    &$38.1^{+9.4}_{-8.6 }$                 & $  103.4  ^{+25.5}_{-23.3}  \pm7.5   $               & 3.4\\
        4.290   &502.4      & 1.05  & 1.19& 27.15    &$9.4^{+5.7}_{-5.1 }~(<18.0)$         & $ 42 .9 ^{+26.0}_{-23.3}  \pm 3.1   ~(<94.0)$       & 1.3\\   
        4.315   &501.2      & 1.05  & 1.23& 27.03    &$12.1^{+6.2}_{-5.4 }~(<20.9)$       & $ 54 .1 ^{+27.7}_{-24.1}  \pm3.9   ~(<106.4)$      & 1.8\\
        4.340   &505.0      & 1.05  & 1.26& 27.60    &$13.7^{+6.9}_{-6.1 }~(<23.2)$       & $ 58 .1 ^{+29.3}_{-25.9}  \pm 4.1   ~(<114.7)$      & 1.5\\
        4.360   &543.9      & 1.05  & 1.28& 29.06    &$15.6^{+6.9}_{-6.2 }~(<24.5)$      & $ 57 .4 ^{+25.4}_{-22.8}   \pm 4.0   ~(<105.0)$      & 1.8\\
        4.380   &522.7      & 1.05  & 1.31& 27.35    &$21.8^{+6.8}_{-6.1 }~(<31.0)$        & $ 87 .1 ^{+27.2}_{-24.4}  \pm 6.2   ~(<146.7)$      & 2.6\\
        4.400   &507.8      & 1.05  & 1.32& 27.27    &$21.8^{+7.5}_{-6.8 }~(<32.3)$       & $ 89 .2 ^{+30.7}_{-27.8}  \pm 6.2   ~(<155.5)$      & 2.4\\
        4.420   &1090.7     & 1.05  & 1.33& 28.91    &$20.3^{+8.6}_{-7.9 }~(<30.9)$       & $ 36 .1 ^{+15.3}_{-14.0} \pm 2.5   ~(<65.8)$       & 1.5\\
        4.440   &569.9      & 1.05  & 1.35& 27.57    &$4.4^{+5.0  }_{-4.4 }~(<11.4)$       & $ 15 .5 ^{+17.6}_{-15.5}  \pm 1.1   ~(<48.5)$       & 0.7\\
        4.600   &586.9      & 1.06  & 1.51& 28.51    &$11.4^{+6.6}_{-5.8 }~(<19.2)$         & $ 33 .7 ^{+19.5}_{-17.1}  \pm 2.4   ~(<70.7)$       & 1.2\\
        4.640   &552.5      & 1.05  & 1.54& 26.63    &$2.9^{+6.7}_{-5.9 }~(<13.2)$        & $ 9  .7 ^{+22.5}_{-19.8}  \pm  0.7   ~(<56.4)$       & 0.3\\
        4.660   &529.4      & 1.05  & 1.60& 26.41    &$1.5^{+5.1}_{-4.5}~(<8.7)$            & $ 5  .0 ^{+17.0}_{-14.8}  \pm 0.4   ~(<38.6)$       & 0.2\\
        4.680   &1667.4     & 1.05  & 1.60& 26.23    &$15.5^{+8.8}_{-8.0   }~(<24.3)$     & $ 16 .8 ^{+9.6}_{-8.7}  \pm 1.2   ~(<33.6)$       & 1.0\\
        4.700   &536.5      & 1.06  & 1.61& 26.20    &$16.7^{+5.9}_{-5.0   }~(<24.3)$     & $ 55 .5 ^{+19.6}_{-16.6}  \pm 3.9   ~(<103.6)$      & 2.1\\
        4.750   &366.6      & 1.06  & 1.67& 28.06    &$5.9^{ +5.0   }_{-4.7 }~(<12.8)$    & $ 25 .3 ^{+21.4}_{-20.1}  \pm 1.8   ~(<72.8)$       & 1.0\\
        4.780   &511.5      & 1.06  & 1.72& 27.94    &$6.2^{ +6.1 }_{-5.4 }~(<15.1)$        & $ 19 .1 ^{+18.8}_{-16.6}  \pm 1.4   ~(<60.5)$       & 0.9\\
        4.914   &207.8      & 1.06  & 1.91& 27.11     &$6.0^{ +4.4 }_{-3.5 }~(<11.9)$       & $ 42 .2 ^{+30.9}_{-24.6}   \pm 3.0   ~(<113.2)$      & 1.1\\
    \hline\hline
    \end{tabular}
    }
    \label{tab:NUM_BCSL}
    \end{table}

    \begin{table}[!htp]
        \centering
        \caption{\small
        Numerical results for $\EE\ar K^-\bar\Xi^+\Sigma^0$, where $\frac{1}{|1 - \prod|^{2}}$ is the VP correction factor, $1 + \delta$ is the ISR correction factor, $\epsilon$ is the detection efficiency, $N_{\rm obs}$ denotes the number of the observed signal event, $N^{\rm UL}$ is the upper limit of the signal event, $\sigma^{B}$ represents the Born cross section, and $\sigma^{\rm UL}$ is the upper limit of Born cross section, which take multiplicative and additive systematic uncertainties into account. The first and second uncertainties for $\sigma^{B}$ are statistical and systematic, respectively.
        }
    \resizebox{\textwidth}{!}{
         \renewcommand\arraystretch{1.1}
    \begin{tabular}{cccccllc}   \hline \hline
    \multicolumn{1}{c}{$\sqrt{s}$ (GeV)}  &\multicolumn{1}{c}{$\int{\cal L}dt$ (pb$^{-1})$}  & \multicolumn{1}{c}{$\frac{1}{|1 - \prod|^{2}}$}&\multicolumn{1}{c}{$1+\delta$ }&\multicolumn{1}{c}{$\epsilon$ (\%) } & \multicolumn{1}{l}{$N_{\rm obs}$ ($N^{\rm UL}$)}&\multicolumn{1}{c}{$\sigma^{B}$ $(\sigma^{\rm UL}$) (fb)}&$S~(\sigma)$\\ \hline
        3.510   &405.4     & 1.04  & 0.85 & 24.18  & $144.3^{+15.2}_{-14.5}$          & $ 1536.4^{+161.8}_{-154.4}\pm     104.0     $      & 9.3\\
        3.581   &85.7      & 1.05  & 0.75 & 25.09  & $29.0^{+7.3}_{-6.6}$             & $ 1243.0^{+312.9}_{-282.9}\pm     84.1     $       & 3.3\\
        3.650   &410.0     & 1.02  & 0.85 & 23.40  & $102.0^{+13.4}_{-12.6}$          & $ 979.8^{+128.7}_{-121.0}\pm      66.3     $       & 7.3\\
        3.670   &84.7      & 1.02  & 0.90 & 25.59  & $36.0^{+7.0}_{-6.6}$             & $ 1445.7^{+281.1}_{-265.0}\pm     97.8     $       & 4.2\\
        3.773   &2931.8    & 1.06  & 1.01 & 26.83  & $868.1^{+36.1}_{-35.7}$          & $ 856.0^{+35.6}_{-35.2}  \pm        57.9  $          & 19.7\\
        3.867   &108.9   & 1.06  & 0.95 & 27.67    & $41.4^{+7.7}_{-5.4}$             & $ 1138.4^{+201.7}_{-146.9}\pm       77.0    $        & 4.5\\
        3.871   &110.3   & 1.05  & 1.01 & 26.05    & $40.9^{+8.2}_{-7.5}$             & $ 1104.0^{+222.4}_{-203.8}\pm       74.7    $        & 4.1\\
        4.009   &482.0   & 1.05  & 1.00 & 27.71    & $65.9^{+11.4}_{-11.0}$           & $ 386.5^{+66.9}_{-64.5}  \pm          26.0  $          & 4.4\\
        4.130   &401.5   & 1.05  & 1.01 & 28.87    & $83.6^{+11.4}_{-10.8}$           & $ 559.4^{+76.3}_{-72.3}  \pm          37.8  $          & 5.6\\
        4.160   &408.7   & 1.05  & 1.02 & 26.95    & $72.7^{+12.3}_{-11.7}$           & $ 506.9^{+85.8}_{-81.6}  \pm          34.2  $          & 4.0\\
        4.180   &3189.0  & 1.05  & 1.02 & 27.00    & $570.1^{+32.0}_{-31.8}$          & $ 508.5^{+28.5}_{-28.4}  \pm          34.3  $          & 15.1\\
        4.190   &526.7   & 1.06  & 1.01 & 28.54    & $59.6^{+11.5}_{-10.8}$           & $ 307.9^{+59.4}_{-55.8}  \pm          20.9  $          & 3.8\\
        4.200   &526.0   & 1.06  & 1.01 & 28.97    & $68.9^{+12.4}_{-11.7}$           & $ 350.7^{+63.1}_{-59.6}  \pm          23.6  $          & 4.0\\
        4.210   &517.1   & 1.06  & 1.00& 29.16     & $69.3^{+11.5}_{-10.9}$           & $ 360.1^{+59.8}_{-56.6}  \pm          25.7  $          & 4.4\\
        4.220   &514.6   & 1.06  & 0.99 & 28.80    & $99.1^{+13.0}_{-12.4}$           & $ 529.1^{+69.4}_{-66.2}  \pm          35.7  $          & 6.2\\
        4.230   &1100.9  & 1.06  & 0.99 & 28.54    & $173.6^{+18.1}_{-17.4}$          & $ 437.2^{+45.6}_{-43.8}  \pm          29.6  $          & 7.5\\
        4.237   &530.3   & 1.06  & 0.98 & 29.71    & $80.5^{+10.8}_{-10.7}$           & $ 408.4^{+54.8}_{-54.3}  \pm          27.7  $          & 4.6\\
        4.246   &538.1   & 1.06  & 0.97 & 29.61    & $85.9^{+12.7}_{-12.1}$           & $ 435.4^{+64.4}_{-61.3}  \pm          30.7  $          & 5.0\\
        4.260   &828.4   & 1.05  & 0.97 & 29.37    & $161.9^{+9.4}_{-15.8}$           & $ 537.4^{+31.2}_{-52.4}  \pm          37.7  $          & 7.7\\
        4.290   &502.4   & 1.05  & 0.97 & 29.73    & $83.2^{+11.7}_{-11.0}$           & $ 449.8^{+63.3}_{-59.5}  \pm          31.6  $          & 5.7\\
        4.315   &501.2   & 1.05  & 1.00 & 27.10    & $82.4^{+10.6}_{-10.0}$           & $ 475.2^{+61.1}_{-57.7}  \pm          33.4  $          & 5.6\\
        4.340   &505.0   & 1.05  & 1.02 & 27.22    & $90.9^{+11.7}_{-11.0}$           & $ 507.9^{+65.4}_{-61.5}  \pm          34.2  $          & 4.7\\
        4.360   &543.9   & 1.05  & 1.05 & 27.67    & $57.2^{+11.6}_{-11.0}$           & $ 283.6^{+57.5}_{-54.5}  \pm          19.1  $          & 3.2\\
        4.380   &522.7   & 1.05  & 1.07 & 30.30    & $77.5^{+11.8}_{-11.4}$           & $ 358.2^{+54.5}_{-52.7}  \pm          24.1  $          & 4.6\\
        4.400   &507.8   & 1.05  & 1.09 & 27.73    & $61.2^{+11.2}_{-10.6}$           & $ 312.4^{+57.2}_{-54.1}  \pm          21.1  $          & 3.8\\
        4.420   &1090.7  & 1.05  & 1.09 & 27.86    & $122.3^{+16.1}_{-15.5}$          & $ 289.3^{+38.1}_{-36.7}  \pm          19.6  $          & 5.7\\
        4.440   &569.9   & 1.05  & 1.1  & 29.80    & $35.1^{+9.2}_{-8.6}~(<51.0)$     & $ 147.2^{+38.6}_{-36.1}   \pm  9.9  ~(<213.5)$   & 2.6\\
        4.600   &586.9   & 1.06  & 1.11 & 27.89    & $60.9^{+11.6}_{-11.0}$           & $ 262.6^{+50.0}_{-47.4}  \pm          18.3  $          & 3.6\\
        4.640   &552.5   & 1.05  & 1.15 & 30.23    & $17.2^{+9.4}_{-8.7}~(<31.6)$      &$ 71.6^{+39.1}_{-36.2}  \pm          5.0  ~(<136.2)$  & 0.7\\
        4.660   &529.4   & 1.05  & 1.16 & 26.54    & $44.6^{+9.2}_{-8.9}$             & $ 214.3^{+44.2}_{-42.8}  \pm          14.4  $          & 3.3\\
        4.680   &1667.4  & 1.05  & 1.16 & 26.40    & $128.9^{+16.3}_{-15.9}$          & $ 202.1^{+25.6}_{-24.9}  \pm          14.2  $          & 5.3\\
        4.700   &536.5   & 1.06  & 1.17 & 26.33    & $37.9^{+8.2}_{-7.5}~(<59.9)$       & $183.2^{+39.6}_{-36.2}  \pm   12.8 ~(<296.8)$   & 1.9\\
        4.750   &366.6   & 1.06  & 1.17 & 26.50    & $37.1^{+7.4}_{-6.9}~(<47.9)$      & $255.3^{+50.9}_{-47.5}  \pm         17.8  ~(<272.3)$ & 2.2\\
        4.780   &511.5   & 1.06  & 1.18 & 29.31    & $28.2^{+8.6}_{-8.1}~(<40.9)$      & $127.7^{+39.0}_{-36.7}  \pm         8.9  ~(<163.9)$  & 1.9\\
        4.914   &207.8   & 1.06  & 1.19 & 29.20    & $20.6^{+5.5}_{-4.8}~(<28.2)$      & $228.8^{+61.1}_{-53.3}  \pm         15.8  ~(<268.4)$ & 1.6\\
        \hline\hline
        \end{tabular}
        }
        \label{tab:NUM_BCSS}
        \end{table}
 \section{Systematic uncertainty}\label{SYS_UN}
The systematic uncertainties in the measurement of the Born cross-section arise from various sources, which are categorized as multiplicative and additive. Multiplicative terms refer on uncertainties due to kaon tracking and PID efficiencies, $\Xib$ reconstruction, MC simulation sample size, branching fraction and input line shape. The additive terms include signal shape and background shape in fit method.
\subsection{Luminosity}
The luminosities at all energy points are measured using  Bhabha events, with uncertainties of 1.0\% below 4.0 GeV, 0.7\% from 4.0 to 4.6 GeV, and 0.6\% above $4.6\rm~GeV$~\cite{llll,BESIII:2022dxl}.
\subsection{Kaon tracking and PID efficiencies}
The systematic uncertainties associated with the kaon tracking and PID are estimated with a control sample of $J/\psi   \rightarrow K^{*}K$~\cite{ksys} decays The difference in tracking or PID efficiencies between data and MC simulation is $1.0\%$. 
The total systematic uncertainty from these sources is assigned to be $1.4\%$ by adding the tracking and PID uncertainties in quadrature. 
\subsection{$\bar\Xi^{+}$ reconstruction}
The systematic uncertainty due to the $\bar\Xi^{+}$ reconstruction arises from the knowledge of the tracking and PID, and $\Lambda$ reconstruction efficiencies,  and possible biases associated with the required decay length of the $\Lambda/\Xi$, and the $\Lambda/\Xi$ mass window.
The combined uncertainty is estimated with a control sample of  $\psi(3686)\ar\Xi^{-}\bar{\Xi}^{+}$ decays  using the same method described in refs.~\cite{BESIII:2016nix,BESIII:2016ssr, BESIII:2019dve, BESIII:2020ktn, BESIII:2021aer, BESIII:2021gca, BESIII:2022mfx, BESIII:2022lsz, BESIII:2023lkg}.  The efficiency difference between data and MC simulation is found to be $5.1\%$, which is assigned  as the systematic uncertainty.

\subsection{MC simulation sample size}
The systematic uncertainty arising from the MC simulation sample size is calculated as $\sqrt{\frac{\epsilon(1-\epsilon)}{N}}/\epsilon$,  where $\epsilon$ is the detection efficiency and $N$ is the number of generated signal MC events. 

\subsection{MC modeling}
The systematic uncertainty arising from the MC modeling is estimated by comparing the difference in detection efficiencies between the PHSP and HypWK models. The efficiencies are 25.6\% for the HypWK model and 25.7\% for the PHSP model. The difference of signal modeling can be negligible.

\subsection{Fit method}
The sources of the systematic uncertainty in the fit of the $M^{\rm  recoil}_{K^-\Xib}$ spectrum include the signal shape and background shape. The uncertainty due to the signal shape is studied by varying the default signal shape convolved with a Gaussian function, and the yield difference is taken as the systematic uncertainty, which is $1.8\%$ for the $\Lambda$ signal shape and negligible for $\Sigma^0$ signal shape. The uncertainty due to the background modeling is estimated to be $4.0\%$ by alternative fit with a second-order Chebyshev function.

\subsection{Branching fraction} 
The uncertainty of the branching fraction of $\bar\Lambda\ar \bar p\pi^{+}$ is 0.8\% from the PDG~\cite{PDG2020}. The uncertainty on the branching fraction of $\bar\Xi^{+}\ar\pi^{+}\bar\Lambda$ is negligible in the analysis.

\subsection{Input line shape}
The ISR correction and the detection efficiency depend on the line shape of the cross section. The associated systematic uncertainty arises from the statistical uncertainty of the cross sections, which is estimated by varying the central value of the cross section within $\pm1\sigma$ of the statistical uncertainty. Then, the $(1 + \delta)\epsilon$ values for each energy point are recalculated. This process is repeated 3000 times, and a Gaussian function is used to fit the distribution of the 3000 values of  $(1 + \delta)\epsilon$. The deviation of the Gaussian function is taken as the corresponding systematic uncertainty. 

\subsection{Total systematic uncertainty}
The various systematic uncertainties on the Born cross section measurement for $\EE\ar K^-\bar{\Xi}^+\Lambda/\Sigma^{0}$ are summarized in tables~\ref{systematic1} and~\ref{systematic2}.
Assuming all sources are independent, the total systematic uncertainty is determined by adding these values in quadrature.

\begin{table}[!hpt]
	\begin{center}
	{\caption{Systematic uncertainties (in \%) and their sources for each energy point on the Born cross section measurement for $\KXL$. Here, L denotes luminosity, TP denotes tracking and PID, $\Xi^{-}$ Rec denotes $\Xi^{-}$ reconstruction, MC  denotes MC sample size, ${\cal{B}}$ denotes branching fraction, BS and SS denote background shape and signal shape, respectively, and ILS denotes input line shape.}
        \label{systematic1}
	}
	   \resizebox{.95\columnwidth}{!}{
   \begin{tabular}{cccccccccc} \hline\hline
	$\sqrt{s}$ (GeV) & L & Kaon TP& $\bar\Xi^{+}$ Rec. & MC & ${\cal{B}}$ & BS & SS & ILS & Total    \\ \hline 
        3.510  & 1.0 & 1.4 & 5.1 & 0.4 & 0.8 & 4.0 & 1.8 & 0.3 & 7.0\\
        3.581  & 1.0 & 1.4 & 5.1 & 0.4 & 0.8 & 4.0 & 1.8 & 0.1 & 7.0\\
        3.650  & 1.0 & 1.4 & 5.1 & 0.4 & 0.8 & 4.0 & 1.8 & 0.3 & 7.0\\
        3.670  & 1.0 & 1.4 & 5.1 & 0.4 & 0.8 & 4.0 & 1.8 & 0.4 & 7.0\\
        3.773  & 1.0 & 1.4 & 5.1 & 0.4 & 0.8 & 4.0 & 1.8 & 0.3 & 7.0\\
        3.867  & 1.0 & 1.4 & 5.1 & 0.4 & 0.8 & 4.0 & 1.8 & 2.3 & 7.4\\
        3.871  & 1.0 & 1.4 & 5.1 & 0.4 & 0.8 & 4.0 & 1.8 & 2.2 & 7.3\\
        4.009  & 0.7 & 1.4 & 5.1 & 0.4 & 0.8 & 4.0 & 1.8 & 1.9 & 7.2\\
        4.130  & 0.7 & 1.4 & 5.1 & 0.4 & 0.8 & 4.0 & 1.8 & 1.9 & 7.2\\
        4.160  & 0.7 & 1.4 & 5.1 & 0.4 & 0.8 & 4.0 & 1.8 & 1.3 & 7.1\\
        4.180  & 0.7 & 1.4 & 5.1 & 0.4 & 0.8 & 4.0 & 1.8 & 0.1 & 7.0\\
        4.190  & 0.7 & 1.4 & 5.1 & 0.4 & 0.8 & 4.0 & 1.8 & 1.3 & 7.1\\
        4.200  & 0.7 & 1.4 & 5.1 & 0.4 & 0.8 & 4.0 & 1.8 & 2.1 & 7.3\\
        4.210  & 0.7 & 1.4 & 5.1 & 0.4 & 0.8 & 4.0 & 1.8 & 1.2 & 7.1\\
        4.220  & 0.7 & 1.4 & 5.1 & 0.4 & 0.8 & 4.0 & 1.8 & 1.7 & 7.2\\
        4.230  & 0.7 & 1.4 & 5.1 & 0.4 & 0.8 & 4.0 & 1.8 & 0.4 & 7.0\\
        4.237  & 0.7 & 1.4 & 5.1 & 0.4 & 0.8 & 4.0 & 1.8 & 1.3 & 7.1\\
        4.246  & 0.7 & 1.4 & 5.1 & 0.4 & 0.8 & 4.0 & 1.8 & 1.9 & 7.2\\
        4.260  & 0.7 & 1.4 & 5.1 & 0.4 & 0.8 & 4.0 & 1.8 & 2.0 & 7.2\\
        4.290  & 0.7 & 1.4 & 5.1 & 0.4 & 0.8 & 4.0 & 1.8 & 1.5 & 7.1\\
        4.315  & 0.7 & 1.4 & 5.1 & 0.4 & 0.8 & 4.0 & 1.8 & 2.0 & 7.2\\
        4.340  & 0.7 & 1.4 & 5.1 & 0.4 & 0.8 & 4.0 & 1.8 & 0.3 & 7.0\\
        4.360  & 0.7 & 1.4 & 5.1 & 0.3 & 0.8 & 4.0 & 1.8 & 0.4 & 7.0\\
        4.380  & 0.7 & 1.4 & 5.1 & 0.4 & 0.8 & 4.0 & 1.8 & 1.2 & 7.1\\
        4.400  & 0.7 & 1.4 & 5.1 & 0.4 & 0.8 & 4.0 & 1.8 & 0.3 & 7.0\\
        4.420  & 0.7 & 1.4 & 5.1 & 0.4 & 0.8 & 4.0 & 1.8 & 0.5 & 7.0\\
        4.440  & 0.7 & 1.4 & 5.1 & 0.4 & 0.8 & 4.0 & 1.8 & 1.4 & 7.1\\
        4.600  & 0.7 & 1.4 & 5.1 & 0.3 & 0.8 & 4.0 & 1.8 & 0.4 & 7.0\\
        4.640  & 0.6 & 1.4 & 5.1 & 0.4 & 0.8 & 4.0 & 1.8 & 1.3 & 7.1\\
        4.660  & 0.6 & 1.4 & 5.1 & 0.4 & 0.8 & 4.0 & 1.8 & 0.9 & 7.0\\
        4.680  & 0.6 & 1.4 & 5.1 & 0.4 & 0.8 & 4.0 & 1.8 & 0.6 & 7.0\\
        4.700  & 0.6 & 1.4 & 5.1 & 0.4 & 0.8 & 4.0 & 1.8 & 0.7 & 7.0\\
        4.750  & 0.6 & 1.4 & 5.1 & 0.4 & 0.8 & 4.0 & 1.8 & 1.2 & 7.1\\
        4.780  & 0.6 & 1.4 & 5.1 & 0.4 & 0.8 & 4.0 & 1.8 & 1.4 & 7.1\\
        4.914  & 0.6 & 1.4 & 5.1 & 0.4 & 0.8 & 4.0 & 1.8 & 0.9 & 7.0\\
			\hline\hline
			\end{tabular}
				}
	\end{center}
\end{table}

\begin{table}[!hpt]
	\begin{center}
	{\caption{Systematic uncertainties (in \%) and their sources for each energy point on the Born cross section measurement for $\KXS$. Here, L denotes luminosity, TP denotes tracking and PID, $\Xi^{-}$ Rec denotes $\Xi^{-}$ reconstruction, MC denotes MC sample size, ${\cal{B}}$ denotes branching fraction, BS and SS denote background shape and signal shape, respectively, and ILS denotes input line shape.}
        \label{systematic2}
	}
	   \resizebox{.95\columnwidth}{!}{
   \begin{tabular}{cccccccccc} \hline\hline
	$\sqrt{s}$ (GeV) & L & Kaon TP & $\bar\Xi^{+}$ Rec. & MC  & ${\cal{B}}$ &  BS &SS& ILS & Total    \\ \hline 
        3.510  &1.0 &1.4 &5.1 & 0.4 & 0.8 & 4.0 &0.0& 0.1& 6.8\\
        3.581  &1.0 &1.4 &5.1 & 0.4 & 0.8 & 4.0 &0.0& 0.1& 6.8\\
        3.650  &1.0 &1.4 &5.1 & 0.4 & 0.8 & 4.0 &0.0& 0.1& 6.8\\
        3.670   &1.0 &1.4 &5.1 & 0.4 & 0.8 & 4.0 &0.0& 0.1& 6.8\\
        3.773  &1.0 &1.4 &5.1 & 0.4 & 0.8 & 4.0 &0.0& 0.1& 6.8\\
        3.867  &1.0 &1.4 &5.1 & 0.4 & 0.8 & 4.0 &0.0& 0.2& 6.8\\
        3.871  &1.0 &1.4 &5.1 & 0.4 & 0.8 & 4.0 &0.0& 0.2& 6.8\\
        4.009  &0.7 &1.4 &5.1 & 0.4 & 0.8 & 4.0 &0.0& 0.4& 6.7\\
        4.130  &0.7 &1.4 &5.1 & 0.4 & 0.8 & 4.0 &0.0& 0.6& 6.8\\
        4.160  &0.7 &1.4 &5.1 & 0.4 & 0.8 & 4.0 &0.0& 0.4& 6.7\\
        4.180  &0.7 &1.4 &5.1 & 0.4 & 0.8 & 4.0 &0.0& 0.5& 6.7\\
        4.190  &0.7 &1.4 &5.1 & 0.4 & 0.8 & 4.0 &0.0& 0.9& 6.8\\
        4.200  &0.7 &1.4 &5.1 & 0.4 & 0.8 & 4.0 &0.0& 0.3& 6.7\\
        4.210  &0.7 &1.4 &5.1 & 0.4 & 0.8 & 4.0 &0.0& 2.4& 7.1\\
        4.220  &0.7 &1.4 &5.1 & 0.4 & 0.8 & 4.0 &0.0& 0.4& 6.7\\
        4.230  &0.7 &1.4 &5.1 & 0.4 & 0.8 & 4.0 &0.0& 0.7& 6.8\\
        4.237  &0.7 &1.4 &5.1 & 0.4 & 0.8 & 4.0 &0.0& 0.9& 6.8\\
        4.246  &0.7 &1.4 &5.1 & 0.4 & 0.8 & 4.0 &0.0& 2.1& 7.0\\
        4.260  &0.7 &1.4 &5.1 & 0.4 & 0.8 & 4.0 &0.0& 2.0& 7.0\\
        4.290  &0.7 &1.4 &5.1 & 0.4 & 0.8 & 4.0 &0.0& 2.0& 7.0\\
        4.315  &0.7 &1.4 &5.1 & 0.4 & 0.8 & 4.0 &0.0& 2.0& 7.0\\
        4.340  &0.7 &1.4 &5.1 & 0.4 & 0.8 & 4.0 &0.0& 0.3& 6.7\\
        4.360  &0.7 &1.4 &5.1 & 0.3 & 0.8 & 4.0 &0.0& 0.5& 6.7\\
        4.380  &0.7 &1.4 &5.1 & 0.4 & 0.8 & 4.0 &0.0& 0.4& 6.7\\
        4.400  &0.7 &1.4 &5.1 & 0.4 & 0.8 & 4.0 &0.0& 0.6& 6.8\\
        4.420  &0.7 &1.4 &5.1 & 0.4 & 0.8 & 4.0 &0.0& 0.7& 6.8\\
        4.440  &0.7 &1.4 &5.1 & 0.4 & 0.8 & 4.0 &0.0& 0.4& 6.7\\
        4.600  &0.7 &1.4 &5.1 & 0.3 & 0.8 & 4.0 &0.0& 1.8& 7.0\\
        4.640  &0.6 &1.4 &5.1 & 0.4 & 0.8 & 4.0 &0.0& 2.0& 7.0\\
        4.660  &0.6 &1.4 &5.1 & 0.4 & 0.8 & 4.0 &0.0& 0.4& 6.7\\
        4.680  &0.6 &1.4 &5.1 & 0.4 & 0.8 & 4.0 &0.0& 2.0& 7.0\\
        4.700  &0.6 &1.4 &5.1 & 0.4 & 0.8 & 4.0 &0.0& 1.9& 7.0\\
        4.750  &0.6 &1.4 &5.1 & 0.4 & 0.8 & 4.0 &0.0& 1.8& 7.0\\
        4.780  &0.6 &1.4 &5.1 & 0.4 & 0.8 & 4.0 &0.0& 1.8& 7.0\\
        4.914  &0.6 &1.4 &5.1 & 0.4 & 0.8 & 4.0 &0.0& 1.6& 6.9\\
			\hline\hline
			\end{tabular}
				}
	\end{center}
\end{table}
 \section{Fit to the dressed cross section}
The potential resonances in the line shape of the cross section for $\KXX$ are studied by fitting the dressed cross section, $\sigma^{\rm dressed} =\sigma^{B}/|1-\Pi|^2$ (without the VP effect) with the least $\chi^{2}$ method. The fit minimizes
  \begin{equation}
\chi^{2} = \Delta X^{T}V^{-1}\Delta X,
 \end{equation}
where $\Delta X$ is the vector of residuals between measured and fitted cross section. The covariance matrix $V$ incorporates the correlated and uncorrelated uncertainties among different energy points, where the systematic uncertainties due to the luminosity, kaon tracking and PID, $\bar\Xi^+(\Xi^-)$ reconstruction, and branching fraction are assumed to be fully correlated among the CM energies,  and the other sources of uncertainty are taken to be uncorrelated.

\begin{figure}[H]
  \begin{center}
  \includegraphics[width=0.45\textwidth]{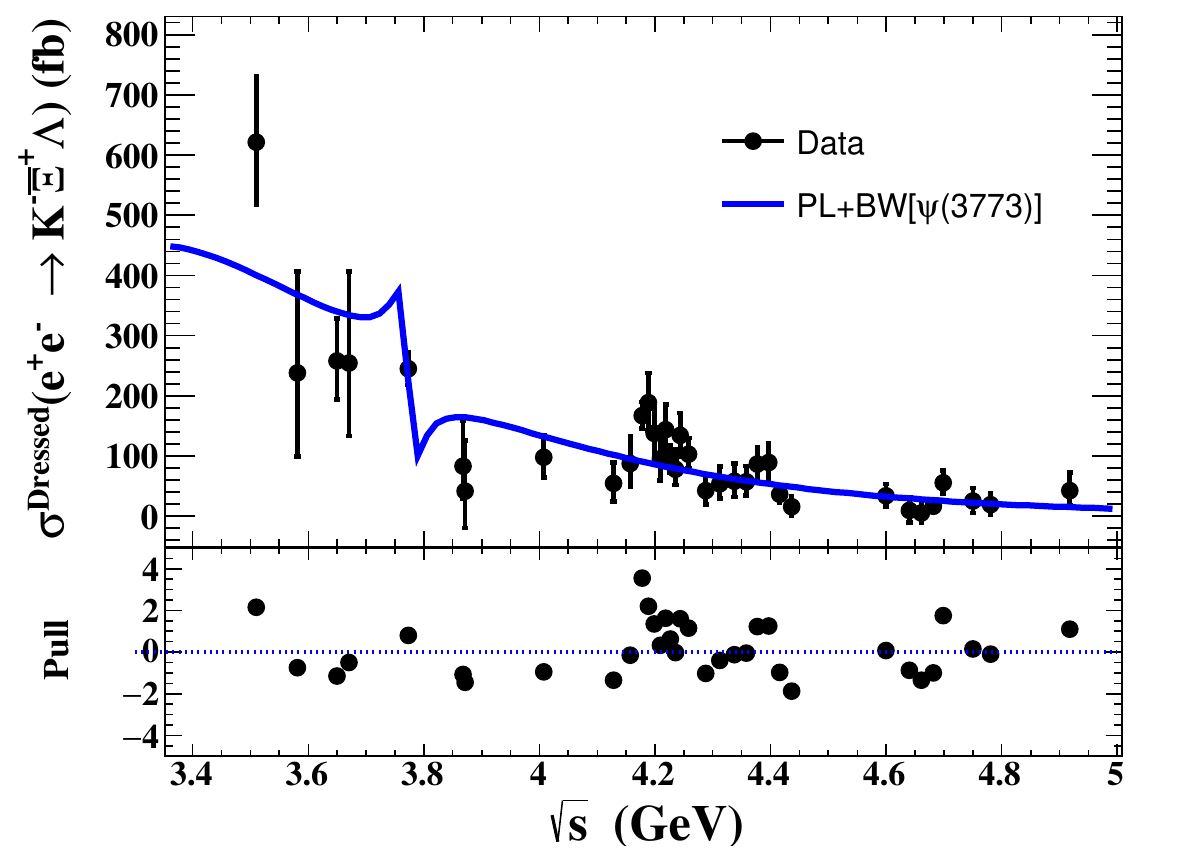}
  \includegraphics[width=0.45\textwidth]{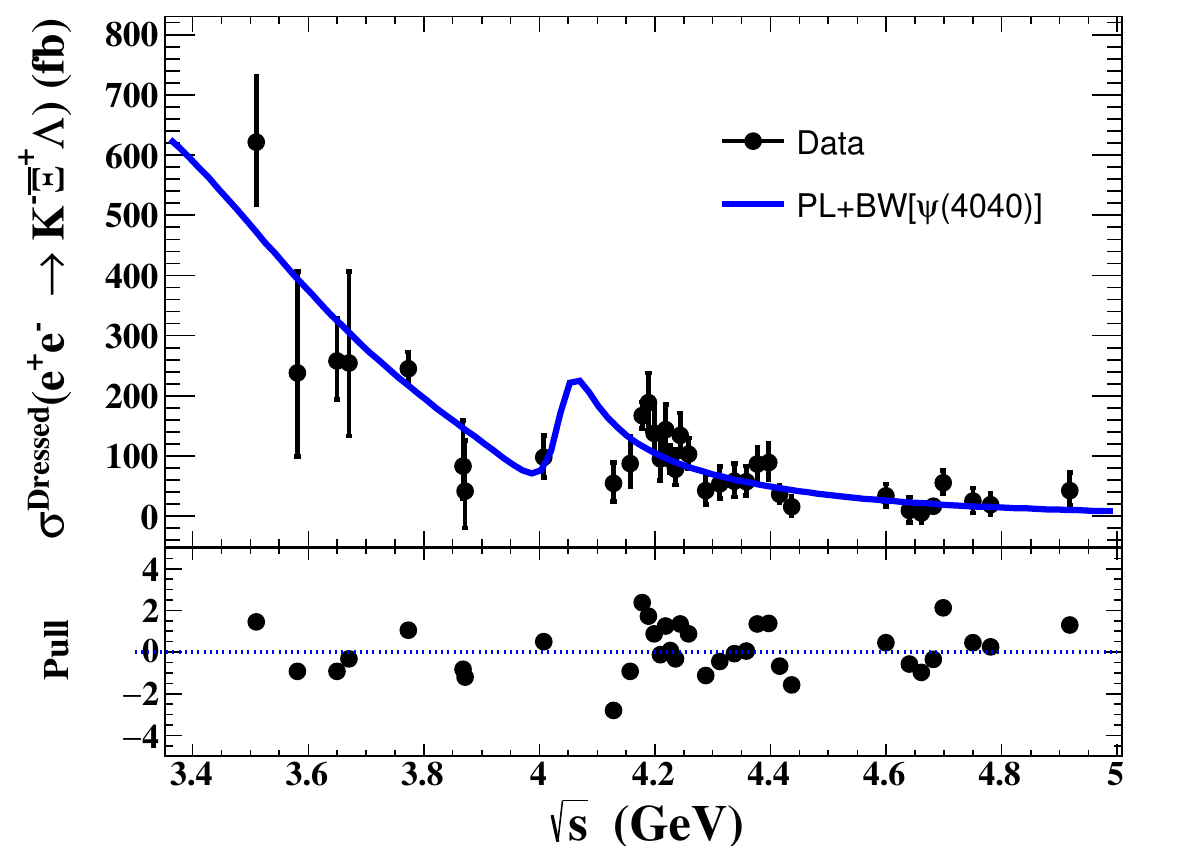}\\
  \includegraphics[width=0.45\textwidth]{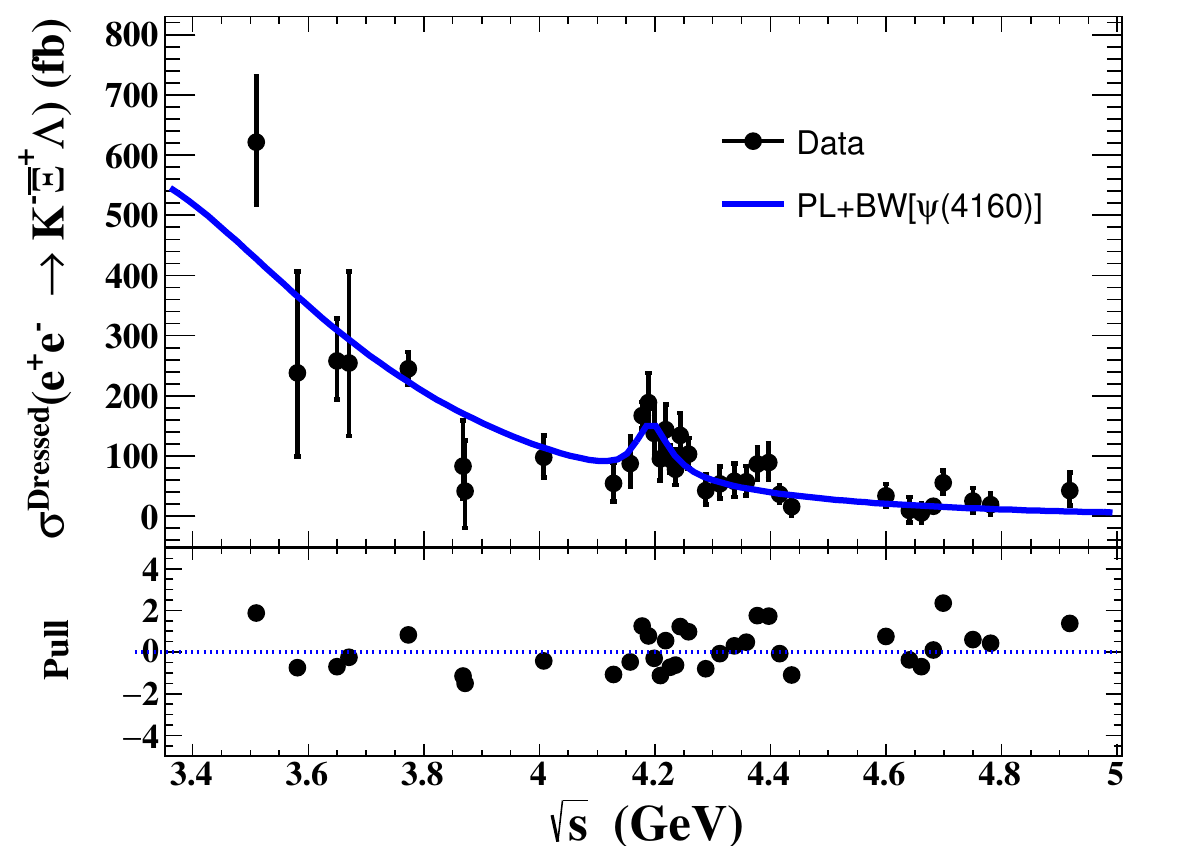}
  \includegraphics[width=0.45\textwidth]{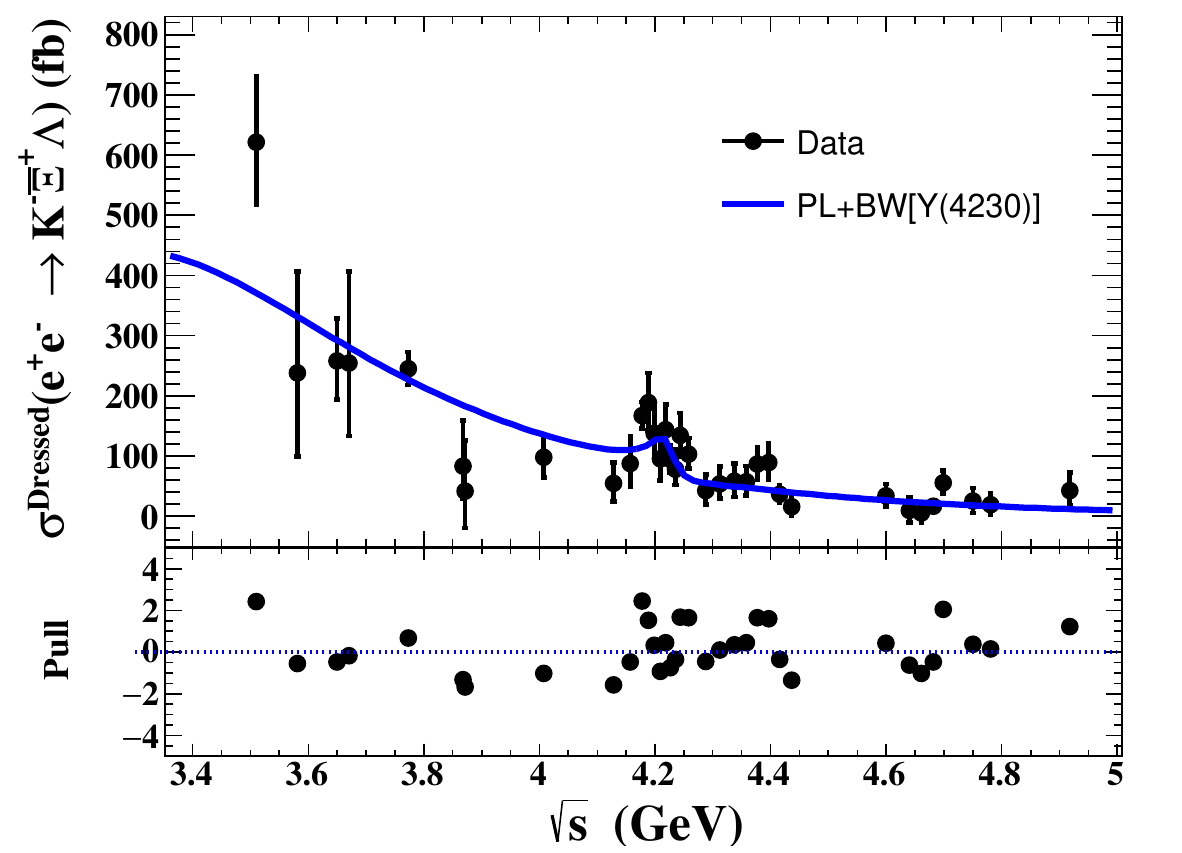}\\
  \includegraphics[width=0.45\textwidth]{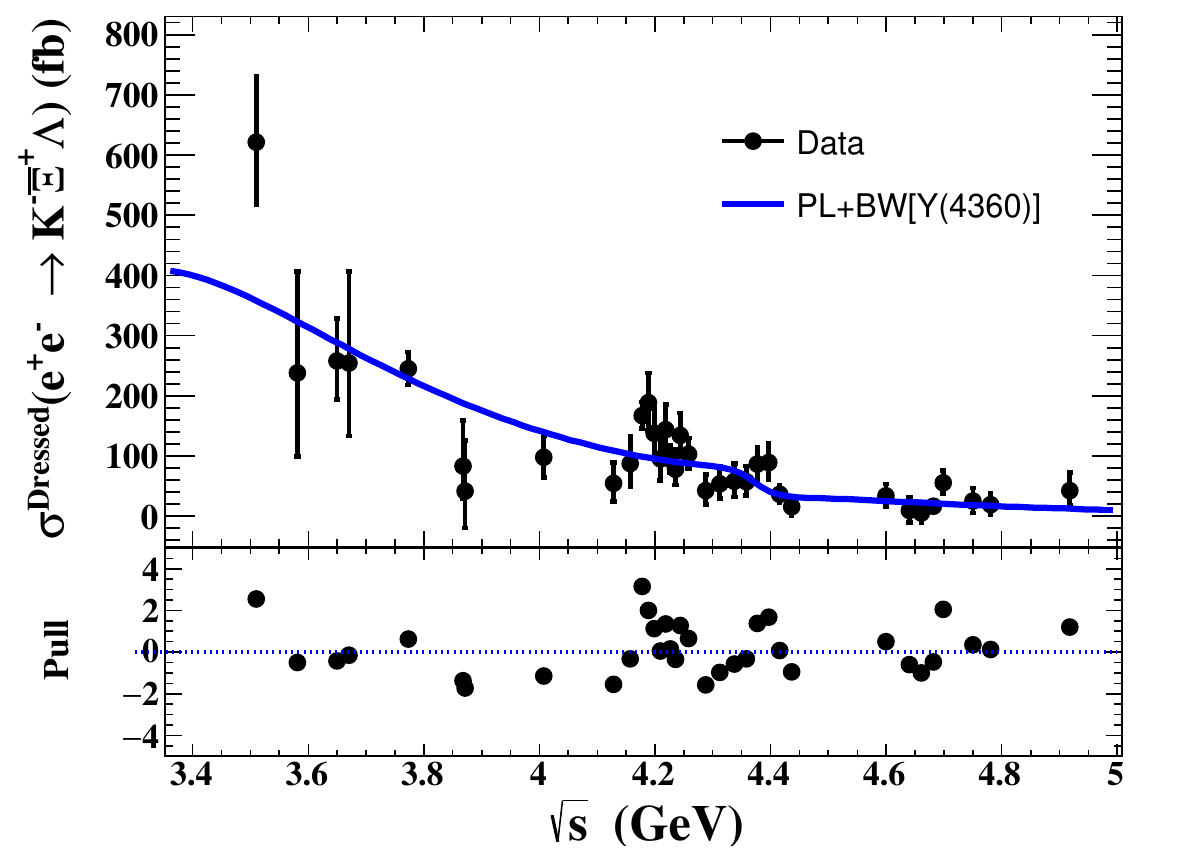}
  \includegraphics[width=0.45\textwidth]{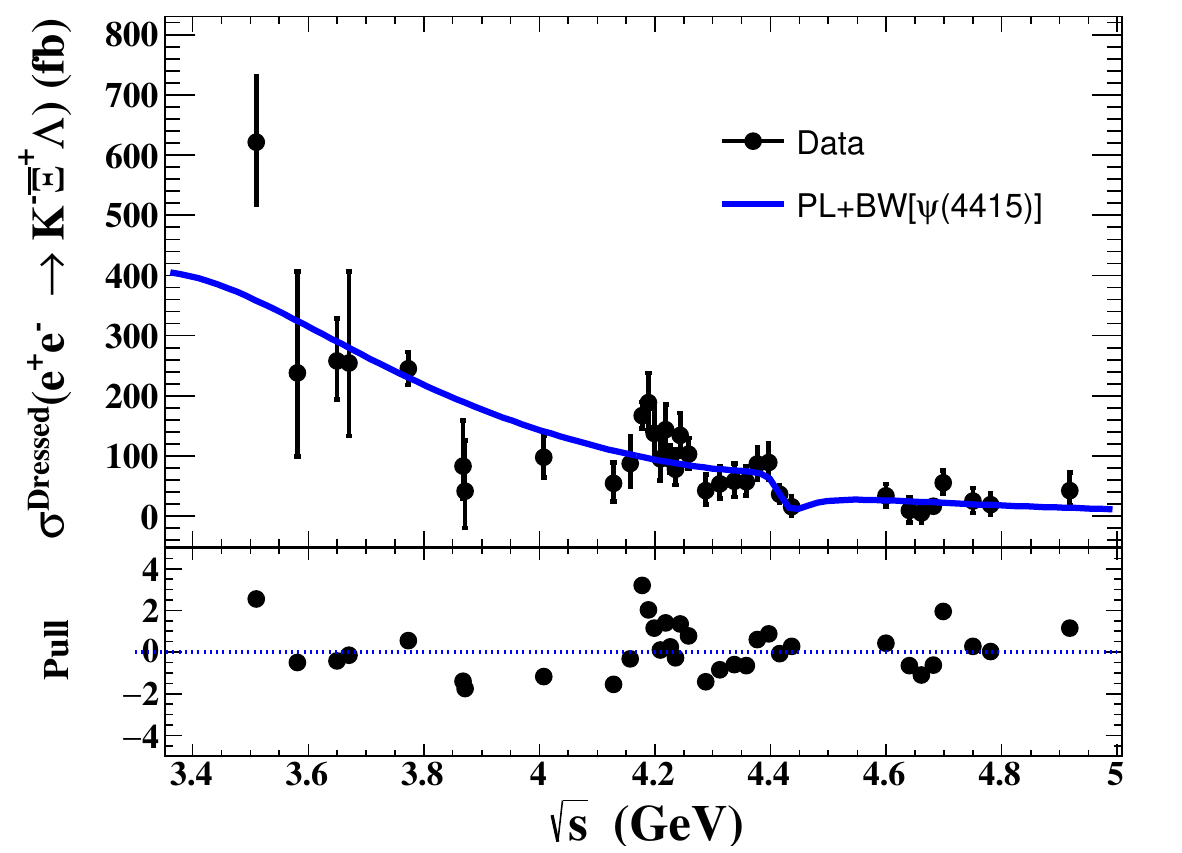}\\
  \includegraphics[width=0.45\textwidth]{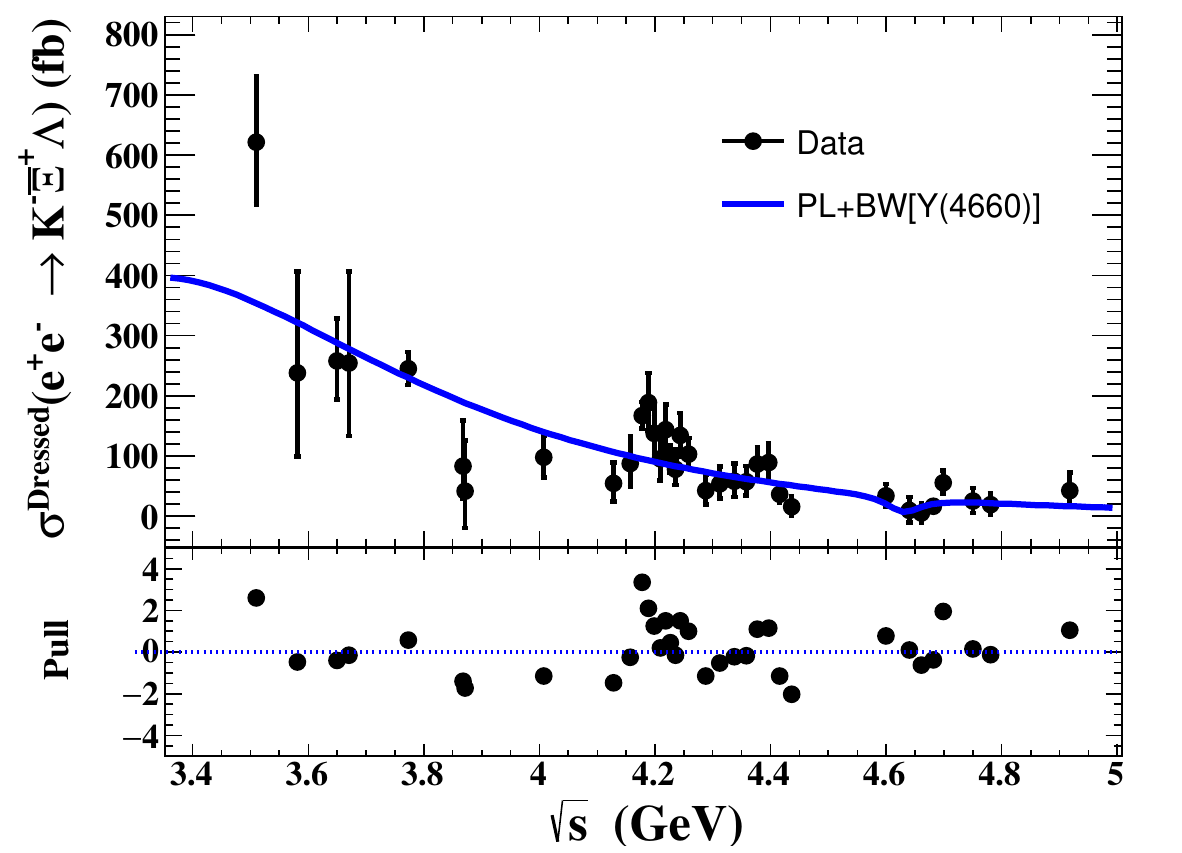}
  \includegraphics[width=0.45\textwidth]{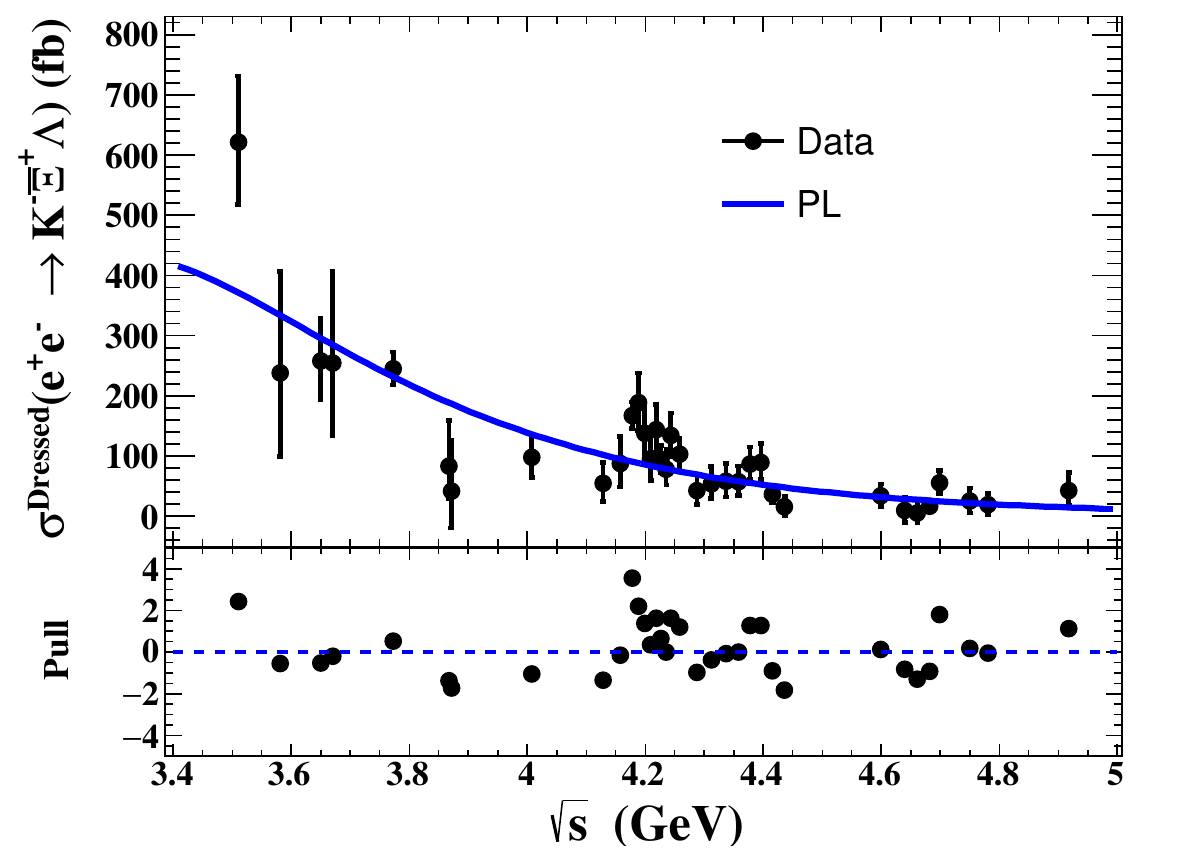}
  \end{center}
  \caption{Fits to the dressed cross sections of $\EE\ar K^-\bar{\Xi}^+\Lambda$ with an assumption of a resonance ($\psi(4040)$, $\psi(4160)$, $Y(4230)$, $Y(4360)$, $\psi(4415)$ or $Y(4660)$) plus a continuum contribution. The blue solid line is the fit result.}
  \label{Fig:Fitanother}
\end{figure}

\begin{figure}[H]
  \begin{center}
  \includegraphics[width=0.45\textwidth]{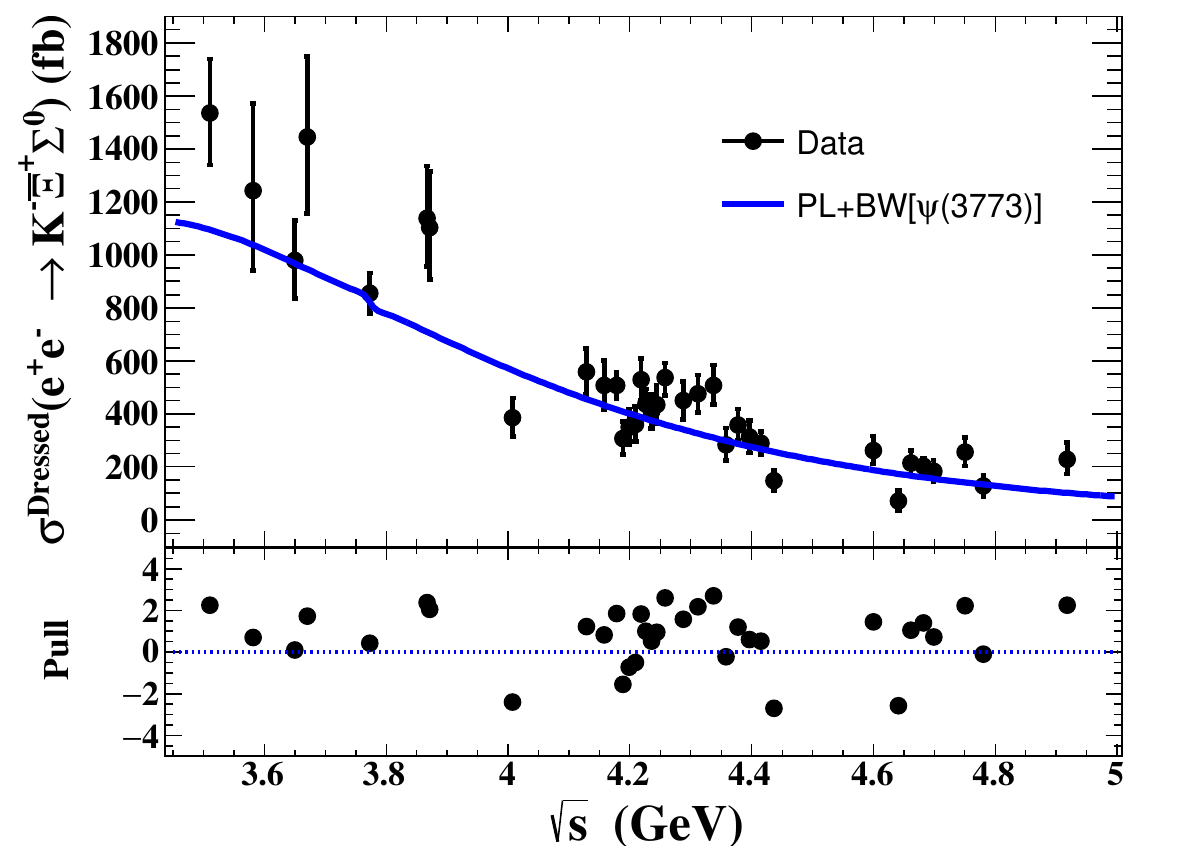}
  \includegraphics[width=0.45\textwidth]{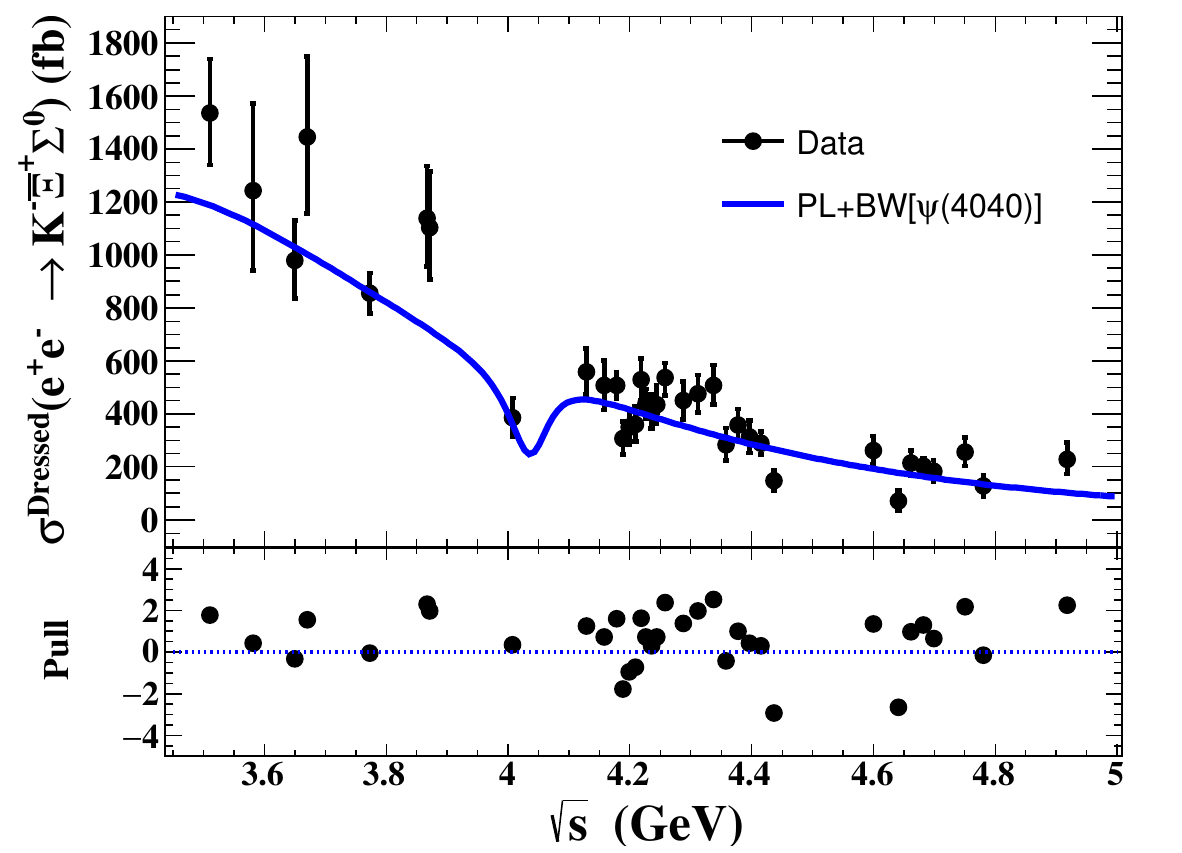}\\
  \includegraphics[width=0.45\textwidth]{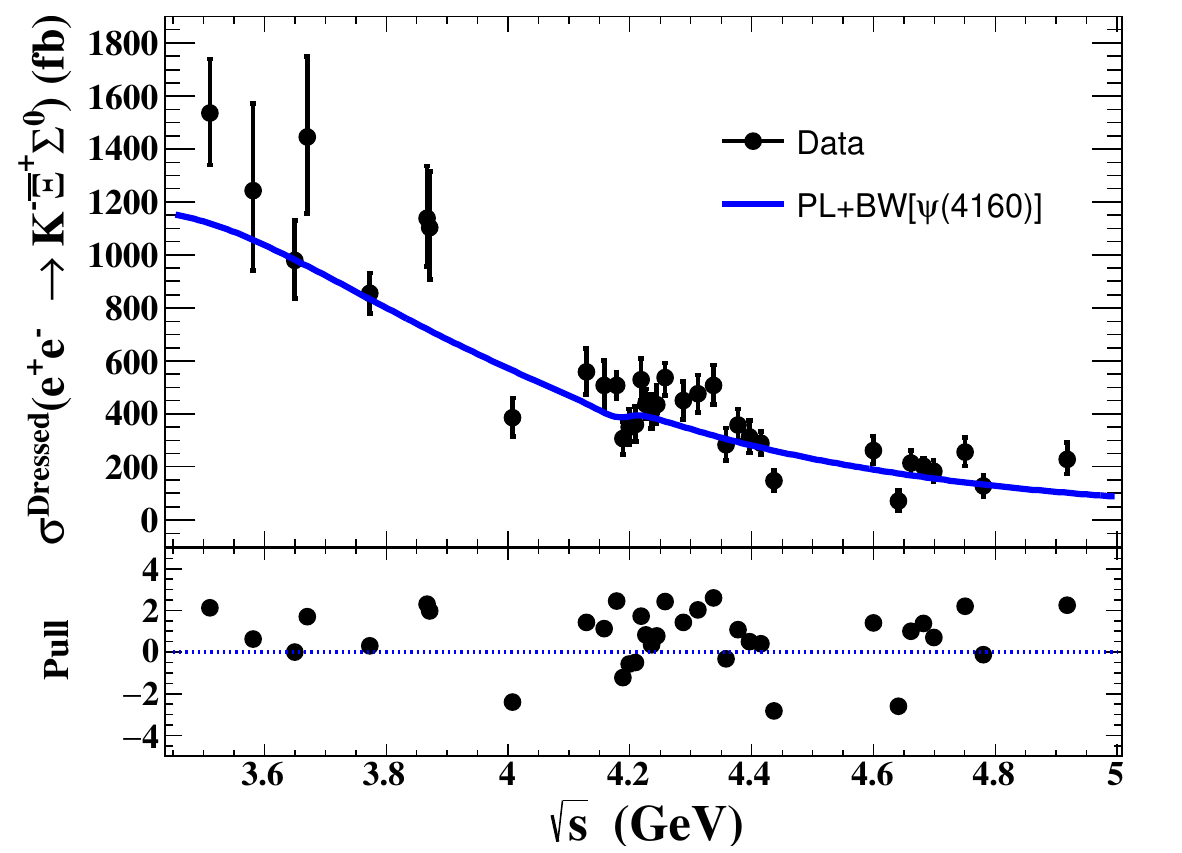}
  \includegraphics[width=0.45\textwidth]{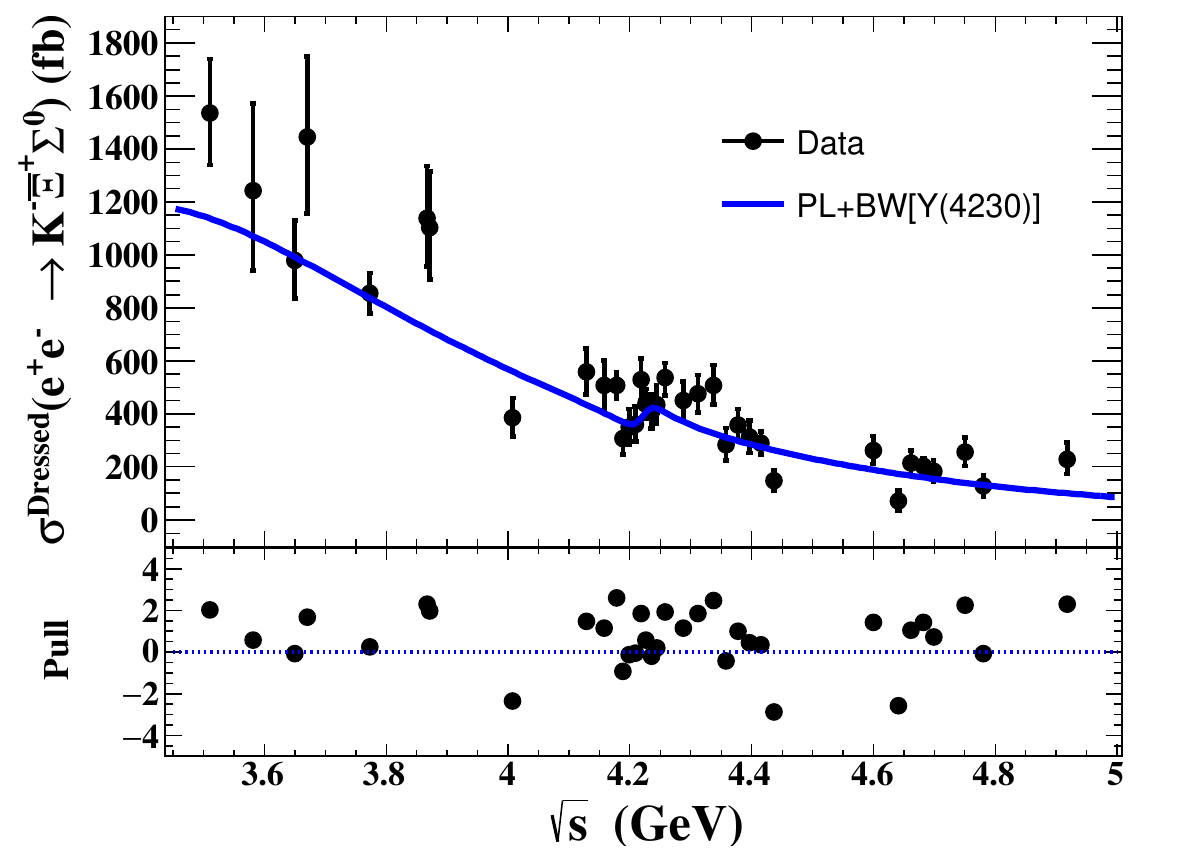}\\
  \includegraphics[width=0.45\textwidth]{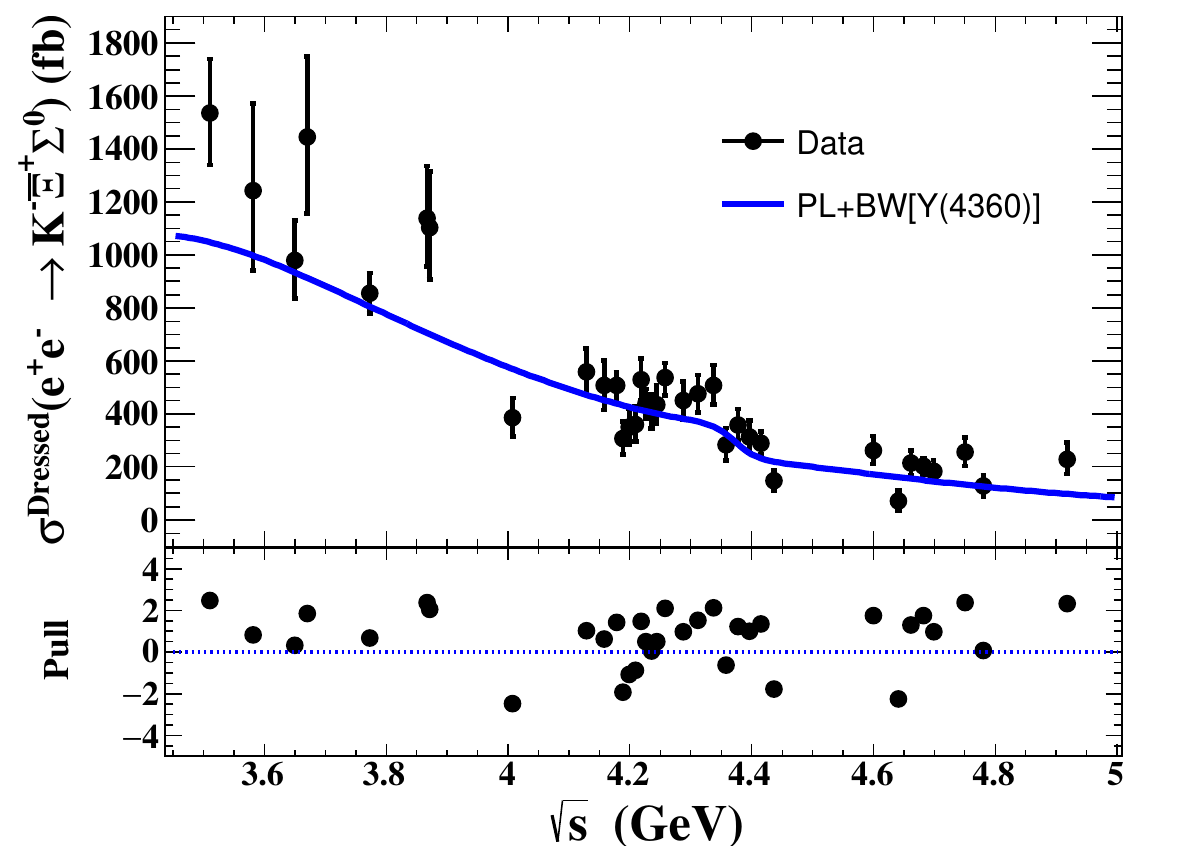}
  \includegraphics[width=0.45\textwidth]{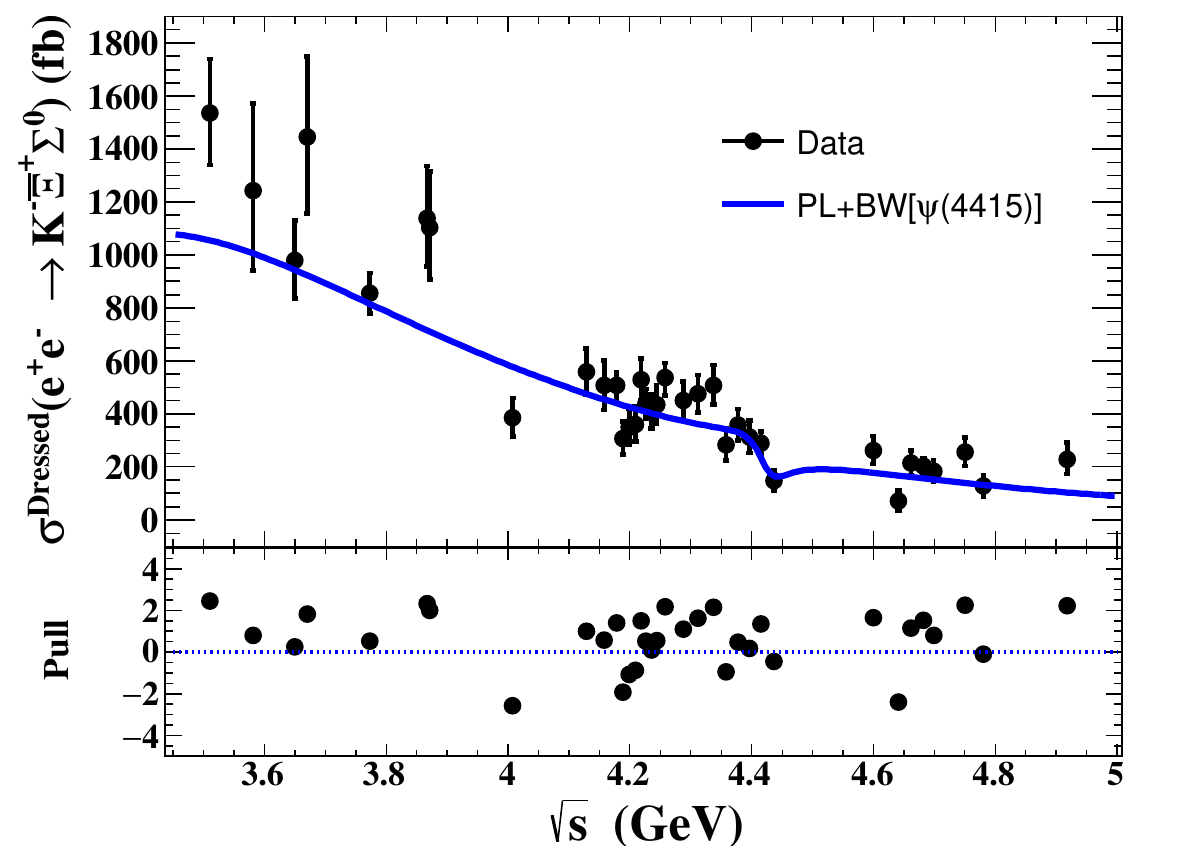}\\
  \includegraphics[width=0.45\textwidth]{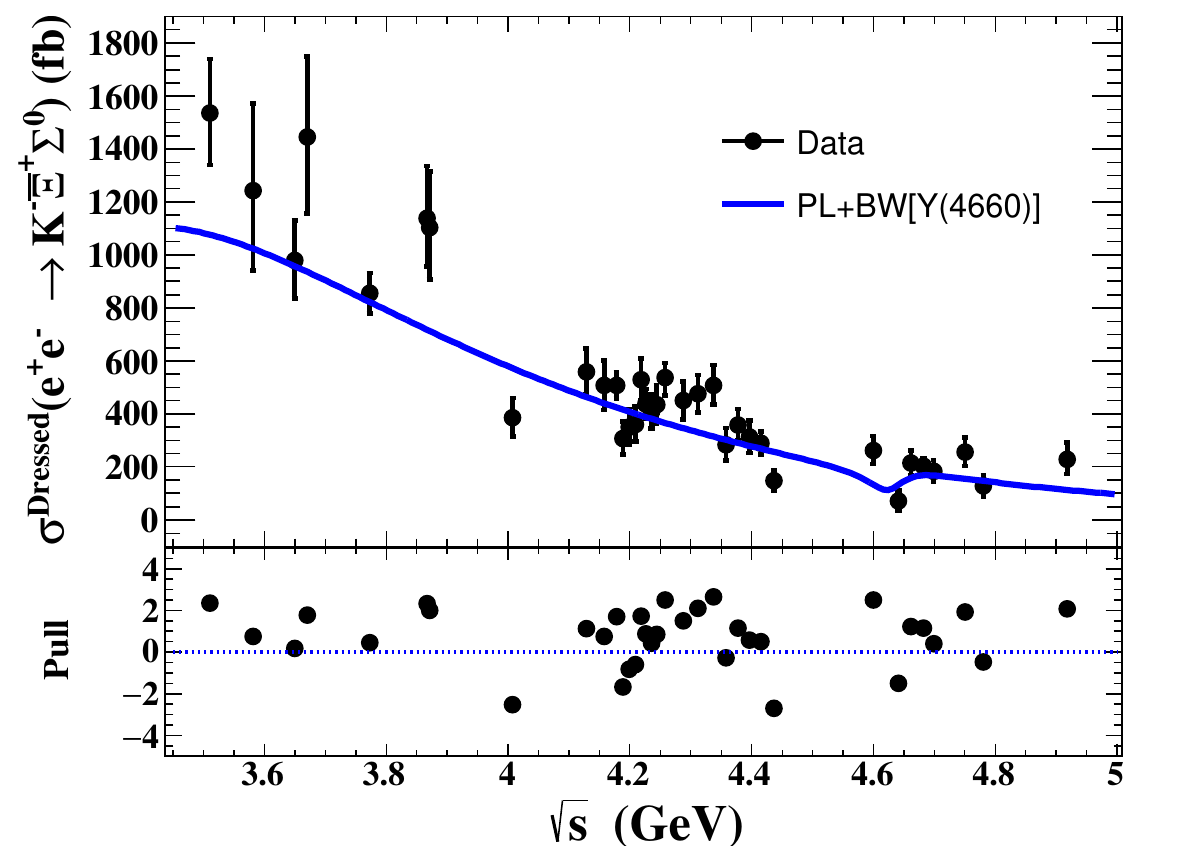}
  \includegraphics[width=0.45\textwidth]{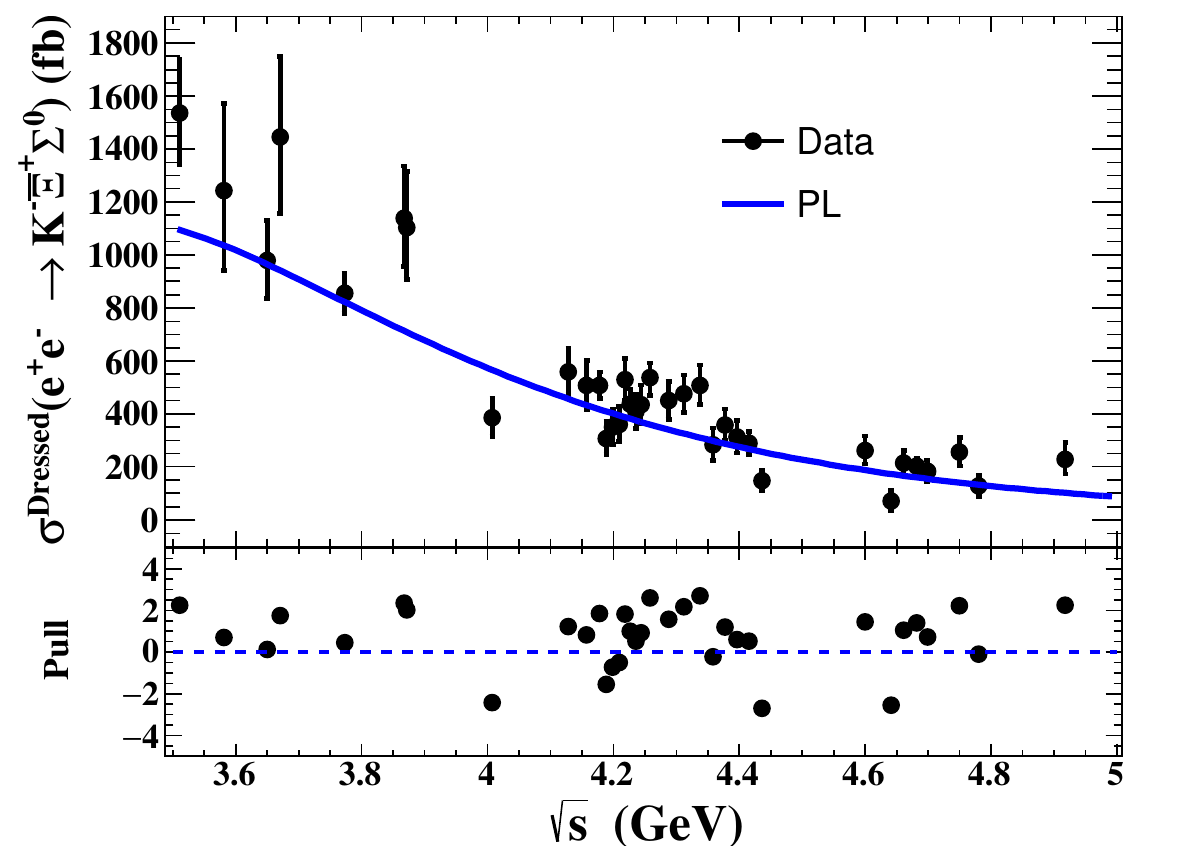}\\
  \end{center}
  \caption{Fits to the dressed cross sections of $\EE\ar K^-\Bar{\Xi}^+\Sigma^0$ with an assumption of a resonance ($\psi(4040)$, $\psi(4160)$, $Y(4230)$, $Y(4360)$, $\psi(4415)$ or $Y(4660)$) plus a continuum contribution. The blue solid line is the fit result.}
  \label{Fig:XiXi::CS::Line-shape_other}
\end{figure}

Assuming the $\KXX$ signals are produced by a resonance decay and the continuum process, a fit to the dressed cross section is applied with the coherent sum of a power-law (PL) function plus a Breit-Wigner (BW) function defined as
 \begin{gather}
	\sigma^{\rm dressed}(\sqrt{s}) = \left|c_{0}\frac{\sqrt{P(\sqrt{s})}}{\sqrt{s}^{n}} + e^{i\phi}{\rm BW}(\sqrt{s})\sqrt{\frac{P(\sqrt{s})}{P(M)}}\right|^{2},\\
     {\rm BW}(\sqrt{s}) = \frac{\sqrt{12\pi\Gamma_{ee}{\cal{B}}\Gamma}}{s-M^{2}+iM\Gamma}.
      \end{gather}
 Here $\phi$ is the relative phase between the BW function and the PL function, $c_0$ and $n$ are free fit parameters, $\sqrt{P(\sqrt{s})}$ is the three-body PHSP factor, the mass $M$ and total width $\Gamma$ are fixed to the assumed resonance with the PDG values~\cite{PDG2020}, and $\Gamma_{ee}{\cal{B}}$ is the product of the electronic partial width, and the branching fraction for the assumed resonance decaying into the $K\Xib\Lambda/\Sigma^0$ final state. The significance for each resonance, after considering the systematic uncertainty, is calculated by comparing the change of $\chi^{2}/n.d.f$ with and without the resonance hypothesis.  
Evidence for the $ \psi(4160)\ar K^-\Xib\Lambda$ decay with a significance of 4.4$\sigma$ is found.  Additional possible charmonium (-like) states are included in the fit, but no significant signal is found for any other contribution.
Thus, the upper limits of the products of branching fraction and the electronic partial width for these charmonium(-like) states decaying into the $K\Xib\Lambda/\Sigma^0$ final state including systematic uncertainty are provided at the 90\% C.L. using the Bayesian approach~\cite{Zhu:2008ca}.
Figures~\ref{Fig:Fitanother} and \ref{Fig:XiXi::CS::Line-shape_other}
show the fit to the dressed cross section of $K^-\bar\Xi^+\Lambda$ and $K^-\bar\Xi^+\Sigma^0$ with resonances included [i.e. $\psi(3770)$, $\psi(4040)$, $\psi(4160)$, $Y(4230)$, $Y(4360)$, $\psi(4415)$, or $Y(4660)$], and without. The possible multi-solutions of resonances parameters for the fit of dressed cross sections are obtained based on a two dimensional scan method which scans all the pairs of $\Gamma_{ee}\mathcal{B}$ and $\phi$ in parameter space. And the fit results are summarized in table~\ref{tab:multisolution}.

\begin{table}[]
\centering
\caption{The fitted parameters to the dressed cross section for the $\KXL$ and $K^-\bar\Xi^+\Sigma^0$ processes with two solutions (\uppercase\expandafter{\romannumeral1} and \uppercase\expandafter{\romannumeral2}). The fit procedure includes both statistical and systematic uncertainties except for the CM energy calibration. The relative phase is given by $\phi$. ${\cal{B}}$ is the branching fraction of the assumed resonance decaying into the final state. Note that the values in the brackets represent the upper limit at 90$\%$ C.L. with a most conservative evaluation.}
\resizebox{15.cm}{!}{
\renewcommand{\arraystretch}{1.2}
\begin{tabular}{c|c|c|c|c|c|c}
\hline
\hline
\multicolumn{7}{c}{$K^-\Bar{\Xi}^+\Lambda$}\\
\hline
% \hline
\multirow{2}{*}{Resonance}  & \multicolumn{2}{|c|}{$\Gamma_{ee}\mathcal{B}$ ($10^{-3}$ eV)}& \multicolumn{2}{|c|}{$\phi$ (rad)} &\multirow{2}{*}{$ \chi^2/n.d.f$}&\multirow{2}{*}{$S $ $(\sigma)$}\\
\cline{2-5}
& \uppercase\expandafter{\romannumeral1} & \uppercase\expandafter{\romannumeral2}&  \uppercase\expandafter{\romannumeral1} & \uppercase\expandafter{\romannumeral2} &   &\\
\hline
$\psi(3770)$ &$21.0$ $\pm$ 3.7 $(<25.0)$& $1.7$ $\pm$ 0.5 & $-1.9$ $\pm$ 0.3 &$-2.8$ $\pm$ 0.2& $1.8$     & 0.5 \\
% \hline
$\psi(4040)$ & $45.0 $	$\pm$ $6.3 ~(< 62.0) $   & $5.1 $	$\pm$ 2.3 & $-1.3 $	$\pm$ 0.1  &$-1.3 $	$\pm$ 0.1& 1.4     & 2.8 \\

$\psi(4160)$ &$2.1  $	$\pm$ 0.2    & $1.5  $	$\pm$ 0.4 &$-1.6 $	$\pm$ 0.1   &$-1.3$	$\pm$ 0.2 & 1.1     & 4.4 \\
% \hline
$\psi(4230)$ &$21.3  $	$\pm$ 1.5 $(< 24.9)$   & $0.6  $	$\pm$ 0.3 &$-1.8 $	$\pm$ 0.1 &$2.5 $	$\pm$ 0.3 & 1.5     & 2.8 \\
% \hline
$\psi(4360)$ & $28.9  $	$\pm$ 2.7 $(< 35.8)$  & $0.6  $	$\pm$ 0.1 &$-1.8 $	$\pm$ 0.1 &$-2.9$	$\pm$ 0.1& 1.6    & 1.7 \\
% \hline
$\psi(4415)$ & $9.3 $	$\pm$ 2.3 $(<14.3)$  & $1.7 $	$\pm$ 1.1 & $-1.9 $	$\pm$ 0.1  &$-2.3 $	$\pm$ 0.2  & 1.6    & 1.2 \\
% \hline
$\psi(4660)$ & $6.8  $	$\pm$ 3.5 $(<13.0)$ &  $0.8  $	$\pm$ 1.5 &$-1.6 $	$\pm$ 0.1 &$-1.6 $	$\pm$ 0.1 & 1.7    & 1.2 \\
\hline
% \hline
\multicolumn{7}{c}{$K^-\Bar{\Xi}^+\Sigma^0$}\\
\hline
% \hline
\multirow{2}{*}{Resonance}  & \multicolumn{2}{|c|}{$\Gamma_{ee}\mathcal{B}$ ($10^{-3}$ eV)}& \multicolumn{2}{|c|}{$\phi$ (rad)} &\multirow{2}{*}{$ \chi^2/n.d.f$}&\multirow{2}{*}{$S $ $(\sigma)$}\\
\cline{2-5}
& \uppercase\expandafter{\romannumeral1} & \uppercase\expandafter{\romannumeral2}&  \uppercase\expandafter{\romannumeral1} & \uppercase\expandafter{\romannumeral2} &   &\\
\hline
$\psi(3770)$ &$83.1$ $\pm$ 3.2 $(<89.5)$& $0.3$ $\pm$ 3.3 & $-1.6$ $\pm$ 0.2 &$-2.7$ $\pm$ 1.7& 2.2     & 1.5 \\
% \hline
$\psi(4040)$ &$5.3 $	$\pm$ $2.5 ~(< 12.5) $   & $4.2 $	$\pm$ $2.3$ & $-1.3 $	$\pm$ 0.3  & $-1.1 $	$\pm$ 0.3& 2.0     & 2.0 \\
% \hline
$\psi(4160)$ &$0.4  $	$\pm$ 0.7 $(<1.5)$   & $0.1  $	$\pm$ 0.9&$-0.1 $	$\pm$ 0.5   &$0.1 $	$\pm$ 0.4 & 2.3    & 0.9 \\
% \hline
$\psi(4230)$ &$0.6  $	$\pm$ 0.2 $(<1.6)$   & $0.2  $	$\pm$ 0.1 &$0.2 $	$\pm$ 0.3 &$0.3 $	$\pm$ 0.3& 2.3    & 0.9 \\
% \hline
$\psi(4360)$ & $1.1  $	$\pm$ 0.6 $(<2.8)$  & $0.9  $	$\pm$ 0.4  &$2.9 $	$\pm$ 0.3 &$2.9 $	$\pm$ 0.3& 2.0    & 1.0 \\
% \hline
$\psi(4415)$ & $77.0 $	$\pm$ 4.5 $(<87.0)$  & $1.8$	$\pm$ 0.8& $-1.7 $	$\pm$ 0.2  &$-2.5 $	$\pm$ 0.2& 2.0   & 2.7 \\
% \hline
$\psi(4660)$ & $62.5  $	$\pm$ 6.2 $(<77.3)$ & $0.6  $	$\pm$ 1.0 &$-1.6 $	$\pm$ 0.2 &$-1.3 $	$\pm$ 0.2& 2.2 & 1.5 \\
\hline
\hline
\end{tabular}
}
\label{tab:multisolution}
\end{table}

 \section{Summary}
Using a total of 25 fb$^{-1}$ of $e^{+}e^{-}$ collision data collected at $\sqrt s$ between 3.510 and \SI{4.914}{GeV} with the BESIII detector at the BEPCII collider, the exclusive Born cross sections for $\EE\ar K^-\bar{\Xi}^+\Lambda/\Sigma^0$ at thirty-five energy points are measured with the partial reconstruction strategy.
A fit to the dressed cross sections for $ \EE\ar K^-\bar{\Xi}^+\Lambda/\Sigma^0$ with the assumption of one resonance plus a continuum contribution is performed. The fitted parameter, $\Gamma_{ee}{\cal{B}}$, for each assumed resonance are summarized in table~\ref{tab:multisolution}. Evidence is found for the $\psi(4160)\ar K^-\Xib\Lambda$ decay with a significance of 4.4$\sigma$ including systematic uncertainty.  No significant signal of any state decaying into the $K\Xib\Lambda/\Sigma^0$ final state is found for other charmonium (-like) resonances.
The upper limits for the product of the electronic partial width and branching fraction for all assumed resonances decaying into the $K^-\bar{\Xi}^+\Lambda/\Sigma^0$ final state are determined. 
These results are valuable as they add to the  experimental information regarding the three-body baryonic decay of charmonium (-like) states, which may provide important  insights into the nature of baryonic production above the open-charm region.
 
\acknowledgments
The BESIII Collaboration thanks the staff of BEPCII and the IHEP computing center for their strong support. This work is supported in part by National Key R\&D Program of China under Contracts Nos. 2020YFA0406400, 2020YFA0406300,  2023YFA1606000; National Natural Science Foundation of China (NSFC) under Contracts Nos. 
12075107, 12247101,
11635010, 11735014, 11835012, 11935015, 11935016, 11935018, 11961141012, 12025502, 12035009, 12035013, 12061131003, 12192260, 12192261, 12192262, 12192263, 12192264, 12192265, 12221005, 12225509, 12235017; 
the 111 Project under Grant No. B20063; 
the Chinese Academy of Sciences (CAS) Large-Scale Scientific Facility Program; the CAS Center for Excellence in Particle Physics (CCEPP); Joint Large-Scale Scientific Facility Funds of the NSFC and CAS under Contract No. U1832207; CAS Key Research Program of Frontier Sciences under Contracts Nos. QYZDJ-SSW-SLH003, QYZDJ-SSW-SLH040; 100 Talents Program of CAS; The Institute of Nuclear and Particle Physics (INPAC) and Shanghai Key Laboratory for Particle Physics and Cosmology; European Union's Horizon 2020 research and innovation programme under Marie Sklodowska-Curie grant agreement under Contract No. 894790; German Research Foundation DFG under Contracts Nos. 455635585, Collaborative Research Center CRC 1044, FOR5327, GRK 2149; Istituto Nazionale di Fisica Nucleare, Italy; Ministry of Development of Turkey under Contract No. DPT2006K-120470; National Research Foundation of Korea under Contract No. NRF-2022R1A2C1092335; National Science and Technology fund of Mongolia; National Science Research and Innovation Fund (NSRF) via the Program Management Unit for Human Resources \& Institutional Development, Research and Innovation of Thailand under Contract No. B16F640076; Polish National Science Centre under Contract No. 2019/35/O/ST2/02907; The Swedish Research Council; U. S. Department of Energy under Contract No. DE-FG02-05ER41374.

\newpage
{\bf \noindent The BESIII collaboration}\\
\\
{\small
M.~Ablikim$^{1}$, M.~N.~Achasov$^{4,c}$, P.~Adlarson$^{75}$, O.~Afedulidis$^{3}$, X.~C.~Ai$^{80}$, R.~Aliberti$^{35}$, A.~Amoroso$^{74A,74C}$, Q.~An$^{71,58,a}$, Y.~Bai$^{57}$, O.~Bakina$^{36}$
% ...
\\
%\vspace{0.2cm}
%(BESIII Collaboration)
%\vspace{0.2cm} 
{\it
$^{1}$ Institute of High Energy Physics, Beijing 100049, People's Republic of China\\
% ...
$^{3}$ Bochum  Ruhr-University, D-44780 Bochum, Germany\\
$^{4}$ Budker Institute of Nuclear Physics SB RAS (BINP), Novosibirsk 630090, Russia\\
% ...
$^{35}$ Johannes Gutenberg University of Mainz, Johann-Joachim-Becher-Weg 45, D-55099 Mainz, Germany\\
$^{36}$ Joint Institute for Nuclear Research, 141980 Dubna, Moscow region, Russia\\
% ...
$^{57}$ Southeast University, Nanjing 211100, People's Republic of China\\
$^{58}$ State Key Laboratory of Particle Detection and Electronics, Beijing 100049, Hefei 230026, People's Republic of China\\
% ...
$^{71}$ University of Science and Technology of China, Hefei 230026, People's Republic of China\\
% ...
$^{74}$ University of Turin and INFN, (A)University of Turin, I-10125, Turin, Italy; (B)University of Eastern Piedmont, I-15121, Alessandria, Italy; (C)INFN, I-10125, Turin, Italy\\
$^{75}$ Uppsala University, Box 516, SE-75120 Uppsala, Sweden\\
% ...
$^{80}$ Zhengzhou University, Zhengzhou 450001, People's Republic of China\\
\vspace{0.2cm}
$^{a}$ Deceased\\
% ...
$^{c}$ Also at the Novosibirsk State University, Novosibirsk, 630090, Russia\\
% ...
}}

\begin{thebibliography}{99}

\bibitem{Brambilla:2010cs}
N.~Brambilla, \textit{et al.}
\textit{Heavy quarkonium: progress, puzzles, and opportunities},
\textcolor{blue}{\href{https://link.springer.com/article/10.1140/epjc/s10052-010-1534-9} {\textit{Eur. Phys. J. C} \textbf{71} (2011) 1534}} 
[\textcolor{blue}{\href{https://arxiv.org/abs/1010.5827}{arXiv:hep-ph/1010.5827}}]
[\textcolor{blue}{\href{https://inspirehep.net/literature/874793}{\textsc{inSPIRE}}}].




\bibitem{Briceno:2015rlt}
R.~A.~Briceno, \textit{et al.}
\textit{Issues and opportunities in exotic hadrons}, 
\textcolor{blue}{\href{https://iopscience.iop.org/article/10.1088/1674-1137/40/4/042001} {\textit{Chin. Phys. C} \textbf{40} (2016) 042001}} 
[\textcolor{blue}{\href{https://arxiv.org/abs/1511.06779}{arXiv:1511.06779}}]
[\textcolor{blue}{\href{https://inspirehep.net/literature/1405969}{\textsc{inSPIRE}}}].






\bibitem{Barnes:2005pb}
T.~Barnes, S.~Godfrey and E.~S.~Swanson,
\textit{Higher charmonia}, 
\textcolor{blue}{\href{https://doi.org/10.1103/PhysRevD.72.054026} {\textit{Phys. Rev. D} \textbf{72} (2005) 054026}} 
[\textcolor{blue}{\href{https://doi.org/10.48550/arXiv.hep-ph/0505002}{arXiv:hep-ph/0505002}}]
[\textcolor{blue}{\href{https://inspirehep.net/search?p=find+EPRINT\%2BarXiv\%3Ahep-ph/0505002}{\textsc{inSPIRE}}}].

\bibitem{BES:2001ckj}
BES Collaboration,
\textit{Measurements of the cross section for $e^+ e^- \to hadrons$  at center-of-mass energies from 2 GeV to 5 GeV}, 
\textcolor{blue}{\href{https://doi.org/10.1103/PhysRevLett.88.101802} {\textit{Phys. Rev. Lett.} \textbf{88} (2002) 101802}}
[\textcolor{blue}{\href{https://arxiv.org/abs/hep-ex/0102003}{arXiv:hep-ex/0102003}}] 
[\textcolor{blue}{\href{https://inspirehep.net/search?p=find+EPRINT\%2BarXiv\%3Ahep-ex/0102003}{\textsc{inSPIRE}}}].

\bibitem{BaBar:2005hhc}
\textsc{BaBar}  Collaboration,
\textit{Observation of a broad structure in the $\pi^+ \pi^- J/\psi$ mass spectrum around 4.26 GeV/c$^2$}, 
\textcolor{blue}{\href{https://doi.org/10.1103/PhysRevLett.95.142001} {\textit{Phys. Rev. Lett.} \textbf{95} (2005) 142001}}
[\textcolor{blue}{\href{https://arxiv.org/abs/hep-ex/0506081}{arXiv:hep-ex/0506081}}]
[\textcolor{blue}{\href{https://inspirehep.net/search?p=find+EPRINT\%2BarXiv\%3Ahep-ex/0506081}{\textsc{inSPIRE}}}].

\bibitem{BaBar:2006ait}
\textsc{BaBar}  Collaboration,
\textit{Evidence of a broad structure at an invariant mass of 4.32 $GeV/c^{2}$ in the reaction $e^{+} e^{-} \to \pi^{+} \pi^{-} \psi(2S)$ measured at \textsc{BaBar}}, 
\textcolor{blue}{\href{https://doi.org/10.1103/PhysRevLett.98.212001} {\textit{Phys. Rev. Lett.} \textbf{98} (2007) 212001}} 
[\textcolor{blue}{\href{https://arxiv.org/abs/hep-ex/0610057}{arXiv:hep-ex/0610057}}] 
[\textcolor{blue}{\href{https://inspirehep.net/search?p=find+EPRINT\%2BarXiv\%3Ahep-ex/0610057}{\textsc{inSPIRE}}}].

\bibitem{Belle:2007umv}
Belle Collaboration,
\textit{Observation of two resonant structures in $e^+e^-\to\pi^+\pi^-\psi(2S)$ via initial state radiation at Belle}, 
\textcolor{blue}{\href{https://doi.org/10.1103/PhysRevLett.99.142002} {\textit{Phys. Rev. Lett.} \textbf{99} (2007) 142002}} 
[\textcolor{blue}{\href{https://arxiv.org/abs/0707.3699}{arXiv: 0707.3699}}] 
[\textcolor{blue}{\href{https://inspirehep.net/search?p=find+EPRINT\%2BarXiv\%3A 0707.3699 }{\textsc{inSPIRE}}}].

\bibitem{Belle:2007dxy}
Belle Collaboration,
\textit{Measurement of $e^+e^-\to \pi^+\pi^- J/\psi$ cross section via initial state radiation at Belle}, 
\textcolor{blue}{\href{https://doi.org/10.1103/PhysRevLett.99.182004} {\textit{Phys. Rev. Lett.} \textbf{99} (2007) 182004}} 
[\textcolor{blue}{\href{https://arxiv.org/abs/0707.2541}{arXiv:0707.2541}}] 
[\textcolor{blue}{\href{https://inspirehep.net/search?p=find+EPRINT\%2BarXiv\%3A0707.2541 }{\textsc{inSPIRE}}}].

\bibitem{Belle:2008xmh}
Belle Collaboration,
\textit{Observation of a near-threshold enhancement in the $e^+e^-\to\Lambda^+_{c}\bar\Lambda^{-}_{c}$ cross section using initial-state radiation}, 
\textcolor{blue}{\href{https://doi.org/10.1103/PhysRevLett.101.172001} {\textit{Phys. Rev. Lett.} \textbf{101} (2008) 172001}} 
[\textcolor{blue}{\href{https://arxiv.org/abs/0807.4458}{arXiv:0807.4458}}] 
[\textcolor{blue}{\href{https://inspirehep.net/search?p=find+EPRINT\%2BarXiv\%3A0807.4458 }{\textsc{inSPIRE}}}].

\bibitem{BaBar:2012vyb}
\textsc{BaBar}  Collaboration,
\textit{Study of the reaction $e^{+}e^{-} \to J/\psi\pi^{+}\pi^{-}$ via initial-state radiation at \textsc{BaBar}}, 
\textcolor{blue}{\href{https://doi.org/10.1103/PhysRevD.86.051102} {\textit{Phys.Rev.D} \textbf{86} (2012) 051102}} 
[\textcolor{blue}{\href{https://arxiv.org/abs/1204.2158}{arXiv:1204.2158}}] 
[\textcolor{blue}{\href{https://inspirehep.net/search?p=find+EPRINT\%2BarXiv\%3A1204.2158 }{\textsc{inSPIRE}}}].

\bibitem{Belle:2013yex}
Belle Collaboration,
\textit{Study of $e^+e^-\to\pi^+\pi^-J/\psi$ and observation of a charged charmonium-like state at Belle}, 
\textcolor{blue}{\href{https://doi.org/10.1103/PhysRevLett.111.019901} {\textit{Phys. Rev. Lett.} \textbf{111} (2013) 019901}} 
[\textcolor{blue}{\href{https://arxiv.org/abs/1304.0121}{arXiv:1304.0121}}] 
[\textcolor{blue}{\href{https://inspirehep.net/search?p=find+EPRINT\%2BarXiv\%3A1304.0121 }{\textsc{inSPIRE}}}].

\bibitem{BaBar:2012hpr}
\textsc{BaBar}  Collaboration,
\textit{Study of the reaction $e^{+}e^{-}\to \psi(2S)\pi^{+}\pi^{-}$ via initial-state radiation at \textsc{BaBar}}, 
\textcolor{blue}{\href{https://doi.org/10.1103/PhysRevD.89.111103} {\textit{Phys. Rev. D} \textbf{89} (2014)  111103}} 
[\textcolor{blue}{\href{https://arxiv.org/abs/1211.6271}{arXiv:1211.6271}}] 
[\textcolor{blue}{\href{https://inspirehep.net/search?p=find+EPRINT\%2BarXiv\%3A1211.6271 }{\textsc{inSPIRE}}}].

\bibitem{Belle:2014wyt}
Belle Collaboration,
\textit{Measurement of $e^+e^- \to \pi^+\pi^-\psi(2S)$ via Initial State Radiation at Belle}, 
\textcolor{blue}{\href{https://doi.org/10.1103/PhysRevD.91.112007} {\textit{Phys. Rev. D} \textbf{91} (2015) 112007}} 
[\textcolor{blue}{\href{https://arxiv.org/abs/1410.7641}{arXiv:1410.7641}}] 
[\textcolor{blue}{\href{https://inspirehep.net/search?p=find+EPRINT\%2BarXiv\%3A1410.7641 }{\textsc{inSPIRE}}}].

\bbt{CLEO}
CLEO Collaboration,
\textit{Charmonium decays of $Y(4260)$, $\psi(4160)$ and $\psi(4040)$}, 
\textcolor{blue}{\href{https://doi.org/10.1103/PhysRevLett.96.162003} {\textit{Phys. Rev. Lett.} \textbf{96} (2006) 162003}} 
[\textcolor{blue}{\href{https://arxiv.org/abs/hep-ex/0602034}{arXiv:hep-ex/0602034}}] 
[\textcolor{blue}{\href{https://inspirehep.net/search?p=find+EPRINT\%2BarXiv\%3Ahep-ex/0602034 }{\textsc{inSPIRE}}}].

\bbt{BESIIIAB}
BESIII Collaboration,
\textit{Study of $e^+e^-\to\omega\chi_{cJ}$ at center-of-mass energies from 4.21 to 4.42 GeV}, 
\textcolor{blue}{\href{https://doi.org/10.1103/PhysRevLett.114.092003} {\textit{Phys. Rev. Lett.} \textbf{114} (2015)  092003}} 
[\textcolor{blue}{\href{https://arxiv.org/abs/1410.6538}{arXiv:1410.6538}}] 
[\textcolor{blue}{\href{https://inspirehep.net/search?p=find+EPRINT\%2BarXiv\%3A1410.6538 }{\textsc{inSPIRE}}}].

\bibitem{BESIII:2023cmv}
BESIII Collaboration,
\textit{Observation of three charmonium-like states with ${J}^{PC}={1}^{\ensuremath{-}\ensuremath{-}}$ in ${e}^{+}{e}^{\ensuremath{-}}\ensuremath{\rightarrow}{D}^{*0}{D}^{*\ensuremath{-}}{\ensuremath{\pi}}^{+}$},
\textcolor{blue}{\href{https://doi.org/10.1103/PhysRevLett.130.121901} {\textit{Phys. Rev. Lett.} \textbf{130} (2023) 121901}} 
[\textcolor{blue}{\href{https://arxiv.org/abs/2301.07321}{arXiv:2301.07321}}]
[\textcolor{blue}{\href{https://inspirehep.net/search?p=find+EPRINT\%2BarXiv\%3A2301.07321}{\textsc{inSPIRE}}}].

\bibitem{Close:2005iz}
F.~E.~Close and P.~R.~Page,
\textit{Gluonic charmonium resonances at \textsc{BaBar} and BELLE}, 
\textcolor{blue}{\href{https://doi.org/10.1016/j.physletb.2005.09.016} {\textit{Phys. Lett. B} \textbf{628} (2005) 215}} 
[\textcolor{blue}{\href{https://arxiv.org/abs/hep-ph/0507199}{arXiv:hep-ph/0507199}}] 
[\textcolor{blue}{\href{https://inspirehep.net/search?p=find+EPRINT\%2BarXiv\%3Ahep-ph/0507199 }{\textsc{inSPIRE}}}].

\bbt{Farrar} 
N.~Brambilla, \textit{et al.}
\textit{Heavy quarkonium: progress, puzzles, and opportunities}, 
\textcolor{blue}{\href{https://doi.org/10.1140/epjc/s10052-010-1534-9} {\textit{Eur. Phys. J. C} \textbf{71} (2011) 1534}} 
[\textcolor{blue}{\href{https://arxiv.org/abs/1010.5827}{arXiv:1010.5827}}] 
[\textcolor{blue}{\href{https://inspirehep.net/search?p=find+EPRINT\%2BarXiv\%3A1010.5827 }{\textsc{inSPIRE}}}].

\bbt{Briceno}
R.~A.~Briceno, \textit{et al.}
\textit{Issues and opportunities in exotic hadrons}, 
\textcolor{blue}{\href{https://doi.org/10.1088/1674-1137/40/4/042001} {\textit{Chin. Phys. C} \textbf{40} (2016)  042001}} 
[\textcolor{blue}{\href{https://arxiv.org/abs/1511.06779}{arXiv:1511.06779}}] 
[\textcolor{blue}{\href{https://inspirehep.net/search?p=find+EPRINT\%2BarXiv\%3A1511.06779  }{\textsc{inSPIRE}}}].

\bibitem{Chen:2016qju}
H.~X.~Chen, W.~Chen, X.~Liu and S.~L.~Zhu,
\textit{The hidden-charm pentaquark and tetraquark states}, 
\textcolor{blue}{\href{https://doi.org/10.1016/j.physrep.2016.05.004} {\textit{Phys. Rept.} \textbf{639} (2016) 1-121}} 
[\textcolor{blue}{\href{https://arxiv.org/abs/1601.02092}{arXiv:1601.02092 }}]
[\textcolor{blue}{\href{https://inspirehep.net/search?p=find+EPRINT\%2BarXiv\%3A1601.02092  }{\textsc{inSPIRE}}}].

\bibitem{Wang:2019mhs}
J.~Z.~Wang, D.Y.~Chen, X.~Liu and T.~Matsuki,
\textit{Constructing $J/\psi$ family with updated data of charmoniumlike $Y$ states},
\textcolor{blue}{\href{https://doi.org/10.1103/PhysRevD.99.114003} {\textit{Phys. Rev. D} \textbf{99} (2019) 114003}} 
[\textcolor{blue}{\href{https://arxiv.org/abs/1903.07115}{arXiv:1903.07115}}]
[\textcolor{blue}{\href{https://inspirehep.net/literature?sort=mostrecent&size=25&page=1&q=Phys.Rev.D\%2099\%20\%282019\%29\%2011\%2C\%20114003}{\textsc{inSPIRE}}}].

\bibitem{Qian:2021neg} 
R.~Q.~Qian, Q.~Huang and X.~Liu, 
\textit{Predicted $\Lambda\bar\Lambda$ and $\Xi^-\bar\Xi^{+}$ decay modes of the charmoniumlike $Y(4230)$}, 
\textcolor{blue}{\href{https://doi.org/10.1016/j.physletb.2022.137292} {\textit{Phys. Lett. B} \textbf{833} (2022) 137292}} 
[\textcolor{blue}{\href{https://arxiv.org/abs/2111.13821}{arXiv:2111.13821}}] 
[\textcolor{blue}{\href{https://inspirehep.net/search?p=find+EPRINT\%2BarXiv\%3A2111.13821 }{\textsc{inSPIRE}}}].



\bibitem{Ablikim:2013pgf} 
BESIII Collaboration,
\textit{Search for baryonic decays of $\psi(3770)$ and $\psi(4040)$}, 
\textcolor{blue}{\href{https://doi.org/10.1103/PhysRevD.87.112011} {\textit{Phys. Rev. D} \textbf{87} (2013)  112011}} 
[\textcolor{blue}{\href{https://arxiv.org/abs/1305.1782}{arXiv:1305.1782}}] 
[\textcolor{blue}{\href{https://inspirehep.net/search?p=find+EPRINT\%2BarXiv\%3A1305.1782 }{\textsc{inSPIRE}}}].

\bibitem{BESIII:2017kqg}
BESIII Collaboration,
\textit{Precision measurement of the $e^{+}e^{-}~\rightarrow~\Lambda_{c}^{+} \bar{\Lambda}_{c}^{-}$ cross section near threshold}, 
\textcolor{blue}{\href{https://doi.org/10.1103/PhysRevLett.120.132001} {\textit{Phys. Rev. Lett.} \textbf{120} (2018)  132001}} 
[\textcolor{blue}{\href{https://arxiv.org/abs/1710.00150}{arXiv:1710.00150}}] 
[\textcolor{blue}{\href{https://inspirehep.net/search?p=find+EPRINT\%2BarXiv\%3A1710.00150 }{\textsc{inSPIRE}}}].

\bibitem{Ablikim:2019kkp} 
BESIII Collaboration,
\textit{Measurement of the Cross Section for $e^+e^-\to\Xi^-\bar{\Xi}^+$ and Observation of an Excited $\Xi$ Baryon}, 
\textcolor{blue}{\href{https://doi.org/10.1103/PhysRevLett.124.032002} {\textit{Phys. Rev. Lett.} \textbf{124} (2020)  032002}} 
[\textcolor{blue}{\href{https://arxiv.org/abs/1910.04921}{arXiv:1910.04921}}] 
[\textcolor{blue}{\href{https://inspirehep.net/search?p=find+EPRINT\%2BarXiv\%3A1910.04921 }{\textsc{inSPIRE}}}].

\bibitem{Wang:2021lfq}
BESIII Collaboration,
\textit{Study of baryon pair production at BESIII}, 
\textcolor{blue}{\href{https://doi.org/10.22323/1.385.0026} {\textit{PoS} \textbf{CHARM2020} 026 (2021)}} 
[\textcolor{blue}{\href{https://inspirehep.net/literature/1926588}{\textsc{inSPIRE}}}].

\bibitem{BESIII:2021ccp} 
BESIII Collaboration,
\textit{Measurement of the cross section for $e^+e^-\to\Lambda\bar{\Lambda}$ and evidence of the decay $\psi(3770)\to\Lambda\bar{\Lambda}$}, 
\textcolor{blue}{\href{https://doi.org/10.1103/PhysRevD.104.L091104} {\textit{Phys. Rev. D} \textbf{104} (2021)  L091104}} 
[\textcolor{blue}{\href{https://arxiv.org/abs/2108.02410}{arXiv:2108.02410}}] 
[\textcolor{blue}{\href{https://inspirehep.net/search?p=find+EPRINT\%2BarXiv\%3A2108.02410 }{\textsc{inSPIRE}}}].

\bibitem{Wang:2022bzl}
BESIII Collaboration,
\textit{Measurement of Energy-Dependent Pair-Production Cross Section and Electromagnetic Form Factors of a Charmed Baryon}, 
\textcolor{blue}{\href{https://journals.aps.org/prl/abstract/10.1103/PhysRevLett.131.191901} {\textit{Phys. Rev. Lett.}  \textbf{131} (2023) 91901}} 
[\textcolor{blue}{\href{https://arxiv.org/abs/2307.07316}{arXiv:2307.07316}}] 
[\textcolor{blue}{\href{https://inspirehep.net/literature/2090070}{\textsc{inSPIRE}}}].



\bibitem{BESIII:2023rse}
BESIII Collaboration,
\textit{Measurement of the cross section of $e^+e^-\rightarrow\Xi^{-}\bar\Xi^{+}$ at center-of-mass energies between 3.510 and 4.843 GeV}, 
\textcolor{blue}{\href{https://link.springer.com/article/10.1007/JHEP11(2023)228} {\textit{JHEP} \textbf{11} (2023)  228}} 
[\textcolor{blue}{\href{https://arxiv.org/abs/2309.04215}{arXiv:2309.04215}}] 
[\textcolor{blue}{\href{https://inspirehep.net/literature/2695411}{\textsc{inSPIRE}}}].




\bibitem{BESIII:2024umc}
BESIII Collaboration,
\textit{Measurement of Born cross section of $ {e}^{+}{e}^{-}\to {\Sigma}^{+}{\overline{\Sigma}}^{-} $ at center-of-mass energies between 3.510 and 4.951 GeV},
\textcolor{blue}{\href{https://link.springer.com/article/10.1007/JHEP05(2024)022} {\textit{JHEP} \textbf{05} (2024) 022}}[\textcolor{blue}{\href{https://arxiv.org/abs/2401.09468}{arXiv:2401.09468}}][\textcolor{blue}{\href{https://inspirehep.net/literature/2748736}{\textsc{inSPIRE}}}].


\bibitem{BESIII:2022kzc}
BESIII Collaboration,
\textit{Study of $e^+e^-\rightarrow\Omega^{-}\bar\Omega^{+}$ at center-of-mass energies from 3.49 to 3.67 GeV}, 
\textcolor{blue}{\href{https://doi.org/10.1103/PhysRevD.107.052003} {\textit{Phys. Rev. D} \textbf{107} (2023)  052003}} 
[\textcolor{blue}{\href{https://arxiv.org/abs/2212.03693}{arXiv:2212.03693}}]
[\textcolor{blue}{\href{https://inspirehep.net/search?p=find+EPRINT\%2BarXiv\%3A2212.03693  }{\textsc{inSPIRE}}}].


\bibitem{Wang:2022zyc}
X.~Wang and G.~Huang,
\textit{Electromagnetic Form Factor of Doubly-Strange Hyperon},
\textcolor{blue}{\href{https://www.mdpi.com/2073-8994/14/1/65} {\textit{Symmetry} \textbf{14} (2022) 65}}
[\textcolor{blue}{\href{https://inspirehep.net/literature/2037456}{\textsc{inSPIRE}}}].






\bibitem{ene1}
BESIII Collaboration,
\textit{Precision measurement of the integrated luminosity of the data taken by BESIII at center of mass energies between 3.810 GeV and 4.600 GeV}, 
\textcolor{blue}{\href{https://doi.org/10.1088/1674-1137/39/9/093001} {\textit{Chin. Phys. C} \textbf{39} (2015)  093001}} 
[\textcolor{blue}{\href{https://arxiv.org/abs/1503.03408}{arXiv:1503.03408 }}]
[\textcolor{blue}{\href{https://inspirehep.net/search?p=find+EPRINT\%2BarXiv\%3A1503.03408  }{\textsc{inSPIRE}}}].

\bibitem{ene3}
BESIII Collaboration,
\textit{Luminosities and energies of $e^+ e^-$ collision data taken between 4.61 GeV and 4.95 GeV at BESIII}, 
\textcolor{blue}{\href{https://doi.org/10.1088/1674-1137/ac84cc} {\textit{Chin. Phys. C} \textbf{46} (2022)  113003}} 
[\textcolor{blue}{\href{https://arxiv.org/abs/2205.04809}{arXiv:2205.04809}}]
[\textcolor{blue}{\href{https://inspirehep.net/search?p=find+EPRINT\%2BarXiv\%3A2205.04809 }{\textsc{inSPIRE}}}].

\bbt{besiii} 
BESIII Collaboration,
\textit{Design and construction of the BESIII detector}, 
\textcolor{blue}{\href{https://doi.org/10.1016/j.nima.2009.12.050} {\textit{Nucl. Instrum. Meth. A} \textbf{614} (2010) 345-399}} 
[\textcolor{blue}{\href{https://arxiv.org/abs/0911.4960}{arXiv: 0911.4960}}] 
[\textcolor{blue}{\href{https://inspirehep.net/search?p=find+EPRINT\%2BarXiv\%3A 0911.4960 }{\textsc{inSPIRE}}}].

\bibitem{BEPCII} 
C. Yu \textit{et al.},
\textit{BEPCII performance and beam dynamics studies on luminosity}, 
\textcolor{blue}{\href{https://doi.org/10.18429/JACoW-IPAC2016-TUYA01} {\textit{ Proceedings, 7th International Particle Accelerator Conference (IPAC 2016)} May 8-13 2016 }} 
[\textcolor{blue}{\href{https://inspirehep.net/literature/1469857}{\textsc{inSPIRE}}}].



\bibitem{Ablikim:2019hff} 
BESIII Collaboration, 
\textit{Future physics programme of BESIII}, 
\textcolor{blue}{\href{https://doi.org/10.1088/1674-1137/44/4/040001} {\textit{Chin. Phys. C} \textbf{44} (2020)  040001} }
[\textcolor{blue}{\href{https://arxiv.org/abs/1912.05983}{arXiv:1912.05983}}] 
[\textcolor{blue}{\href{https://inspirehep.net/search?p=find+EPRINT\%2BarXiv\%3A1912.05983 }{\textsc{inSPIRE}}}].

\bibitem{EcmsMea}
J.~Lu, Y.~Xiao, and X.~Ji,
\textit{Online monitoring of the center-of-mass energy from real data at BESIII}, 
\textcolor{blue}{\href{https://doi.org/10.1007/s41605-020-00188-8} {\textit{Radiat Detect Technol Methods} \textbf{4}  (2020) 337–344}}. 

\bibitem{EventFilter}
J.~W.~Zhang, L.~H.~Wu, and S.~S.~Sun \textit{et al.},
\textit{Suppression of top-up injection backgrounds with offline event filter in the BESIII experiment}, 
\textcolor{blue}{\href{https://doi.org/10.1007/s41605-022-00331-7} {\textit{Radiat. Detect. Technol. Methods} {\textbf 6}  (2022) 289–293} } 
[\textcolor{blue}{\href{https://inspirehep.net/literature/2145899}{\textsc{inSPIRE}}}].

\bibitem{etof1}
X.~Li \textit{et al.},
\textit{Study of MRPC technology for BESIII endcap-TOF upgrade}, 
\textcolor{blue}{\href{https://doi.org/10.1007/s41605-017-0014-2} {\textit{Radiat. Detect. Technol. Methods} {\textbf 1} 13 (2017)}}. 

\bibitem{etof2}
Y.~X.~Guo \textit{et al.},
\textit{The study of time calibration for upgraded end cap TOF of BESIII}, 
\textcolor{blue}{\href{https://doi.org/10.1007/s41605-017-0012-4} {\textit{Radiat. Detect. Technol. Methods} {\textbf 1} 15 (2017)}}. 

\bibitem{etof3}
 P.~Cao \textit{et al.},
\textit{Design and construction of the new BESIII endcap Time-of-Flight system with MRPC Technology}, 
\textcolor{blue}{\href{https://doi.org/10.1016/j.nima.2019.163053} {\textit{Nucl. Instrum. Meth. A} \textbf{953} (2020) 163053 }} 
[\textcolor{blue}{\href{https://inspirehep.net/literature/1775466}{\textsc{inSPIRE}}}].


\bibitem{GEANT4} 
GEANT4 Collaboration,  
\textit{GEANT4---a simulation toolkit}, 
\textcolor{blue}{\href{https://doi.org/10.1016/S0168-9002(03)01368-8} {\textit{Nucl. Instrum. Meth. A} \textbf {506} (2003) 250}}.

\bibitem{Huang:2022wuo}
K.~X.~Huang, Z.~J.~Li, Z.~Qian, J.~Zhu, H.~Y.~Li, Y.~M.~Zhang, S.~S.~Sun and Z.~Y.~You,
\textit{Method for detector description transformation to unity and application in BESIII}, 
\textcolor{blue}{\href{https://doi.org/10.1007/s41365-022-01133-8} {\textit{Nucl. Sci. Tech.} \textbf{33} (2022) 142}} 
[\textcolor{blue}{\href{https://arxiv.org/abs/2206.10117}{arXiv:2206.10117}}]
[\textcolor{blue}{\href{https://inspirehep.net/search?p=find+EPRINT\%2BarXiv\%3A2206.10117 }{\textsc{inSPIRE}}}].

\bibitem{KKMC} 
S. Jadach, B. F. L. Ward and Z. Was,
\textit{Coherent exclusive exponentiation for precision Monte Carlo calculations}, 
\textcolor{blue}{ \href{https://doi.org/10.1103/PhysRevD.63.113009} {\textit{Phys. Rev. D} \textbf{63} (2001) 113009}} 
[\textcolor{blue}{\href{https://arxiv.org/abs/hep-ph/0006359}{arXiv:hep-ph/0006359}}] 
[\textcolor{blue}{\href{https://inspirehep.net/search?p=find+EPRINT\%2BarXiv\%3Ahep-ph/0006359 }{\textsc{inSPIRE}}}].

\bibitem{EVTGEN}
D. J. Lange,
\textit{The EvtGen particle decay simulation package}, 
\textcolor{blue}{\href{https://doi.org/10.1016/S0168-9002(01)00089-4} {\textit{Nucl. Instrum. Meth. A} \textbf{462} (2001) 152-155} }.

\bibitem{evtgen2} 
R. G. Ping,
\textit{Event generators at BESIII}, 
\textcolor{blue}{ \href{https://doi.org/10.1088/1674-1137/32/8/001} {\textit{Chin. Phys. C} \textbf{32} (2008) 599}}.



\bibitem{BESIII:2012ghz}
BESIII Collaboration,
\textit{Measurements of baryon pair decays of $\chi_{cJ}$ mesons}, 
\textcolor{blue}{ \href{https://doi.org/10.1103/PhysRevD.87.032007} {\textit{Phys. Rev. D} \textbf{87} (2013)  059901} }
[\textcolor{blue}{\href{https://arxiv.org/abs/1211.2283}{arXiv:1211.2283}}]
[\textcolor{blue}{\href{https://inspirehep.net/search?p=find+EPRINT\%2BarXiv\%3A1211.2283}{\textsc{inSPIRE}}}].

\bibitem{BESIII:2021cvv}
BESIII Collaboration,
\textit{Measurement of $\Lambda$ baryon polarization in $e^+e^-\rightarrow\Lambda\bar\Lambda$ at $\sqrt{s} = 3.773$ GeV}, 
\textcolor{blue}{\href{https://doi.org/10.1103/PhysRevD.105.L011101} {\textit{Phys. Rev. D} \textbf{105} (2022)  L011101}}
[\textcolor{blue}{\href{https://arxiv.org/abs/2111.11742}{arXiv:2111.11742}}] 
[\textcolor{blue}{\href{https://inspirehep.net/search?p=find+EPRINT\%2BarXiv\%3A2111.11742}{\textsc{inSPIRE}}}].


\bibitem{BESIII:2023euh}
BESIII Collaboration,
\textit{Measurement of $\Lambda$ transverse polarization in $e^{+}e^{-}$ collisions at $\sqrt{s}= 3.68-3.71$ GeV}, 
\textcolor{blue}{ \href{https://doi.org/10.1007/JHEP06(2022)074} {\textit{JHEP} \textbf{10} (2023) 81} }
[\textcolor{blue}{\href{https://arxiv.org/abs/2303.00271}{arXiv:2303.00271}}]
[\textcolor{blue}{\href{https://inspirehep.net/search?p=find+EPRINT\%2BarXiv\%3A2303.00271 }{\textsc{inSPIRE}}}].


\bibitem{vtxfit} 
M. Xu  \textit{et al.},
\textit{Decay vertex reconstruction and 3-dimensional lifetime determination at BESIII}, 
\textcolor{blue}{\href{https://doi.org/10.1088/1674-1137/33/6/005} {\textit{Chin. Phys. C} \textbf{33} (2009) 428}} 
[\textcolor{blue}{\href{https://inspirehep.net/literature/1122428}{\textsc{inSPIRE}}}].


\bbt{PDG2020} 
Particle Data Group,
\textit{Review of particle physics}, 
\textcolor{blue}{\href{https://doi.org/10.1093/ptep/ptac097} {\textit{PTEP} \textbf{2022} (2022) 083C01}}
[\textcolor{blue}{\href{https://inspirehep.net/literature/2106994}{\textsc{inSPIRE}}}].

\bibitem{Zhu:2008ca}
Y.~S.~Zhu,
\textit{Bayesian credible interval construction for Poisson statistics}, 
\textcolor{blue}{\href{https://doi.org/10.1088/1674-1137/32/5/007} {\textit{Chin. Phys. C} \textbf{32} (2008) 363}} 
[\textcolor{blue}{\href{https://arxiv.org/abs/0812.2705}{arXiv:0812.2705}}]
[\textcolor{blue}{\href{https://inspirehep.net/search?p=find+EPRINT\%2BarXiv\%3A0812.2705 }{\textsc{inSPIRE}}}].


 \bibitem{Kuraev:1985hb}
E.~A.~Kuraev and V.~S.~Fadin,
\textit{On radiative corrections to $e^+e^-$ single photon annihilation at high-energy}, 
\textcolor{blue}{ {\textit{Sov. J. Nucl. Phys.} \textbf{41} (1985) 466-472}} 
[\textcolor{blue}{\href{https://inspirehep.net/literature/217313}{\textsc{inSPIRE}}}].

\bibitem{Jegerlehner:2011ti}
F.~Jegerlehner and R.~Szafron,
\textit{$\rho^0-\gamma$ mixing in the neutral channel pion form factor $F_{\pi}^{e}$ and its role in comparing $e^+ e^-$ with $\tau$ spectral functions}, 
\textcolor{blue}{\href{https://doi.org/10.1140/epjc/s10052-011-1632-3} {\textit{Eur. Phys. J. C} \textbf{71} (2011) 1632}} 
[\textcolor{blue}{\href{https://arxiv.org/abs/1101.2872}{arXiv:1101.2872}}]
[\textcolor{blue}{\href{https://inspirehep.net/search?p=find+EPRINT\%2BarXiv\%3A1101.2872 }{\textsc{inSPIRE}}}].

\bibitem{Sun:2020ehv}
W.~Sun, T.~Liu, M.~Jing, L.~Wang, B.~Zhong, and W.~Song,
\textit{An iterative weighting method to apply ISR correction to $e^+ e^-$ hadronic cross section measurements}, 
\textcolor{blue}{\href{https://doi.org/10.1007/s11467-021-1085-6} {\textit{Front. Phys. (Beijing)} \textbf{16} (2021)  64501}} 
[\textcolor{blue}{\href{https://arxiv.org/abs/2011.07889}{arXiv:2011.07889}}]
[\textcolor{blue}{\href{https://inspirehep.net/search?p=find+EPRINT\%2BarXiv\%3A2011.07889  }{\textsc{inSPIRE}}}]. 



\bibitem{BESIII:2022dxl}
BESIII Collaboration,
\textit{Measurement of integrated luminosities at BESIII for data samples at center-of-mass energies between 4.0 and 4.6 GeV}, 
\textcolor{blue}{\href{https://doi.org/10.1088/1674-1137/ac80b4} {\textit{Chin. Phys. C} \textbf{46} (2022)  113002}} 
[\textcolor{blue}{\href{https://arxiv.org/abs/2203.03133}{arXiv:2203.03133}}]
[\textcolor{blue}{\href{https://inspirehep.net/search?p=find+EPRINT\%2BarXiv\%3A2203.03133  }{\textsc{inSPIRE}}}].

\bibitem{llll}
BESIII Collaboration,
\textit{Luminosities and energies of $e^+ e^-$ collision data taken between 4.61 GeV and 4.95 GeV at BESIII}, 
\textcolor{blue}{\href{https://doi.org/10.1088/1674-1137/ac84cc} {\textit{Chin. Phys. C} \textbf{46} (2022)  113003}} 
[\textcolor{blue}{\href{https://arxiv.org/abs/2205.04809}{arXiv:2205.04809}}]
[\textcolor{blue}{\href{https://inspirehep.net/search?p=find+EPRINT\%2BarXiv\%3A2205.04809 }{\textsc{inSPIRE}}}].



\bibitem{ksys}
BESIII Collaboration,
\textit{Measurements of $\psi(3686)\to K^-\Lambda\bar\Xi^+ + c.c.$ and $\psi(3686)\to \gamma K^-\Lambda\bar\Xi^+ + c.c.$}
\textcolor{blue}{\href{https://doi.org/10.1103/PhysRevD.91.092006} {\textit{Phys. Rev. D} \textbf{91} (2015)  092006}} 
[\textcolor{blue}{\href{https://arxiv.org/pdf/1504.02025.pdf}{arXiv:1504.02025}}]
[\textcolor{blue}{\href{https://inspirehep.net/literature/1358401}{\textsc{inSPIRE}}}].

\bibitem{BESIII:2016ssr}
BESIII Collaboration,
\textit{Study of $\psi$ decays to the $\Xi^{-}\bar\Xi^{+}$ and $\Sigma(1385)^{\mp}\bar\Sigma(1385)^{\pm}$ final states}, 
\textcolor{blue}{\href{https://doi.org/10.1103/PhysRevD.93.072003} {\textit{Phys. Rev. D} \textbf{93} (2016)  072003} }
[\textcolor{blue}{\href{https://arxiv.org/abs/1602.06754}{arXiv:1602.06754}}]
[\textcolor{blue}{\href{https://inspirehep.net/search?p=find+EPRINT\%2BarXiv\%3A1602.06754 }{\textsc{inSPIRE}}}].

\bibitem{BESIII:2016nix}
BESIII Collaboration,
\textit{Study of $J/\psi$ and $\psi(3686)\rightarrow\Sigma(1385)^{0}\bar\Sigma(1385)^{0}$ and $\Xi^0\bar\Xi^{0}$}, 
\textcolor{blue}{\href{https://doi.org/10.1016/j.physletb.2017.04.048} {\textit{Phys. Lett. B} \textbf{770} (2017) 217} }
[\textcolor{blue}{\href{https://arxiv.org/abs/1612.08664}{arXiv:1612.08664}}]
[\textcolor{blue}{\href{https://inspirehep.net/search?p=find+EPRINT\%2BarXiv\%3A1612.08664  }{\textsc{inSPIRE}}}].

\bibitem{BESIII:2019dve}
BESIII Collaboration,
\textit{Observation of $\psi(3686)\rightarrow\Xi(1530)^{-}\bar\Xi(1530)^{+}$ and $\Xi(1530)^{-}\bar\Xi^{+}$}, 
\textcolor{blue}{\href{https://doi.org/10.1103/PhysRevD.100.051101} {\textit{Phys. Rev. D} \textbf{100} (2019)  051101} }
[\textcolor{blue}{\href{https://arxiv.org/abs/1907.13041}{arXiv:1907.13041}}]
[\textcolor{blue}{\href{https://inspirehep.net/search?p=find+EPRINT\%2BarXiv\%3A1907.13041 }{\textsc{inSPIRE}}}].
 
\bibitem{BESIII:2020ktn} 
BESIII Collaboration,
\textit{Measurement of cross section for $e^+e^-\to\Xi^-\bar{\Xi}^+$ near threshold at BESIII}, 
\textcolor{blue}{\href{https://doi.org/10.1103/PhysRevD.103.012005} {\textit{Phys. Rev. D} \textbf{103} (2021)  012005} }
[\textcolor{blue}{\href{https://arxiv.org/abs/2010.08320}{arXiv:2010.08320}}]
[\textcolor{blue}{\href{https://inspirehep.net/search?p=find+EPRINT\%2BarXiv\%3A2010.08320  }{\textsc{inSPIRE}}}].

\bibitem{BESIII:2021aer} 
BESIII Collaboration,
\textit{Measurement of cross section for $e^+e^-\to\Xi^0\bar{\Xi}^0$ near threshold}, 
\textcolor{blue}{ \href{https://doi.org/10.1016/j.physletb.2021.136557} {\textit{Phys. Lett. B} \textbf{820} (2021) 136557} }
[\textcolor{blue}{\href{https://arxiv.org/abs/2105.14657}{arXiv:2105.14657}}]
[\textcolor{blue}{\href{https://inspirehep.net/search?p=find+EPRINT\%2BarXiv\%3A2105.14657 }{\textsc{inSPIRE}}}].

\bibitem{BESIII:2021gca}
BESIII Collaboration, 
\textit{Observation of $\psi(3686)\to\Xi(1530)^{0}\bar{\Xi}(1530)^{0}$ and $\Xi(1530)^{0}\bar{\Xi}^0$}, 
\textcolor{blue}{ \href{https://doi.org/10.1103/PhysRevD.104.092012} {\textit{Phys. Rev. D} \textbf{104} (2021)  092012} }
[\textcolor{blue}{\href{https://arxiv.org/abs/2109.06621}{arXiv:2109.06621}}]
[\textcolor{blue}{\href{https://inspirehep.net/search?p=find+EPRINT\%2BarXiv\%3A2109.06621 }{\textsc{inSPIRE}}}].



\bibitem{BESIII:2022mfx}
BESIII Collaboration,
\textit{Study of the processes $\chi_{cJ} \to \Xi^- \bar{\Xi}^+$ and $\Xi^0 \bar{\Xi}^0$}, 
\textcolor{blue}{ \href{https://doi.org/10.1007/JHEP06(2022)074} {\textit{JHEP} \textbf{06} (2022) 74} }
[\textcolor{blue}{\href{https://arxiv.org/abs/2202.08058}{arXiv:2202.08058}}]
[\textcolor{blue}{\href{https://inspirehep.net/search?p=find+EPRINT\%2BarXiv\%3A2202.08058 }{\textsc{inSPIRE}}}].

\bibitem{BESIII:2022lsz}
BESIII Collaboration,
\textit{Observation of $\Xi^-$ hyperon transverse polarization in $\psi(3686)\to\Xi^{-}\bar\Xi^{+}$}, 
\textcolor{blue}{\href{https://doi.org/10.1103/PhysRevD.106.L091101} {\textit{Phys. Rev. D} \textbf{106} (2022)  L091101}} 
[\textcolor{blue}{\href{https://arxiv.org/abs/2206.10900}{arXiv:2206.10900}}]
[\textcolor{blue}{\href{https://inspirehep.net/search?p=find+EPRINT\%2BarXiv\%3A2206.10900 }{\textsc{inSPIRE}}}].

\bibitem{BESIII:2023lkg}
BESIII Collaboration,
\textit{First simultaneous measurement of $\Xi^0$ and $\bar{\Xi}^0$ asymmetry parameters in $\psi(3686)$ decay}, 
\textcolor{blue}{\href{https://doi.org/10.1103/PhysRevD.106.L091101} {\textit{Phys. Rev. D} \textbf{108} (2023)  L011101}} 
[\textcolor{blue}{\href{https://arxiv.org/abs/2302.09767}{arXiv:2302.09767}}]
[\textcolor{blue}{\href{https://inspirehep.net/search?p=find+EPRINT\%2BarXiv\%3A2302.09767}{\textsc{inSPIRE}}}].








% \bibitem{Lundberg:2009iu}
% J.~Lundberg, J.~Conrad, W.~Rolke, and A.~Lopez,
% \textit{Limits, discovery and cut optimization for a Poisson process with uncertainty in background and signal efficiency: TRolke 2.0}, 
% \textcolor{blue}{\href{https://doi.org/10.1016/j.cpc.2009.11.001} {\textit{Comput. Phys. Commun.} \textbf{181} (2010) 683}} 
% [\textcolor{blue}{\href{https://arxiv.org/abs/0907.3450}{arXiv:0907.3450}}]
% [\textcolor{blue}{\href{https://inspirehep.net/search?p=find+EPRINT\%2BarXiv\%3A0907.3450 }{\textsc{inSPIRE}}}].











\end{thebibliography}
\end{document}